\documentclass[final,5p,times,twocolumn]{elsarticle}

\usepackage{amssymb}
\usepackage{amsmath}
\usepackage{graphicx} 
\usepackage{setspace}             
\usepackage{etoolbox}             

\usepackage{amssymb}
\usepackage[utf8]{inputenc}
\usepackage{amsmath}
\usepackage[dvipsnames]{xcolor}
\usepackage{adjustbox} 
\usepackage{makecell}  
\usepackage{float}     
\usepackage{siunitx}   
\usepackage{float}
\usepackage{graphicx} 
\usepackage{subcaption}
\usepackage{soul}
\usepackage{url}
\usepackage{siunitx}
\usepackage{physics}
\usepackage{stfloats}
\usepackage{tabularray}
\usepackage{cuted}
\usepackage{booktabs}
\usepackage{array}
\usepackage{ragged2e}
\usepackage{hyperref}

\newcommand{\blue}[1]{\textcolor{blue}{#1}}

\journal{Nuclear Physics B}

\usepackage{placeins}

\begin{document}
	\raggedbottom
	
	\begin{frontmatter}
		
		\title{Spatial effects of Kuramoto modeling on the power-grid frequencies} 
		
		\author[ek,bme]{Krist\'of Benedek}
		
		\author[ek]{Michelle Cirunay}
		
		\author[ek]{Sangita Dutta}
		
		\author[de]{Attila Balogh }
		
		\author[ekbi]{B\'alint Hartmann}
		
		\author[ek]{G\'eza \'Odor}
		
		\affiliation[ek]{organization={Institute of Technical Physics and Materials Science, Centre for Energy Research},
			addressline={P.O. Box 49, Konkoly-Thege Miklós út 29-33, 1121}, 
			city={Budapest},
			postcode={H-1525}, 
			country={Hungary}}
		\affiliation[bme]{organization={Department of Theoretical Physics, Institute of Physics, Budapest University of Technology and Economics},
			addressline={Műegyetem rkp. 3.}, 
			city={Budapest},
			postcode={H-1111}, 
			country={Hungary}}
		
		\affiliation[de]{organization={Department of Physics, Institute of Physics, University of Debrecen,},
			addressline={Bem t\'er 18.}, 
			city={Debrecen},
			postcode={H-4026}, 
			country={Hungary}}
		
		\affiliation[ekbi]{organization={Institute of Energy Security and Environmental Safety, HUN-REN Centre for Energy Research}, 
			addressline={P.O. Box 49, Konkoly-Thege Miklós út 29-33, 1121},
			city={Budapest}, 
			postcode={ H-1525},
			country={Hungary},
		}%
		
		\begin{abstract}
			Spatial coupling and heterogeneity of transmission models can generate complex power-grid frequency dynamics. In this work, we investigate which features can be attributed to the spatial aspects of the system, and which to noise or other processes (for eg. market changes and power dispatches) when we use the Kuramoto model (KM) with and without uncorrelated noise as a mathematical framework to model the real power grid. Since the KM has inherent chaoticity, generated by the nonlinearities, it is expected to introduce nontrivial behavior of local frequency dynamics that could not be described by an aggregated single-oscillator description with colored noise~\cite{KraljicIEEETRANS2022, Gorjao_2021, schaefer2018nongaussian,lonardi2026understandingcomplexityfrequencyphase}.
			
			To support our findings with real data, we analyse one month of \SI{100}{\milli\s} frequency measurements from sixteen Hungarian transmission-grid sites and compare them to the KM on a 387-node transmission network~\cite{hartmann_12models} and on two-dimensional lattices. The measured spectra exhibit two scaling regimes, with exponents of approximately $1.63$ and $2.23$. The one-second frequency increments decorrelate rapidly, whereas the correlations of their local inverse variance decay slowly, initially as a power law, $\tau^{-0.26(2)}$, up to lags of $15$ minutes. Using superstatistics~\cite{BeckCONTMECH2004}, we show that a Gaussian mixture constructed from the measured local variances captures much of the non-Gaussian increment distribution, but leaves the sparse extreme events unexplained.
			
			We also study the autocorrelation functions of the frequency time series and provide an estimate for the non-stationary, long-time autocorrelation exponent of the Hungarian power grid. Solutions of the KM on low-dimensional networks provide numerical evidence for the algebraic decay of long-time autocorrelations and power spectra, similar to the measurements, with exponents satisfying the Wiener-Khinchin relation in the steady state. On the heterogeneous transmission network, KM also develops persistent frequency patterns and multi-peaked distributions through partial synchronization and frequency locking. However, neither deterministic dynamics nor the tested independent Gaussian driving reproduces exactly the measured spectra and increment statistics simultaneously. In the case of the empirical data, the measured frequency variance is shared across substations, while simulated fluctuations predominantly reflect relative motion between nodes. These results provide complementary reasons for the observed nontrivial behavior of empirical frequency data, arising from the dynamical fluctuations of the swing equations on low-dimensional spatial structures besides the external driving due to the 15-minute dispatches.
		\end{abstract}
		
		
		
		
	\end{frontmatter}

	\begin{table*}[!t]
		\centering
		\footnotesize
		\setlength{\tabcolsep}{3pt}
		\renewcommand{\arraystretch}{1.12}
		\begin{minipage}[t]{0.49\textwidth}
			\begin{tabular}[t]{@{}>{\RaggedRight\arraybackslash}p{0.2\textwidth}>{\RaggedRight\arraybackslash}p{0.78\textwidth}@{}}
				\toprule
				\textbf{Symbol} & \textbf{Meaning} \\
				\midrule
				\multicolumn{2}{@{}l}{\textit{Indices, times and windows}} \\
				$i,j$ & node (model) or measurement-site index \\
				$N$, $L$ & number of model nodes ($387$ for HU387, $L^2$ for an $L\times L$ lattice) \\
				$n$ & number of measurement sites in the mean $\bar f$ ($n=9$) \\
				$t$, $t_0$ & time; reference time of two-time correlations \\
				$\tau$ & time lag of increments ($\Delta f_\tau$, $\Delta\omega_\tau$) and of correlation functions ($C(\tau)$) \\
				$\delta t$, $\Delta t$ & integration step; sampling interval of a record \\
				$w$ & averaging window of $\langle f\rangle_w$ \\
				$T$ & superstatistical coarse-graining window \\
				$d$ & fast time scale: first $1/e$ crossing of the increment ACF \\
				\addlinespace[3pt]
				\multicolumn{2}{@{}l}{\textit{Measured quantities}} \\
				$f_i(t)$ & measured frequency at site $i$ [Hz] \\
				$x_i(t)$ & deviation from nominal, $f_i(t)-50\,\mathrm{Hz}$ \\
				$\bar f(t)$ & mean over the $n$ simultaneously observed sites, Eq.~(\ref{eq:common_local}) \\
				$\langle f\rangle_w$ & frequency averaged over a window $w$ \\
				$\Delta f_{i,\tau}$ & increment $f_i(t+\tau)-f_i(t)$, Eq.~(\ref{eqn:increment}) \\
				$\sigma_f$ & standard deviation of the measured frequency \\
				$\beta$, $\tilde\beta$ & local inverse variance $[\operatorname{Var}_T(\Delta f_\tau)]^{-1}$, Eq.~(\ref{eq:beta_def}); $\tilde\beta=\beta/\max\beta$ \\
				$g(\beta)$ & distribution of $\beta$ (superstatistical kernel) \\
				$\kappa$ & kurtosis ($\kappa=3$ for a Gaussian) \\
				$q$ & $q$-Gaussian parameter \\
				$p(\cdot)$, $P(\cdot)$ & probability density \\
				\addlinespace[3pt]
				\multicolumn{2}{@{}l}{\textit{Multimodality}} \\
				$D_{\rm A}$ & Ashman's bimodality statistic, Eq.~(\ref{eq:ashman_d}) \\
				\bottomrule
			\end{tabular}
		\end{minipage}\hfill
		\begin{minipage}[t]{0.49\textwidth}
			\begin{tabular}[t]{@{}>{\RaggedRight\arraybackslash}p{0.2\textwidth}>{\RaggedRight\arraybackslash}p{0.78\textwidth}@{}}
				\toprule
				\textbf{Symbol} & \textbf{Meaning} \\
				\midrule
				\multicolumn{2}{@{}l}{\textit{Kuramoto model, Eq.~(\ref{model_eq})}} \\
				$\theta_i$ & phase of node $i$ in the co-moving frame \\
				$\omega_i$ & simulated frequency $\dot\theta_i$; values in (m)Hz are $\omega_i/2\pi$ \\
				$\bar\omega(t)$ & network-mean frequency $N^{-1}\sum_i\omega_i(t)$ \\
				$\bar\omega_i$ & late-time mean frequency of node $i$ (second half of a run) \\
				$\Delta\omega_\tau$, $\sigma_\omega$ & simulated increment; standard deviation of simulated frequencies ($\sigma_\omega(t)$: over nodes at time $t$) \\
				$m_i$, $\alpha_i$ & inertia; damping \\
				$P_i$ & quenched power injection (self-frequency) \\
				$K$, $K_c$ & global coupling; synchronization point \\
				$W_{ij}$, $s_i$, $k_i$ & line weight; strength $\sum_jW_{ij}$; degree \\
				$\xi_i$, $D$ & white noise, $\langle\xi_i(t)\xi_j(t')\rangle=2D\delta_{ij}\delta(t-t')$; noise strength \\
				$c$ & initial phase spread, $\theta_i(0)\in(0,2\pi c)$ \\
				$r$, $R$ & Kuramoto order parameter; its realization average \\
				$\Omega(t)$ & frequency variance over nodes \\
				$d_s$ & spectral dimension of the network \\
				\addlinespace[3pt]
				\multicolumn{2}{@{}l}{\textit{Correlations and spectra}} \\
				$C_X(\tau)$ & normalized ACF of a stationary signal $X$, Eq.~(\ref{eq:acf_main}) \\
				$C(t_0,\tau)$ & two-time ACF of a measured frequency \\
				$C_\omega$, $\bar C_\omega$ & two-time ACF of simulated frequencies over nodes, normalized by $\sigma_\omega(t)\sigma_\omega(t_0)$ or $\sigma_\omega^2(t_0)$, Eq.~(\ref{eq:cv}) \\
				$C_\theta$ & phase autocorrelation \\
				$C_{\Delta f}$, $C_\beta$ & ACF of the increments and of $\beta$ \\
				$\rho$, $\rho_\beta$ & ACF decay exponent, $C(\tau)\propto\tau^{-\rho}$ \\
				$\nu$ & spectral frequency [Hz] \\
				$S_f$, $S_\omega$ & PSD of a measured / simulated frequency ($S_{\bar f}$, $S_{\bar\omega}$ of the means) \\
				$\gamma$ & PSD exponent, $S(\nu)\propto\nu^{-\gamma}$ \\
				\bottomrule
			\end{tabular}
		\end{minipage}
		\caption{Notation. Measured frequencies are denoted by $f$, simulated ones by $\omega$. Autocorrelation exponents are denoted by $\rho$ and power-spectral exponents by $\gamma$; for a stationary process with $0<\rho<1$, $\gamma=1-\rho$ and for nonstationary $\gamma=3-\rho$.}
		\label{tab:notation}
	\end{table*}
	
	\section{Introduction}
	\label{sec:intro}
	
	The transition towards electrification of the industrial, commercial, and domestic sectors requires a better understanding and modeling of power grids~\cite{Pere-Damia-Canaria}. The frequency of an AC power grid is the most direct and readily measured indicator of the balance between generation and consumption~\cite{RydinGorjão2020}. Like many previous studies~\cite{GorjaoEUROPHYS2025, WenSCIREP2025, KruseIEEE2020}, we aim to better understand the origin of its fluctuations, but here we also take into account the spatially distributed nature of the underlying system. While standard statistical summaries like probability distributions and power spectra are useful~\cite{HartmannSEGAN2024, KruseIEEE2020}, the same one-point distribution can arise from entirely different microscopic mechanisms~\cite{SchaferPRE2017, KraljicIEEETRANS2022}. Reduced stochastic models often treat the grid as a single effective oscillator driven by white noise~\cite{KraljicIEEETRANS2022, UhlenbeckPR1930, SchaferPRE2017}, yielding an exponential autocorrelation and a Lorentzian spectrum. However, a real transmission grid is not a single oscillator; rather, it is a heterogeneous, spatially embedded network of interacting generators and consumers~\cite{WenSCIREP2025, KraljicIEEETRANS2022}. As ensembles of correlated, fluctuating variables, real power grids may exhibit scale-free spectra and non-Gaussian, heavy-tailed or multi-peaked distributions~\cite{schaefer2018nongaussian, WenSCIREP2025, hartmann_12models}. Furthermore, recent observations show that sub-second rate frequency measurements display a more complex behavior: the power spectral density begins to rise, and the time series display anti-correlations and multifractal properties~\cite{SchaferCOMPLEXITY2023}. Existing stochastic models do not fully capture these behaviors. Although some features can be described by custom noise terms, such models assume that all variability originates externally, from the complex behavior of the environment. This raises the fundamental question: to what extent can the observed complexity be attributed to the dynamics of the second-order Kuramoto model (KM) running on finite-dimensional power-grid networks?

	To investigate this without relying on free stochastic inputs, we employ the second-order Kuramoto model, which is equivalent to the swing equations of electric power systems~\cite{AcebronREVMODPHYS2005, OdorDengENTROPY2023, FilatrellaEtAlEPJB2008, DorflerPNAS2013}. A second-order Kuramoto model retains the phases and frequency deviations fluctuations, from the nominal value of $\SI{50}{\hertz}$, of many coupled nodes and maps, under the usual approximations, to the swing equation~\cite{FilatrellaEtAlEPJB2008,DorflerPNAS2013}. It provides a controlled way to describe which signatures follow from local interactions and from quenched heterogeneity alone. After the system relaxes from the initial state, one can measure these fluctuations in space and time. By varying the initial conditions, we can test for the robustness of the observed behaviors.
	
	We use superstatistics as a diagnostic tool to study the variation of the statistics between short windows~\cite{BeckCONTMECH2004,Gorjao_2021} and show that at a given time lag $\tau$, their statistics indeed can not be described with a single analytical PDF fit. In each short time window, we estimate the variance of the frequency increments and test whether the full increment distribution can be reconstructed as a weighted superposition of local Gaussian distributions (superstatistics) with their measured variances. The usage of superstatistics opens the possibility to account better for the tail behavior, something that most single distribution fits can not describe well. 
	
	Three systems, the empirical HU-HV power grid, KM simulations on the HU387 graph and on 2D lattices, are investigated to identify different physical mechanisms contributing to the observed frequency dynamics. Hungarian measurements provide empirical frequency spectra, temporal correlations, and increment statistics. The periodic square lattice isolates effects that arise from finite-dimensional local coupled
	KM oscillators, whereas the HU387 model additionally incorporates the realistic Hungarian transmission-network topology and heterogeneous static injections
	(input-output terms to KM). Finally, simulations with additive Gaussian noise test the sensitivity of these spatial dynamics to an external time-dependent stochastic perturbation. We compare the same observables across the systems whenever it is possible. The aim is to identify which properties of the empirical measurements can arise from spatial network dynamics alone and which require additional time-dependent forcing.
	
	A useful reference point is the single-oscillator, or mean-field, description of grid-frequency dynamics~\cite{KraljicIEEETRANS2022,UhlenbeckPR1930,Zeraati2022Timescales}. Such a reduction neglects spatial fluctuations and network heterogeneity and, for simple stochastic forcing, produces comparatively simple temporal statistics and single-peaked frequency distributions. In a finite-dimensional or heterogeneous oscillator network, however, the frequencies of individual nodes can fluctuate relative to each other and may develop long-lived spatial structures, Chimera-like states~\cite{10.1063/5.0154581} absent from the mean-field description. One strategy is to represent the omitted fluctuations phenomenologically through external stochastic terms. Here, instead, we retain the spatially resolved Kuramoto dynamics and ask how much of the empirical behavior can already be generated by network structure and heterogeneous coupling. The only additional time-dependent stochastic contribution considered in the present simulations is the additive Gaussian white noise. The notation used throughout the paper is summarized in Table~\ref{tab:notation}.
	
	\section{Methods}
	\label{sec:methods}
	
	\subsection{Empirical frequency data}
	
	Our empirical data consist of synchronous 10-Hz ($\SI{100}{\milli\s}$) frequency measurements from sixteen Hungarian 400-kV transmission-grid sites recorded between 31 January and 28 February 2025. Complete daily records were analysed separately before aggregation, and the number of available days therefore varies between sites. For a measured frequency $f_i(t)$ at site $i$, we define the deviation from the nominal grid frequency as
	\begin{equation}
		x_i(t)=f_i(t)-50\,\mathrm{Hz},
	\end{equation}
	and the lag-$\tau$ frequency increment (where $\tau$ is the time difference between measurements) as
	\begin{equation}
		\label{eqn:increment}
		\Delta f_{i,\tau}(t)=f_i(t+\tau)-f_i(t).
	\end{equation}
	We analysed the distributions of $\Delta f_{i,\tau}$, the kurtosis of $x_i$ distributions, the temporal correlations and the power spectral density (PSD) of $x_i$. For correlation and spectral estimates, the mean values of variables within the analysed segment were removed separately from the subtraction of the nominal frequency above. Data completeness, daily aggregation, and estimator settings are described in the Appendix sections~\ref{supp:data},~\ref{supp:psd}, and Supplementary Material section~\ref{supp:acf}.
	
	\subsection{The second-order Kuramoto model}
	
	The power-grid dynamics can be modelled by the second-order Kuramoto equation~\cite{FilatrellaEtAlEPJB2008,DorflerPNAS2013} 
	based on the swing equations of AC electrical circuits:
	\begin{equation}
		\label{model_eq}
		m_i\ddot{\theta}_i = P_i-\alpha_i\dot{\theta}_i
		+ K\sum_j W_{ij}\sin(\theta_j-\theta_i)
		+ \xi_i(t),
	\end{equation}
	where $\theta_i$ is the phase of node $i$ in a frame rotating at the nominal grid frequency, 
	summation runs over the neighbouring nodes $j$ of $i$; the $m_i$ and $\alpha_i$ denote inertia and 
	damping parameters, respectively, $W_{ij}$ specifies the network coupling, and $K$ controls 
	its overall strength.
	$P_i$ can describe either input or output power or, in the case of lattice simulations $P_i$, is a random variable, 
	drawn from a zero-centered Gaussian distribution with unit variance. In deterministic simulations,  $\xi_i(t)=0$.
	
	It is important to note that even the deterministic KM can display irregular, noise-like frequency fluctuations: in the partially synchronized regime, nonlinearly coupled oscillators produce chaotic dynamics, whose fluctuation variance is maximal near the synchronization point $K_c$~\cite{PhysRevResearch.6.013194,OdorDengENTROPY2023}. Fluctuations decrease to zero as $t\to\infty $ for $K\ge K_c$, but due to the quenched disorder of $P_i$ different realizations may occur, causing a finite spread of variables even in the steady state. 
	In most of our previous power-grid simulations we therefore solved the deterministic KM and averaged over many realizations of the quenched self-frequencies, instead of solving stochastic differential equations. Here we study in more detail the effect of time-dependent Gaussian white noise which is added to the KM, with correlations
	\begin{equation}
		\left\langle\xi_i(t)\xi_j(t')\right\rangle
		= 2D\,\delta_{ij}\delta(t-t'),
	\end{equation}
	where $D$ denotes the noise strength coefficient, $\delta_{ij}$ is the Kronecker delta and $\delta(t)$ is the Dirac delta function. The numerical implementation of the stochastic term is specified in the Supplementary Sec.~\ref{supp:hu387_runs_more} for HU387. Throughout, measured frequencies are denoted by $f$ and simulated ones by $\omega$: $\omega_i=\dot\theta_i$ is the local frequency of node $i$ in the co-moving frame. Where simulated frequencies are quoted in $\mathrm{Hz}$ or $\mathrm{mHz}$, the value given is $\omega_i/(2\pi)$.
	
	To study the dynamics of the KM, at first we focus on the Kuramoto order parameter, which is used to measure the global phase synchronization level of the network defined by
	\begin{equation}
		Z(t)=r(t)e^{i\psi(t)}=\frac{1}{N}\sum_{j=1}^{N}e^{i\theta_j(t)},
	\end{equation}
	where, $N$ is the number of nodes in the network, $Z(t)$ is the complex order parameter and $\psi(t)$ is the average phase value of the oscillators and $r(t)$ is the real-valued global Kuramoto order parameter. When we average over realizations with different quenched self-frequencies denoted by $P_i$ here we get: $R(t)=\langle r(t)\rangle$.
	Next, we measure the frequency variance defined by 
	\begin{equation}
		\Omega(t)=\frac{1}{N} \sum_{j=1}^{N}[\omega_j(t)-\bar \omega(t)]^2,
	\end{equation}
	where $\bar \omega(t)$ is the mean of the frequencies at the same time $t$. We usually also average $\Omega(t)$ over realizations.
	Note that in real power-grid operation we require strong frequency synchronization (small $\Omega$), but the phases
	cannot synchronize fully, because there won't be power flow between nodes, which is proportional to: $\theta_j(t)-\theta_i(t)$.
	We don't have empirical values or $R$ for whole power grids, just some estimates, which means partial phase
	synchronization. Therefore, we allow the global coupling $ K$ to vary in the simulations. In earlier works~\cite{e22060666,PhysRevResearch.6.013194} we argued that PL-tailed outage statistics can be observed because a power supply/demand self-organization tunes the grid close to criticality, near the synchronization $K_c$ point.
	
	In this work, we consider two spatial systems: First, periodic square lattices with nearest-neighbour coupling, Gaussian-distributed quenched nodal frequencies, and uniform inertia and damping provide a dimensionless reference system in which the effects of local coupling, noise, and initial conditions can be isolated. 
	
	To study the synchronization behavior of the noisy Kuramoto model, we have numerically simulated the system Eq.(\ref{model_eq}) on a $2$-dimensional lattice of linear size $L=300$ using the Euler Predictor-Corrector and for $L=2000$ by a stochastic Runge-Kutta algorithm. For the larger lattice, we employed a fixed-step, four-stage stochastic Runge-Kutta scheme~\cite{Kasdin1995}, with independent Gaussian noise increments applied to the angular velocities. To reduce the computational cost, the sine and cosine of each phase were computed once per stage and reused to evaluate the nearest-neighbor coupling terms. These calculations and the Runge-Kutta stage updates were parallelized over the lattice sites using OpenMP on the CPU. We considered random initial self-frequencies, drawn from a unimodal Gaussian distribution, with mean $0$ and standard deviation $1$. 
	
	On the other hand, we have the HU387 model, which involves the Hungarian transmission-network topology together with heterogeneous electrical parameters and static nodal injections. Where dispatch disorder is included, it is drawn once for each realization and subsequently kept fixed, with the total injection rebalanced to zero. The principal control parameters considered here are the global coupling $K$ and, where present, the noise intensity $D$. Integration, initialization, transient removal, sampling, and parameter values are given in the Appendix, Sec.~\ref{supp:simulations}.
	
	Note that the empirical network is observed at sixteen measurement sites rather than at every model node. Additionally, empirical network nodes that have an available corresponding HU387 node are matched to their model counterparts using the known geographical/substation identity of the measurement location. The site-to-node mapping used in the comparison is listed in the Appendix.
	
	\subsection{Temporal correlations and power spectra}
	
	To quantify the temporal dependencies and identify inherent periodicities within the signal, we first calculated the autocorrelation function, $C_X(\tau)$. For a mean-centred stationary segment $X(t)$, we define the normalized temporal autocorrelation function as
	\begin{equation}
		\label{eq:acf_main}
		C_X(\tau)
		=
		\frac{\operatorname{Cov}[X(t),X(t+\tau)]}
		{\operatorname{Var}[X(t)]}
	\end{equation}
	
	To examine non-stationary systems, we additionally use a two-time correlation $C(t_0,\tau)$ that retains the reference time $t_0$. The corresponding estimator is defined by the proper correlator with spatial variances
	\begin{equation}
		\label{eq:cv}
		C_\omega(\tau) = \frac{\left\langle[\omega_i(t)-\bar \omega(t)][\omega_i(t_0)-\bar \omega(t_0)]\right\rangle}{
			\sigma_\omega(t)\,\sigma_\omega(t_0)},
		\qquad t=t_0+\tau ,
	\end{equation}
	where $\bar{\omega}(t)$ and $\sigma_\omega(t)$ denote the instantaneous mean and standard deviation over nodes, respectively. Averaging $\langle\cdot\rangle$ is done for nodes and for different self-frequency realizations.
	The reference time $t_0$ is chosen to be 100 iterations, corresponding to $t_0=1$ as the integration time step size was $\delta t=0.01$. The ACF is calculated as a function of the time lag $\tau = t-t_0$. Note that the ACF calculated in this way should not be the same as the single-site temporal ACF on the measurements. 
	Another way to estimate the autocorrelation can be given by
	\begin{equation}\label{eq:cbar}
		\bar{C}_\omega(\tau) = \frac{\left\langle[\omega_i(t)-\bar \omega(t)][\omega_i(t_0)-\bar \omega(t_0)]\right\rangle}{
			\sigma_\omega^2(t_0)},
		\qquad t=t_0+\tau ,
	\end{equation}
	where $\sigma_\omega^2(t_0)$ is the variance over nodes at the starting time. This definition resembles (\ref{eq:acf_main}) more, as it assumes a stationary state in the normalization, 
	but uses the spatial averages of frequencies at $t_0$. In general, we are not in the stationary state, but
	empirical data analysis assumes it; thus, we also calculated a similar measure with spatial averages. 
	We will show that the decay of (\ref{eq:cbar}) and of (\ref{eq:cv}) exhibit different PL exponents 
	and they are affected by the stochastic noise differently.\\
	For the sake of completeness, the auto-correlations of phases are also calculated, defined as
	\begin{equation}\label{Adef}
		C_{\theta}(t_0,t_0+\tau) =  \langle \cos [ \theta_i(t_0+\tau)-\theta_i(t_0) ]\rangle \  .
	\end{equation}

	While the autocorrelation function maps dependencies across time lags, a complete characterization of the system dynamics requires examining how this power is distributed across distinct frequencies. Therefore, we concurrently computed the Power Spectral Density (PSD) to supplement our analysis in the frequency domain.
	The one-sided PSD $S_f(\nu)$ (or $S_\omega(\nu)$ for simulated frequencies), with spectral frequency $\nu$, is estimated from uniformly sampled records using Welch's method \cite{VanderPlas_2018_lomb_scargle}. Spectral slopes are reported only over explicitly stated finite frequency intervals,
	\begin{equation}
		\label{eq:psd_scaling}
		S_f(\nu)\propto\nu^{-\gamma},
	\end{equation}
	without assuming that a single scaling law applies over the complete spectrum. Irregularly sampled simulation records are analysed separately using a Lomb--Scargle \cite{VanderPlas_2018_lomb_scargle} periodogram. Estimator parameters, aggregation procedures and fitting conventions are given in Appendix Sec.~\ref{supp:psd}.
	
	In a synchronous grid, the frequencies measured at different locations are not independent signals: they share a collective motion, in which the whole synchronous area accelerates and decelerates together. For $n$ sites observed simultaneously, we describe it by the mean frequency
	\begin{equation}
		\label{eq:common_local}
		\bar f(t)=\frac{1}{n}\sum_{j=1}^{n}f_j(t),
	\end{equation}
	and compare its spectrum $S_{\bar f}(\nu)$ with the model. For the measurements, the sites are the $n=9$ substations that also carry a node of the HU387 model, with lines ending at the same substation averaged. In the model, the collective motion is the mean frequency $\bar\omega(t)$ over all nodes, with spectrum $S_{\bar\omega}(\nu)$, which, for equal inertia, is the centre-of-inertia frequency.
	
	Throughout, autocorrelation decay exponents are denoted by $\rho$, $C(\tau)\sim\tau^{-\rho}$, and spectral exponents by $\gamma$, $S(\nu)\sim\nu^{-\gamma}$. For a stationary process with $0<\rho<1$, the corresponding low-frequency spectrum scales with $\gamma=1-\rho$ under the usual Fourier assumptions. Different exponent relations apply to nonstationary self-similar processes with stationary increments. Finally, because the present records are finite and periodically driven, these relations are used only as reference scalings and not imposed in the fits.
	
	\subsection{Superstatistical analysis}
	
	To uncover the long-range autocorrelation, we employ the superstatistics framework to hopefully bypass the 15-minute trading oscillations. Here, we apply the method to the coarse-grained $\beta$ function defined as the inverse variance of the increments within windows of duration $T$,
	\begin{equation}
		\label{eq:beta_def}
		\beta
		=
		\left[\operatorname{Var}_{T}(\Delta f_\tau)\right]^{-1}.
	\end{equation}
	If the increment distribution conditioned on $\beta$ is approximately centred Gaussian, the resulting marginal distribution is
	\begin{equation}
		\label{eqn:superstats}
		p(\Delta f_\tau)
		=
		\int_0^\infty
		\sqrt{\frac{\beta}{2\pi}}
		\exp\!\left[
		-\frac{\beta(\Delta f_\tau)^2}{2}
		\right]
		g(\beta)\,\mathrm{d}\beta ,
	\end{equation}
	where $g(\beta)$ denotes the empirical or fitted distribution of the slowly varying inverse variance.
	
	The fast time scale $d$ is defined as the first $1/e$ crossing of the autocorrelation of $\Delta f_\tau$. The coarse-graining scale $T$ is determined independently from the local-kurtosis criterion described in Supplementary Sec.~\ref{supp:superstats}. The ratio $T/d$ therefore tests whether the local variance evolves on a slower scale than the increment dynamics. The autocorrelation of $\beta(t)$ is analysed separately and is not used to define $T$. Candidate distributions for $g(\beta)$, fitting procedures and robustness tests are described in Supplementary Secs.~\ref{supp:kernels}, \ref{supp:modelsel}, and Appendix~\ref{supp:robust}.
	
	\subsection{Multimodality and frequency locking in the simulations}
	\label{subsec:multimodality_methods}
	
	For the HU387 model, the observable is the late-time mean frequency $\bar\omega_i$ of each node, i.e.\ its time average over the second half of a run (the node index distinguishes it from the instantaneous network mean $\bar\omega(t)$). The runs extend to $t=345600\,\mathrm{s}$ ($96\,\mathrm{h}$) of model time, much longer than the transient (Appendix Sec.~\ref{supp:simulations}). A sample of $\bar\omega_i$, taken either across the nodes of one realization or for one node across the realizations, is fitted with a two-component Gaussian mixture and characterized by Ashman's statistic~\cite{AshmanARXIV1994} $D_{\rm A}$, which is defined as
	\begin{equation}
		\label{eq:ashman_d}
		D_{\rm A}=\frac {\left|\mu _{1}-\mu _{2}\right|}{\sqrt {{\frac {1}{2}}\left(\sigma _{1}^{2}+\sigma _{2}^{2}\right)}},
	\end{equation}
	where $\mu_1$, $\mu_2$ are the means and $\sigma_1$, $\sigma_2$ are the standard deviations of the two Gaussians that create the mixture distribution. A good rule of thumb is $D_A \gtrapprox 2$ indicates bimodality~\cite{AshmanARXIV1994}. A sample is classified as bimodal when $D_{\rm A}$ exceeds $95\%$ obtained from unimodal Gaussian samples of the same size and the smaller component contains at least $5\%$ of the sample. As $D_{\rm A}$ also responds to shoulders and flat-topped peaks, we additionally count the peaks of the fitted mixture density (Appendix Sec.~\ref{supp:multimodality}).
	
	To relate the multimodality to synchronization, each node in each realization is assigned to one of three classes according to $\bar\omega_i$: the \emph{main cluster}, if $\bar\omega_i$ lies within $5\,\mathrm{mHz}$ of the most common late-time frequency of that realization; \emph{free-running}, if $\bar\omega_i$ lies within $10\%$ of its uncoupled frequency $P_i/\alpha$, i.e.\ the coupling barely affects it; and \emph{other} otherwise. The last class contains nodes that lock among themselves into a separate island rotating at its own frequency, and nodes that are only partially entrained by the main cluster. With noise, the $5\,\mathrm{mHz}$ tolerance is replaced by three standard errors of the time average of the node if this is larger.
	
	\section{Results}
	\label{sec:results}
	
	\subsection{Empirical frequency analysis of the Hungarian power grid}
	
	In the following, we present our analyses on the empirical data consisting of synchronous frequency measurements (100ms resolution) from sixteen Hungarian 400-kV transmission-grid sites recorded between 31 January and 28 February 2025.
	
	\subsubsection{Multi-peak behavior of the frequency distributions}

	\begin{figure*}[ht]
		\centering
		\includegraphics[width=0.9\textwidth]{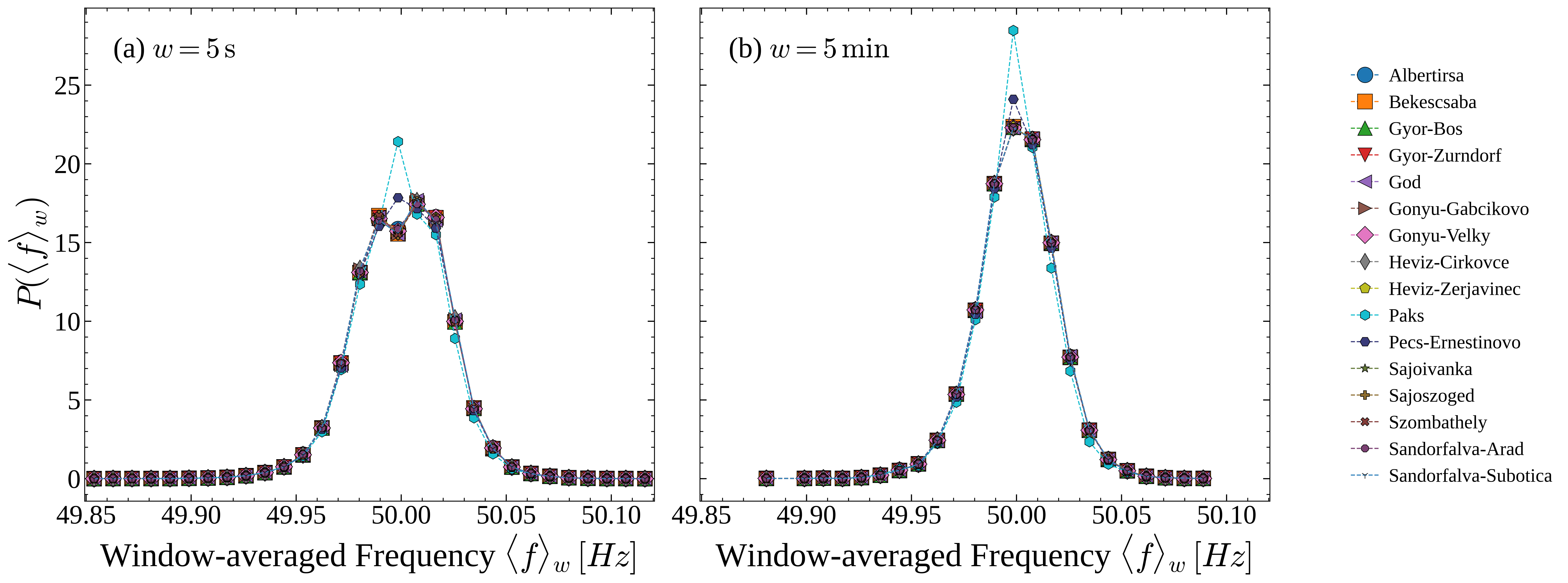}
		\caption{Probability distributions of window-averaged frequency at the 16 monitored 400-kV nodes. The probability density ($P(\langle f\rangle_{w})$) of the window-averaged frequency ($\langle f\rangle_{w}$) is shown for two averaging windows: (a) ($w=5 \mathrm{s}$) and (b) ($w=5 \mathrm{min}$). Each curve represents one monitored transmission node, with distinct markers and colors used to distinguish the nodes. Note the structural transition from bimodal micro-dynamics at fine resolutions to a unimodal Gaussian distribution at the 5-minute scale.}
		\label{fig:stats_window}
	\end{figure*}
	
	In general, depending on the observation window, the shape of the probability density function (PDF) of power grid frequencies varies. In a previous work by Wen et.al.~\cite{WenSCIREP2025}, analyzing grid frequencies across continents demonstrated that distributions exhibit minor differences between day and night, characterized by non-Gaussianity, non-stationarity, and bimodal features in certain regions, attributing this behavior primarily to governor deadbands~\cite{schaefer2018nongaussian,Gorjao_2021}. A governor deadband is the total magnitude of frequency change in a turbine-generator system where no resulting movement of the control valve occurs. In such a day-long observation window, the multi-peaks can be the consequence of daily load balance variations, a temporal heterogeneity. However, considering narrow observation windows i.e., $w \leq 5\mathrm{s}$, across a network of 16 geographically distributed nodes reveals a noticeably different scenario and reasons for multi-peak behavior. In Figure~\ref{fig:stats_window}, the probability density function of the mean grid frequency, $P(\langle f\rangle_{w})$, aggregated across nodes over a one-month period, exhibits a distinct time scale-dependent behavior and transition. At fine time resolutions ($w \leq 5\text{ s}$), $P(\langle f\rangle_{w})$ displays a pronounced bimodal distribution. At the $w = 500\text{ ms}$ timescale, the inertial response is expected to stabilize 
	the node frequencies to nominal values, while for $w = 5\text{s}$ the primary, proportional droop response adjusts the active power. Lack of this suggests that the spatial structure is the main reason for the observed multi-peak behavior. Thus, the observed double peak  can be mainly attributed to the variation of the measurement locations across the 16 stations scattered in space. For example, in Figure~\ref{fig:stats_window}(a), we observe that most nodes have double peaks except for Paks and P\'ecs which have a single peak at $\langle f\rangle_{w} = 49.99 \mathrm{Hz}$, reflecting node-level heterogeneity for short time scales. At a coarser resolution $w = 5\text{ min}$ shown in Figure~\ref{fig:stats_window}(b), this bimodal feature collapses into a smooth, unimodal curve centered at $50.00\text{ Hz}$. This transition probably occurs because a 5-minute window allows for temporal aggregation, via feedback controls, diminishing the spatio-temporal inhomogeneity. 
	
	\subsubsection{Empirical autocorrelation results of the HU power grid frequencies}
	
	\begin{figure*}[!t]
		\centering
		\includegraphics[width=0.9\textwidth]{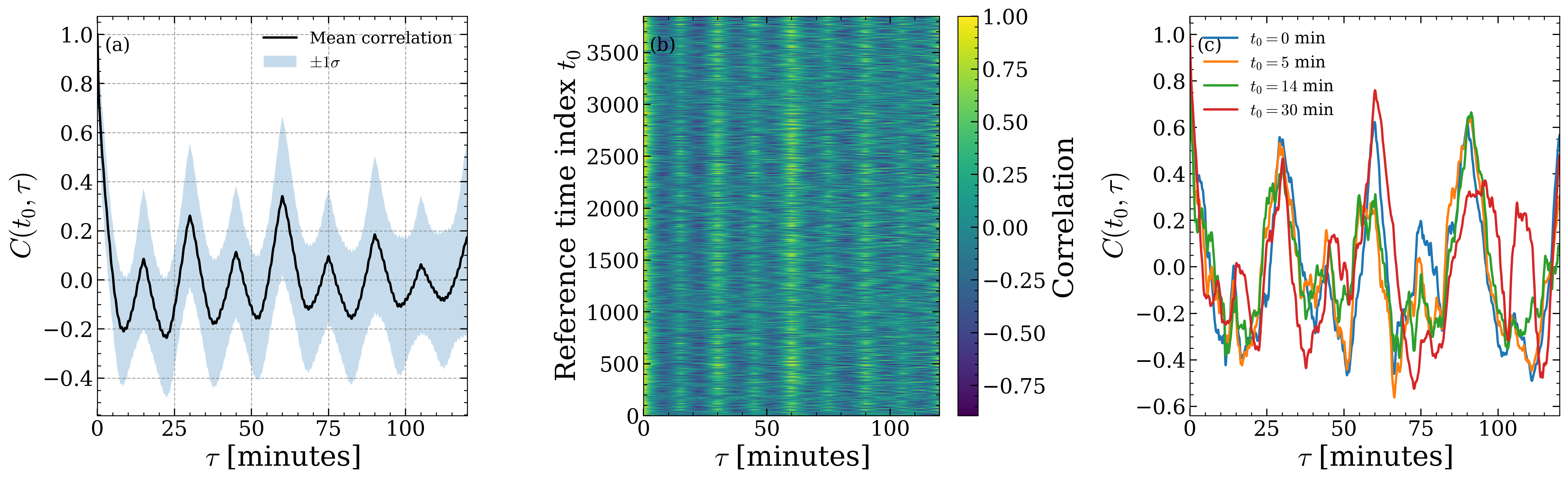}
		\caption{
			Nonstationary auto-correlations across time scales. (a) Average ACF over all $t_0$ (b) full $C(t_0,\tau)$ heatmap (c) Local ACF snapshots for selected $t_0$.}
		\label{fig:acf_nonstationary}
	\end{figure*}
	
	While Gorjão et al.~\cite{GorjaoIEEE2020} capture the periodic frequency signal using a single oscillator model with a step-function supply profile and Gaussian noise reproducing large, negative-diving ACF waves, they do not address long-term decay characteristics. They explain that these 15-minute oscillations stem from rigid market intervals approximating a smooth, monotonic demand curve, yielding a sawtooth power imbalance. Conversely, another work~\cite{KraljicIEEETRANS2022} identifies a slow, power-law (PL) decrease at longer lags. In contrast to those findings, our empirical analysis results shown in Figure~\ref{fig:acf_nonstationary} do not exhibit this long-term power-law decay, aligning instead with the rapid, short-term exponential decay that characterizes the system's primary control loops. Or if there is such decay, the large oscillations hinder seeing it. Note that in our HU data the correlators almost symmetrically oscillate around zero, unlike the UK and South Africa (SA) cases. The measured frequency exhibits a pronounced non-stationary correlation structure. Figure~\ref{fig:acf_nonstationary}(a) shows the autocorrelation averaged over the reference time $t_0$ for B\'ek\'escsaba. As one may observed, the frequency autocorrelation drops exponentially during the first few minutes ($\tau \lesssim 15$min.), aligning with the Ornstein–Uhlenbeck process, describing random walk of the power values of sources and sinks~\cite{Gardiner1985Stochastic,GorjaoIEEE2020}. Although the initial short-range drop in empirical power-grid frequency data follows this exponential decay, real power grids exhibit additional complex features such as periodic peaks at every 15 minutes resulting from the scheduled electricity market trading~\cite{ENTSOe2019dfd}, and heavy-tailed non-Gaussian distributions~\cite{GorjaoIEEE2020}.
	
	This becomes much clearer in Figure~\ref{fig:acf_nonstationary}(b), where the full correlation landscape is shown as a heatmap over both lag $\tau$ and reference time $t_0$. Here, we can observe that the correlation exhibits clear temporal modulation: bands of stronger and weaker correlation shift across time. If the grid frequency were strictly stationary, this heatmap would look uniform top-to-bottom. Instead, we observe shifting brightness and color intensity as $t_0$ progresses. This visually proves non-stationarity: the strength of grid memory/oscillations varies dynamically across almost all reference times ($t_0$).
	
	Finally, Figure~\ref{fig:acf_nonstationary}(c) shows local autocorrelation functions computed at different reference times $t_0$. This highlights how the grid's local memory profile changes over time. Notice how completely different the blue line ($t_0 = 0$) looks compared to the green line ($t_0 = 14$). If the grid were static, all trends would overlap perfectly. At some starting times ($t_0 = 0\text{ or } 30$), strong coherent wave-like oscillations occur but between major market intervals (like $t_0 = 14$ min), the correlation curve is messier and less defined, proving the system's memory changes from minute to minute.
	
\begin{figure}[tbp]
		\centering
		\includegraphics[width=\columnwidth]{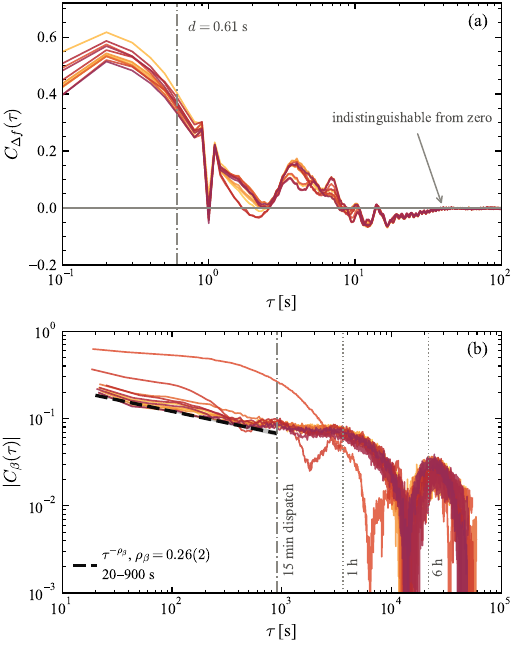}
		\caption{Empirical auto-correlations at twelve measurement sites, one curve per site. (a) Autocorrelation $C_{\Delta f}(\tau)$ of the one-second frequency increments $\Delta f_{1\,\mathrm{s}}$. The vertical line marks the mean $1/e$ time $d=0.61\,\mathrm{s}$; beyond a few tens of seconds the correlation is indistinguishable from zero. (b) Modulus of the autocorrelation $C_\beta(\tau)$ of the window-wise inverse variance $\beta(t)$, with lags converted to physical time using the non-overlapping window duration $T$. The dashed line shows a power law $\propto \tau^{-\rho_\beta}$, $\rho_\beta=0.26(2)$,
			fitted on data over the $20$--$900\,\mathrm{s}$ time lag region; vertical lines mark $15\,\mathrm{min}$, $1\,\mathrm{h}$ and $6\,\mathrm{h}$. 
			The increment correlation decays within seconds, whereas the inverse-variance correlation persists for hours.}
		\label{fig:acf_df_beta}
	\end{figure}

	The rapid dynamics become clearer when frequency increments are considered~\cite{schaefer2018nongaussian,Gorjao_2021}. Figure~\ref{fig:acf_df_beta} compares the autocorrelation of the one-second increments $\Delta f_\tau$, panel (a), with that of the window-wise inverse variance $\beta(t)$, panel (b). The increment ACF loses most of its initial correlation on a sub-second to few-second scale and subsequently oscillates around zero. Its first $1/e$ crossing defines the fast operational time $d$ used in the superstatistical analysis.
	
	In contrast to the increments themselves, the inverse variance retains memory over much longer times~\cite{BeckCONTMECH2004,schaefer2018nongaussian,Gorjao_2021}. Over the intermediate range, its ACF, Fig.~\ref{fig:acf_df_beta}(b), is compatible with an algebraic decay,
	\begin{equation}
		C_{\beta}(\tau)\sim \tau^{-\rho_{\beta}},
	\end{equation}
	with an across-site estimate $\rho_{\beta}\simeq0.26(2)$. The decay changes its character close to the 15-minute scale. Consequently, the fast decorrelation of signed increments differs clearly from the slowly evolving fluctuation intensity. In physical terms, the data contain a rapidly fluctuating component embedded in an operating state whose local variance changes much more slowly
	
	This separation is relevant for the modeling below. A static distribution of nodal injections can generate spatial heterogeneity, but it contains no intrinsic clock corresponding to the slow modulation observed here. Reproducing the empirical correlations therefore requires not only spatial coupling but also a mechanism capable of evolving the effective operating conditions in time.
	
	\subsubsection{Power Spectral Density Analysis of the MAVIR data}
	\label{sec:psd_analysis}
	
	\begin{figure}[tbp]
		\centering
		\includegraphics[width=\columnwidth]{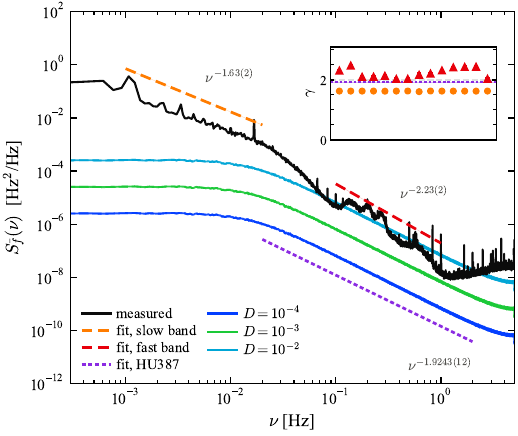}
		\caption{PSD spectrum of the mean frequencies (Eq.~(\ref{eq:common_local})): the measured $S_{\bar f}$ averaged over the measurement substations (black line) and the simulated network-mean $S_{\bar\omega}$ of HU387 (coloured lines) for different noise strengths $D$ (both for $K=0.3,0.5$, as they overlap).
			Dashed lines are up-shifted power laws of the form of Eq.~(\ref{eq:psd_scaling}) for the measured spectrum in the $10^{-3} < \nu < 2\times10^{-2}\,\mathrm{Hz}$ (orange) and in the $0.1 < \nu < 1\,\mathrm{Hz}$ (red) frequency regions. Dotted line: down-shifted PL fit for the simulated spectra for $2\times10^{-2} < \nu < 2\,\mathrm{Hz}$ (identical for all simulated curves). Inset: the fitted exponents for each of the measurement sites. In the low-frequency region (circles) a mean exponent value $\gamma=1.628(5)$ is obtained, while in the high-frequency region (triangles), a mean exponent value $\gamma=2.23(16)$ can be obtained as compared with the simulation exponent $\gamma=1.92$ (dotted line).}
		\label{fig:psd_comparison}
	\end{figure}
	
	We turn our attention to the frequency domain, which provides an independent view of the temporal organization of the fluctuations. The empirical spectrum extends over several decades and cannot be described by a single power law over the complete interval. At low spectral frequencies, approximately $0.001$--$0.02\,\mathrm{Hz}$, the site-averaged fit gives
	\begin{equation}\label{eq:Sf}
		S_f(\nu)\propto \nu^{-\gamma},
		\qquad
		\gamma=1.628(5).
	\end{equation}
	This range corresponds to characteristic times from approximately $50$ to $1000\,\mathrm{s}$ and overlaps the time scales on which the non-stationary ACF and the slowly varying inverse variance are observed. The exponent is distinctly below the Brownian~\cite{UhlenbeckPR1930} value $\gamma=2$, showing that the long-time empirical dynamics are not fully consistent with a single brown noise process~\cite{KraljicIEEETRANS2022,KruseIEEE2020}, corresponding to random walks of
	the I/O power sources.
	
	At higher frequencies, approximately $0.1$--$1\,\mathrm{Hz}$, the measured spectrum is considerably steeper, with a mean exponent $\gamma=2.23(16)$. An approximately inverse-square spectrum over this finite band is compatible with Brownian-like or dynamically filtered broadband forcing~\cite{Pere-Damia-Canaria,KraljicIEEETRANS2022}. However, it does not characterize the complete process: the lower-frequency sector follows a different scaling and contains pronounced periodic and non-stationary structure. The coexistence of the two regimes therefore points to at least two distinct dynamical sectors in the measured fluctuations.
	
	Finally, in the sub-second region we observe a slow rise in the spectrum, compatible with the high-frequency spectrum behavior that can be related to active, high-speed switching frequencies of semiconductor converters used in mass-market devices, modern industrial facilities, and solar  inverters~\cite{RONNBERG20171}. Describing this part of the PSD is out of the scope of our Kuramoto modelling-based approach.
	
	The narrow peaks on top of the broadband spectrum fall into two groups. Resolved with $13\,107\,\mathrm{s}$ segments over one week at B\'ek\'escsaba, the strongest peak below $0.01\,\mathrm{Hz}$ lies at $1/(15\,\mathrm{min})$, one decade above its surroundings, with its harmonics at $2$, $3$ and $4$ per $15\,\mathrm{min}$ and weaker peaks at $1$ and $2$ per hour. These are the quarter-hourly and hourly scheduling periods of the electricity market: generation schedules change in steps at these boundaries, and the resulting deterministic frequency deviations~\cite{ENTSOe2019dfd} appear as spectral lines, consistent with the change of the inverse-variance correlation near $15\,\mathrm{min}$. A second group lies at exact multiples of $1/(60\,\mathrm{s})$, from $1/(60\,\mathrm{s})$ up to $2\,\mathrm{Hz}$, each only one frequency bin wide. Lines at exact multiples of the recording clock point to the measurement chain rather than to grid dynamics, and we do not interpret them physically.
	
	Figure~\ref{fig:psd_comparison} shows the spectrum of the common motion, Eq.~(\ref{eq:common_local}). The variance of the mean frequency's spectrum overlaps by $99.85\%$ with the variance of the individual nodes' frequency spectra, so the site spectrum described above is, to a very good approximation, the spectrum of the collective motion of the synchronous area; this is the part a model is compared with (Sec.~\ref{subsec:hu387}).
	
	We can conclude that spatial  dynamics can generate the functional form of the spectra and long-lived collective modes, but it is not sufficient to reproduce the complete measured spectrum. The empirical data therefore point to a combination of spatial network dynamics and a time-dependent drive. The detailed finite-band fits for an individual measurement site, including the treatment of the high-frequency floor, are given in Appendix Sec.~\ref{supp:bekescsaba_psd}.
	
	\subsubsection{Superstatistical analysis of the measured frequency increments}
	
\begin{figure}[tbp]
		\centering
		\includegraphics[width=\columnwidth]{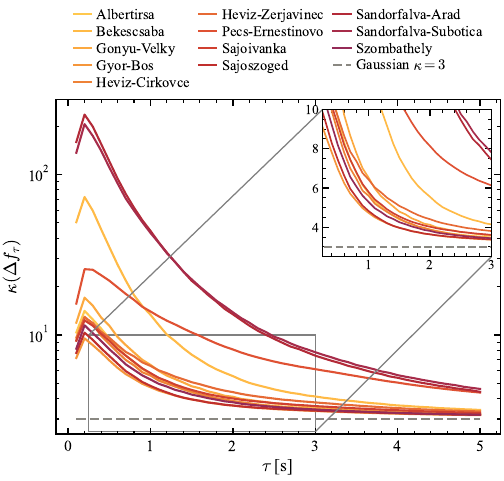}
		\caption{Kurtosis of the measured frequency increments as a function of increment lag. Values greater than three demonstrate non-Gaussian increment statistics, while the largest values are strongly affected by a small number of rare extreme increments.}
		\label{fig:kurtosis_vs_tau}
	\end{figure}

	In this section, we present our results for the superstatistical analysis of the empirical data. Here, the frequency increment statistics provide a direct test of whether the slowly varying fluctuation intensity identified above can account for the observed non-Gaussianity. Figure~\ref{fig:kurtosis_vs_tau} shows that the raw increment distributions are leptokurtic over a broad range of lags. The strongest kurtosis values are, however, dominated by a very small number of extreme increments. The bulk of the distribution is only moderately non-Gaussian, while the outer tail constitutes a distinct sparse population. We therefore use the central distribution for quantitative model comparison and retain the rare extremes as an empirical feature that requires a separate physical or instrumental identification.
	
\begin{figure}[tbp]
		\centering
		\includegraphics[width=\columnwidth]{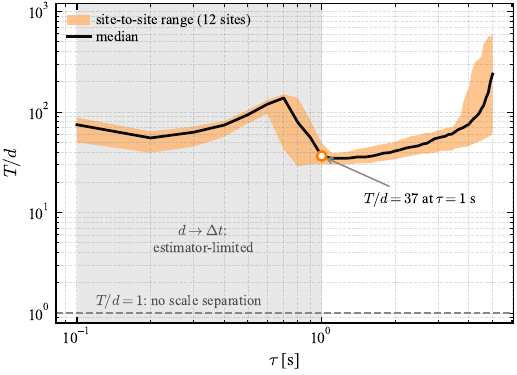}
		\caption{Ratio $T/d$ between the local-Gaussianity window $T$ and the fast increment-correlation scale $d$ for the empirical sites. The large ratio around the one-second increment lag supports a separation between fast increment fluctuations and the slower modulation of their local variance.}
		\label{fig:T_per_d}
	\end{figure}

	A necessary condition for the superstatistical construction is a separation between the fast increment dynamics and the slower variation of the local variance~\cite{BeckCONTMECH2004,schaefer2018nongaussian,Gorjao_2021}. At a one-second increment lag, we obtain
	\begin{equation}
		d\simeq0.61(8)\,\mathrm{s},
		\qquad
		T\simeq21.8(2.3)\,\mathrm{s},
	\end{equation}
	and hence
	\begin{equation}
		\frac{T}{d}\simeq36(5).
	\end{equation}
	The ratio is therefore much larger than unity for the principal analysis scale. Figure~\ref{fig:T_per_d} shows that this separation persists over the useful range of increment lags, apart from the shortest lags where the estimate of $d$ becomes limited by the $0.1$-s acquisition cadence.
	
\begin{figure}[tbp]
		\centering
		\includegraphics[width=\columnwidth]{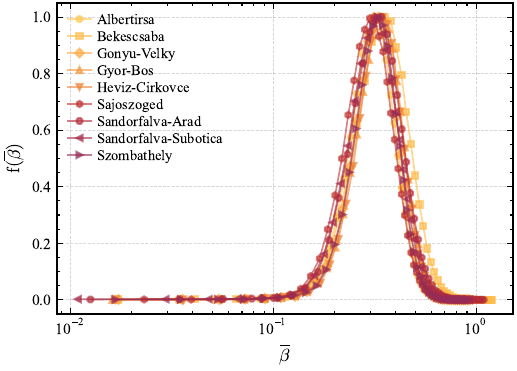}
		\caption{Distributions of the scaled inverse variance $\tilde{\beta}=\beta/\max(\beta)$ for the empirical measurement sites. The similar broad shapes indicate a common component in the slow modulation of the local fluctuation intensity, while the remaining differences reflect site dependence and finite-record effects.}
		\label{fig:scaled_beta}
	\end{figure}

	The inferred inverse-variance $\beta$ distributions are also similar across measurement sites. Figure~\ref{fig:scaled_beta} shows the distributions after scaling each $\beta$ by its sample maximum for visual comparison. Their broad shapes are comparable, supporting the presence of a common system-wide contribution to the slow modulation, although the normalization by the maximum is only a display convention and does not establish an exact universal collapse.
	
	The precise parametric form of $g(\beta)$ is less robust than the existence of the slow variance modulation itself. Under the binned Akaike criterion, Gamma kernels are preferred at ten of the thirteen sites entering the common comparison, whereas a Kolmogorov--Smirnov ranking of the unbinned samples favours an $F$ distribution at ten of thirteen sites.
	\begin{figure*}[tp]
		\centering
		\includegraphics[width=\textwidth]{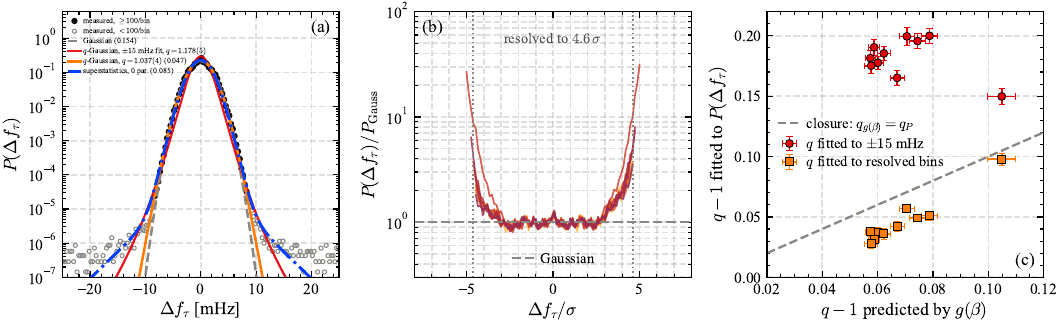}
		\caption{Increment statistics and the superstatistical closure test. (a) One-second increment distribution at Albertirsa; filled symbols are bins with at least $100$ samples, open symbols mark sparser bins. The curves are a Gaussian, a $q$-Gaussian fitted over $\pm15\,\mathrm{mHz}$, a $q$-Gaussian fitted to the resolved bins (two parameters), and the superstatistical prediction, the Gaussian mixture over the measured distribution of $\beta$ with no adjustable parameter; the numbers in brackets are the rms deviation of $\log_{10}P$ over the resolved bins. (b) The increment distributions of eleven sites, each scaled by its own standard deviation $\sigma$ and divided by the unit Gaussian; the dotted lines bound the range resolved at every site, $\pm4.6\,\sigma$.
			(c) Closure test at the ten sites with a Gamma fit of $g(\beta)$: $q-1$ implied by that fit, $q=1+2/(2k+1)$, against $q-1$ fitted to the increment distribution over $\pm15\,\mathrm{mHz}$ (circles) or over the resolved bins (squares); the dashed line is exact closure. The superstatistical mixture accounts for most of the non-Gaussian bulk, whereas the sparse extreme tail remains outside the quantitative closure.}
		\label{fig:supstat_kernel_fit}
	\end{figure*}
	We therefore do not identify a unique superstatistical universality class from the kernel fit alone. Instead, the empirical $g(\beta)$ is used as the primary input to the Gaussian-mixture prediction.
	
	Figure~\ref{fig:supstat_kernel_fit}(a) compares the measured one-second increment PDF at one site with a Gaussian, a directly fitted $q$-Gaussian, and the distribution reconstructed from the measured local inverse variances. The empirical superstatistical mixture reproduces the central distribution substantially better than a single Gaussian. Scaled to their own widths, the distributions of all eleven sites coincide over the resolved range and exceed the Gaussian beyond about $3\sigma$, panel (b). A Gamma approximation to $g(\beta)$ gives a corresponding $q$-Gaussian parameter $q-1\simeq0.069(14)$, while direct fits to the well-populated increment bins give $q-1\simeq0.047(19)$, panel (c); fitting out to $\pm15\,\mathrm{mHz}$, where the bins hold only a few samples, would instead give $0.18(2)$. The agreement is approximate but establishes a non-trivial closure between the independently measured variance fluctuations and the shape of the increment PDF.
	
	The combined empirical picture is therefore internally consistent. Signed frequency increments decorrelate rapidly, but their local variance evolves on a substantially slower and approximately algebraically correlated scale. Mixing the corresponding local Gaussian statistics explains a large part of the measured non-Gaussian increment distribution. At the same time, neither the superstatistical construction nor a single stochastic process accounts for all observations: the rare extreme events remain unresolved, and the PSD exhibits distinct low- and high-frequency scaling regimes. These empirical constraints provide the reference against which the spatial Kuramoto models are tested in the following sections.
	
	\subsection{Kuramoto dynamics on the HU387 and 2D lattice graphs }
	\label{subsec:hu387}
	
	To determine which features of the measured frequency dynamics can arise from spatial network structures and heterogeneous electrical parameters, we consider correlations and the PSD on HU387 and on 2D square lattices with periodic boundary conditions. Scenario 9 of the HU387 power grid~\cite{hartmann_12models} is one of the closest model counterparts to the measured Hungarian system: it retains the transmission-network topology together with heterogeneous nodal injections and electrical parameters. Technical details of the HU387 simulations, including the integration scheme, which is the same for the noiseless and the noisy runs, are given in Appendix Sec.~\ref{supp:hu387_runs}, and the integration scheme, which is the same for the noiseless and the noisy runs, in Supplementary Sec.~\ref{supp:hu387_runs_more}. As this graph has only 387 nodes, possible PL behavior of the KM on it may be hindered; thus, we have extended the comparison to lattices of sizes $L=20, 300, 2000$ to study finite-size effects.
	
	We first test whether the Kuramoto dynamics on the HU387 network may exhibit slow relaxation of the ACF due to fluctuations on low-dimensional networks.
	
	\begin{figure*}[tp]
		\centering
		\includegraphics[width=\textwidth]{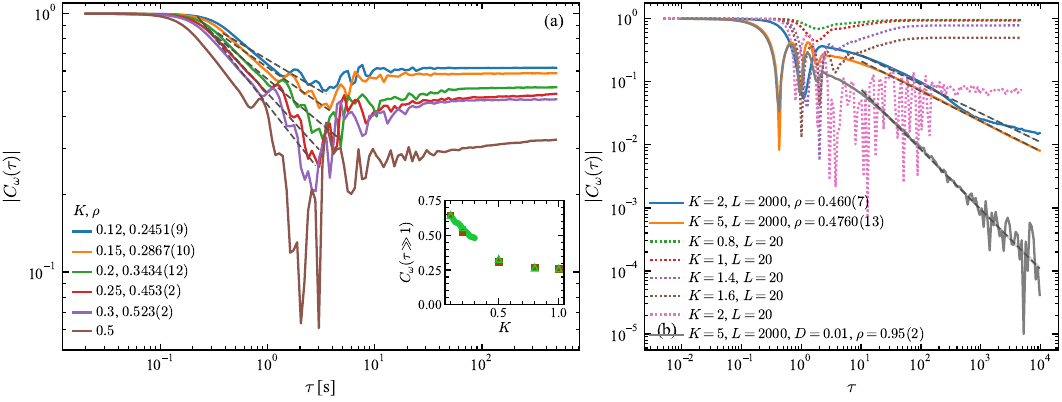}
		\caption{Absolute value of frequency autocorrelation $\lvert C_\omega(\tau)\rvert$ of HU387, Eq.~(\ref{eq:cv}), averaged over $100$ realizations at each coupling.
			(a) $\lvert C_\omega(\tau)\rvert$ on log-log axes for $K=0.12$, $0.15$, $0.2$, $0.25$, $0.3$ and $0.5$ at $D=10^{-3}$, shown up to $\tau=5\times10^{2}\,\mathrm{s}$, before the curves saturate.
			For $D=0$ we found overlapping decay results. The dashed lines are power-law fits $\lvert C_\omega\rvert\propto\tau^{-\rho}$ over the decaying branch of each curve, from where $\lvert C_\omega\rvert$ first falls below $0.9$ to its minimum. The fitted exponents are given in the legend.
			Inset: late-time plateau $\lvert C_\omega(\tau\gg1)\rvert$, the average of $\lvert C_\omega\rvert$ over $\tau\geq10^{3}\,\mathrm{s}$, against $K$. Squares are denote simulations without noise, while triangles correspond to simulations with noise strength $D=10^{-3}$; circles are a finer coupling sweep at $D=10^{-3}$.
			(b) The corresponding correlation for two-dimensional lattices. Solid curves show $L=2000$ and dotted curves show $L=20$; colors denote the coupling values $K$ indicated in the legend. Dashed segments show finite-range power-law fits where a clear decaying regime is observed. Dashed lines are power-law fits $\lvert C_\omega\rvert\propto\tau^{-\rho}$ to the late-time tail,  $\tau\geq10\,\mathrm{s}$, drawn wherever the curve still decays.}
		
		\label{fig:hu387_cv}
	\end{figure*}
	\begin{figure*}[!t]
		\centering
		\includegraphics[width=1\textwidth]{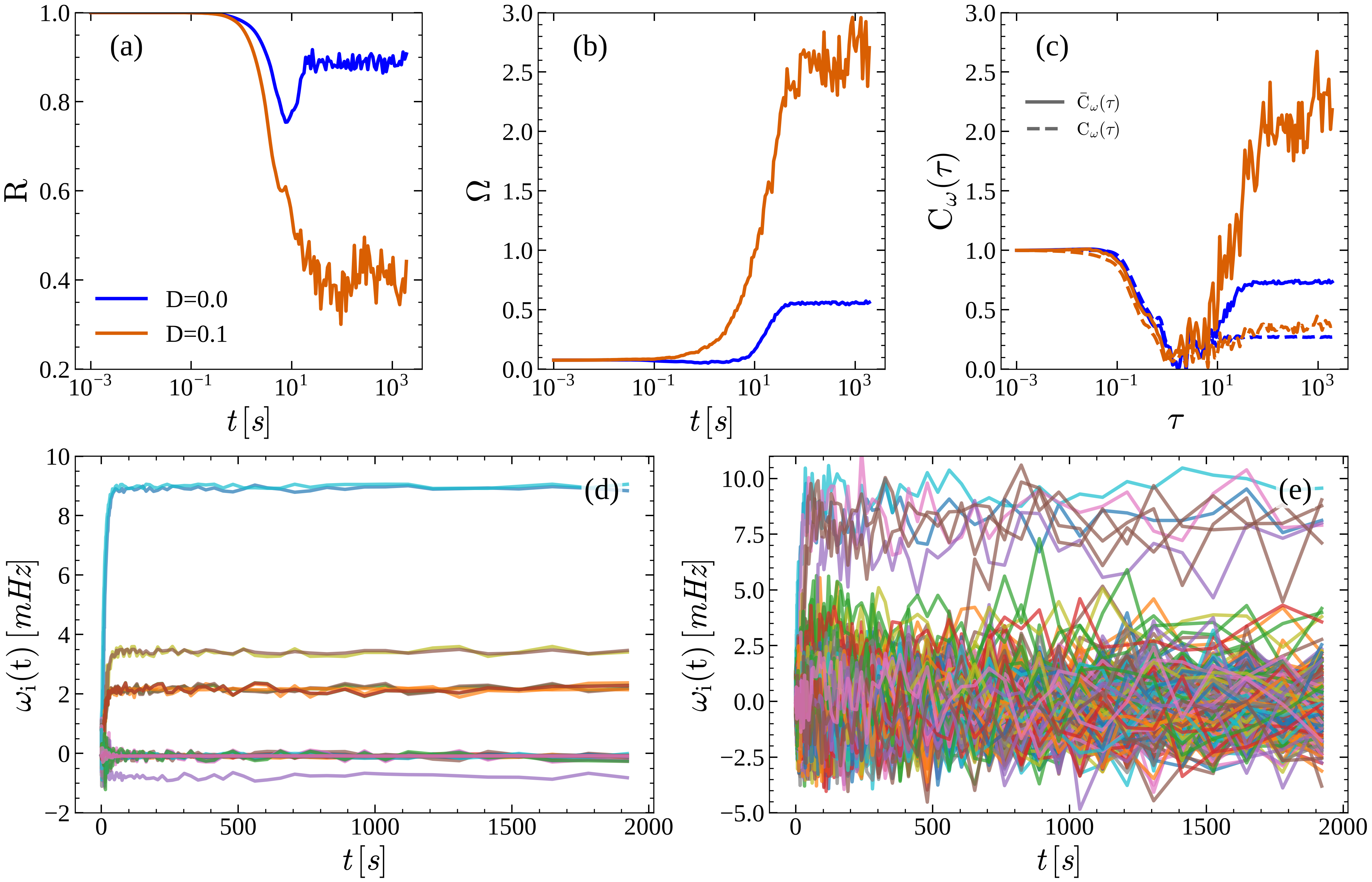}
		\caption{Effects of noise on the dynamics of KM solutions, using HU387 network.  (a) the global phase order parameter, $R$, (b) frequency variance, (c) frequency autocorrelation, and (d)-(e) node frequencies. The blue and brown curves in the upper panel indicate the case for no noise and for $D=0.1$ amplitude stochastic noise. In the bottom panel, the frequency time series for each node has been plotted using different colors for $D=0$ (d) and $D=0.1$ (e). All of these quantities have been measured for one realization of the simulation at the parameter values $K=0.8,~\alpha=0.1$. The outlying, frequency-locked nodes are the same ones as in Fig. \ref{fig:hu387_locking_map}~(d) marked with green and orange: six low degree nodes. Panel (e) demonstrates how stochastic noise resolves frozen states of panel (d)) and removes the complex multi-peak structure. Observe, though, that fluctuations can push one node ($216$) from the central to the peripheral group (brown line for $t>600$).}
		\label{fig:HU_sc9_freq_ts}
	\end{figure*}
	As shown in Fig.~\ref{fig:hu387_cv}(a), the HU387 auto-correlations exhibit an algebraically decaying regime for intermediate couplings. For $0.15\leq K\leq0.3$, finite-range fits of the form
	\begin{equation}
		|C_\omega(\tau)|\sim \tau^{-\rho},
	\end{equation}
	provides exponents approximately in the range $0.25\lesssim\rho\lesssim0.52$. This seems to
	be true both for $D=0$ and $D>0$. The $D=0$ PL decay results coincide with the $D>0$, because the latter happens 
	following a finite crossover time $\tau_c(N,D)$, which makes it invisible on small  graphs, like the HU387, because a 
	saturation or cutoff happens before $\tau_c(N,D)$.
	Similar slow PL decays are observed in the two-dimensional lattice, Fig.~\ref{fig:hu387_cv}(b) indicating that approximately algebraic relaxation can arise from spatially extended Kuramoto dynamics without introducing temporally correlated external noise.
	
	PL fittings for the large $L=2000$, two-dimensional system solutions (Fig.~\ref{fig:hu387_cv}(b)) provide numerical evidence for a universal autocorrelation tail decay $\lvert C_\omega\rvert\propto\tau^{-0.50(2)}$ of KM above the synchronization transition, similar to the $\Omega^{1/2}(t)\propto t^{-d/4}$ scaling~\cite{OdorDengENTROPY2023}, with lattice dimension $d=2$, in the $D=0$ case. The smaller fitted exponents in smaller system are related to finite size effects, that can be confirmed by the controlled 2D lattice simulations. 
	The $D=0$ PL decay tail behavior $\lvert C_\omega\rvert\propto\tau^{-0.50(2) }$ crosses over to $\lvert C_\omega\rvert\propto\tau^{-1.0(1) }$ for any small stochastic noise see Fig.\ref{fig:L2d300_noise_K20}(d). 
	For smaller sizes, the crossover between the $D=0$ and $D>0$ scaling happens at finite $\tau$. Furthermore, the whole PL region shrinks to a narrow $\tau$ region. Note, also that absolute values of the ACF-s are plotted, since they exhibit an oscillation to negative values for small $\tau$-s, in agreement with the sub-second anti-correlations of measured data. 
	The reason here is related to the oscillations of the KM solution in the presence of inertia\cite{xt4m-bkp8}.
	
	At later times the correlations at small $K$-s break down, or for large couplings they approach nonzero plateaus. The lattice results show that the height of this late-time saturation depends strongly on system size thus this is a finite-size effect, instead of nonzero asymptotic correlation in the thermodynamic limit. The occurrence of slow algebraic relaxation in both HU387 and larger sized two-dimensional lattices therefore supports the interpretation that this behavior originates from spatially extended oscillator dynamics above $K_c$. 
	A more detailed, systematic study of this is in preparation.
	
	The frequency-domain comparison reveals an important difference between the measured and simulated systems. In the measurements, approximately $99.85\%$ of the variance of an individual substation frequency is contained in the common motion shared by the observed substations. In the HU387 simulations, by contrast, the variance of the network-mean frequency is only about $0.1$--$0.25\%$ of the variance of a typical node frequency. Thus, most simulated node-frequency fluctuations correspond to relative motion between nodes rather than to coherent motion of the network as a whole.
	
	The two quantities therefore probe different components of the dynamics. Individual-node frequencies contain both the common network motion and relative oscillations between nodes, whereas the network mean suppresses the latter and retains only the coherent component.
	
	We have compared the empirical PSD with those of the simulations of Eq.~(\ref{model_eq}) on the HU387 graph, with $m_i=1$, $\alpha_i=0.1\,\mathrm{s}^{-1}$, and with the noise strength coefficient $D$ in $\mathrm{rad^2\,s^{-3}}$. As Fig.~\ref{fig:psd_comparison} shows, the calculated spectrum of the network-mean frequency resembles a Lorentzian: it is flat below a corner frequency close to $\alpha/(2\pi)\approx0.016\,\mathrm{Hz}$ and decays approximately as $\nu^{-2}$ above it. This form arises even though the added noise is white, i.e.\ without assuming temporally correlated forcing such as an Ornstein--Uhlenbeck process. As shown below, it follows directly from the equation of motion of the network mean.

	For symmetric network couplings, $W_{ij}=W_{ji}$ every interaction term appears once with each 
	sign when Eq.~(\ref{model_eq}) is summed over all nodes. 
	The coupling contribution therefore cancels because $\sin(\theta_j-\theta_i)=-\sin(\theta_i-\theta_j)$.
	
	If the total static injection is balanced, $\sum_i P_i=0$, and all nodes have the same inertia and damping, the mean angular frequency
	$\bar{\omega}=N^{-1}\sum_i\omega_i$
	obeys
	\begin{equation}
		\label{eq:hom}
		\dot{\bar\omega}=-\alpha\bar\omega+\bar\xi,
		\qquad
		\langle\bar\xi(t)\bar\xi(t')\rangle
		=
		\frac{2D}{N}\,\delta(t-t'),
	\end{equation}
	where $\bar{\xi}(t)=N^{-1}\sum_i\xi_i(t)$ is the spatial average of the independent stochastic forcing. Hence the internal coupling redistributes frequency differences among the nodes but does not directly drive the network-mean frequency. This explains why changing $K$ strongly affects the individual-node dynamics while having little effect on the simulated network-mean spectrum.
	
	Above its corner for $2\times10^{-2}< \nu< 2$ Hz the simulated spectrum can be fitted by a power law: with an exponent $\gamma=1.92$ at every $K$ and $D$ investigated, relatively close to the $2.23(16)$ measured exponent value in the $0.1 < \nu < 1$ Hz frequency band. Below the corner, the spectrum is flat and shows none of the measured decay of the $\propto\nu^{-1.628(5)}$. The measured common motion is that of the Continental European synchronous area, of which the Hungarian transmission grid is a small part; an isolated model of that grid has no counterpart to it other than this weakly forced network mean.
	
	The individual-node spectra contain a sequence of narrow peaks associated with oscillatory modes of the network, in which groups of nodes oscillate relative to one another. Above approximately $1\,\mathrm{Hz}$, these modes are sufficiently separated in frequency that they appear as individually resolved peaks. Their characteristic spectral width is set by the damping scale, $\alpha/(2\pi)$; for the present simulations, the measured widths of the peaks are approximately $0.014\,\mathrm{Hz}$. The resulting sequence of peaks forms the fine structure of the simulated node spectra. This structure is unchanged when the integration step is reduced from $0.02$ to $0.005\,\mathrm{s}$ and remains visible after additive noise is introduced, indicating that it is a property of the simulated dynamics rather than a numerical discretization artifact.
	\begin{figure}[tbp]
		\centering
		\includegraphics[width=\columnwidth]{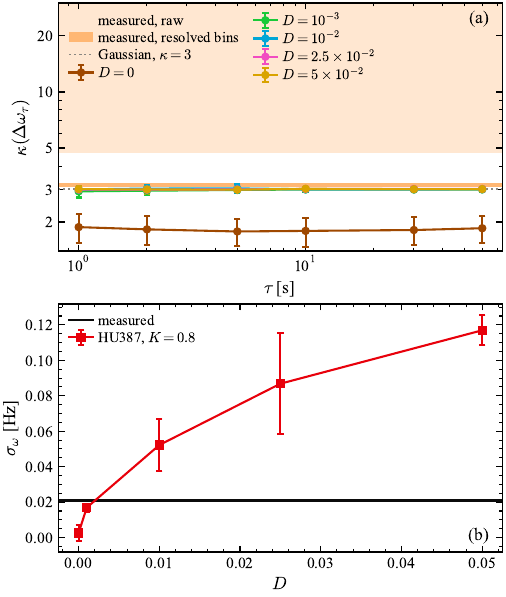}
		\caption{Comparison of distribution width characteristics in HU387 KM simulations for various time lags $\tau$ and different noise strengths $D$ (colored lines) at fixed coupling $K=0.8$, averaged over $20$ realizations and $15$ recorded nodes in the steady state. (a) Kurtosis $\kappa(\Delta\omega_\tau)$ of the frequency increments.
			The shaded regime shows the range of kurtosis for measured data for $\tau=1\,\mathrm{s}$; the Gaussian limit is demarcated by the dotted line at $\kappa=3$. (b) Standard deviation $\sigma_\omega$ of the calculated node-frequency distributions as a function of $D$ (red), compared with the measured value $\sigma_f$ (black line).
		}
		\label{fig:hu387_noise}
	\end{figure}
	
	This behavior differs qualitatively from the measured spectrum. As discussed above, the empirical substation frequencies are dominated by the common motion of the synchronous area, whereas most of the variance of an HU387 node frequency arises from relative motion between nodes. In the present model, the internal coupling strongly influences these relative node dynamics but cancels from the equation governing the network-mean frequency. The discrepancy between the simulated and measured spectra therefore cannot be removed simply by improving the numerical time step or by changing the internal coupling. Reproducing the measured common-mode spectrum requires an additional time-dependent drive that acts coherently, or at least in a correlated manner, across the network.
	
	In Fig.\ref{fig:HU_sc9_freq_ts} we have observed the effect of noise on the collective dynamics of the HU387 power grid network. In the upper panel of the figure, we have plotted the temporal behavior of the order parameter, frequency variance, frequency autocorrelation, and node frequencies for $D=0$ and $D=0.1$. The figure illustrates that in the no-noise case, the system reaches the strong synchronization state characterized by high order parameter values. Even though the simulation of this realization is for $K=0.8 > K_c \simeq  0.2$ the fluctuations do not vanish up to $t=2000$, seemingly in the stationary state, due to the chaotic behavior. Most nodes lock to a common frequency, but a few remain in separate frequency bands/clusters, Fig.~\ref{fig:HU_sc9_freq_ts}(d). For $K\geq0.8$ these are practically always the same $11$ nodes. They form four small groups that are only weakly attached to the rest of the grid. They are marked in Fig.~\ref{fig:hu387_locking_map}(d), and the possible topological reasons why they stay unlocked are discussed in detail in Appendix Sec.~\ref{supp:topinvest}. We conclude that no single measure clearly explains why a node is an outlier; rather, its behavior results from the superposition of many factors. The most significant measure, though, that indicates deviation from the rest of the network is the power-to-strength ratio, $\lvert P_i\rvert/s_i$.
	
	The presence of distinct frequency clusters indicates emergence of partial synchronization akin to Chimera states. This behavior is also reflected in the plotting of frequency variance, the curve saturates near $0.5$ in the steady state. In Fig.\ref{fig:HU_sc9_freq_ts}(c) we have plotted two frequency autocorrelations by two methods: one is normalized by the variance at the starting time ($\sigma_\omega^2(t_0)$), which is $\bar C_\omega$ (solid line), and another is normalized by $\sigma_\omega(t)\sigma_\omega(t_0)$, $C_\omega$, where $\sigma_\omega$ is the standard deviation of frequencies over nodes (dashed dot line). It shows how the time-dependent normalization affects the correlation between frequencies. Next, we have observed that the presence of small stochastic noise causes de-synchronization in the power-grid system (Appendix Sec.~\ref{supp:lattice_noise}). It converts the power grid from a globally phase-locked state into a loosely synchronized state.
	
	By adding stochastic noise, frozen states start to melt; the node frequencies fluctuate strongly and scatter over a much wider range (see Fig.~\ref{fig:HU_sc9_freq_ts}(e)). 
	Groups around the bands of panel (d) are still visible, but the noise can also cause switches between the different frequency groups during simulations.
	Also, the frequency variance and both the autocorrelation increase. However, the autocorrelation normalized by $\sigma_\omega^2(t_0)$ grew rapidly. The fast increase in the frequency variance, autocorrelation, and the broad scattering of node frequency trajectories under small noise account for the upward vertical shift of the PSD curves (Fig.\ref{fig:psd_comparison}).
	
	We have also tested how an additive stochastic white noise generates more fluctuation and broadens the distributions, making them more similar to empirical data. Figure~\ref{fig:hu387_noise} summarizes noise effects on the frequency distribution characteristics at fixed coupling $K=0.8$. At $D=0$, the one-second frequency increment distribution has a kurtosis $\kappa(\Delta\omega_{1\,\mathrm{s}})\simeq1.88$, substantially below the Gaussian value $3$, Fig.~\ref{fig:hu387_noise}(a). Thus, at $K=0.8$, the deterministic chaotic dynamics does not reproduce the measured increment statistics. This conclusion is restricted to the parameter range tested here and does not exclude the possibility that deterministic fluctuations at smaller coupling produce different increment statistics.
	
	Increasing the additive stochastic noise strength $D$ drives the increment kurtosis $\kappa(\Delta\omega_\tau)$ toward the Gaussian value $3$ and simultaneously increases the node-frequency standard deviation $\sigma_\omega$. The simulated value of $\sigma_\omega$ reaches the measured frequency width $\sigma_f$ slightly above $D=10^{-3}$, Fig.~\ref{fig:hu387_noise}(b).
	
	However, agreement between $\sigma_\omega$ and $\sigma_f$ alone does not imply that the simulated and measured fluctuations have the same dynamical origin. In the HU387 simulations, the variance of an individual-node frequency is much larger than the variance of the network-mean frequency, so most of the simulated frequency variability is associated with differences between the instantaneous frequencies of different nodes. In the measurements, by contrast, the simultaneously observed substations show a strong coherent component: a large fraction of their frequency variation is shared across sites rather than arising from independent differences between them (Sec.~\ref{sec:psd_analysis}). This is closely related to the existence of Chimera-like domains in the Kuramoto model. 
	
	For the HU387 simulations considered here, the total static injection is balanced, and the inertia and damping parameters are taken to be homogeneous, with $m_i=1$ and $\alpha_i=0.1\,\mathrm{s}^{-1}$. Under these simulation assumptions, the network-mean frequency obeys Eq.~\eqref{eq:hom}, from which its stationary variance is proportional to $D/(N\alpha)$. At $D=10^{-3}$, the variance of the network-mean frequency is therefore only of order $0.2\%$ of the variance of a typical node frequency. This statement refers specifically to the HU387 simulations and is not intended as an assumption about the real grid. Consequently, although increasing $D$ can reproduce the measured width of the individual-node frequency distribution, it does so mainly by increasing the spread of the node frequencies relative to their network average.
	
	The noise dependence of the increment statistics and that of the spatially normalized autocorrelation are also different. Over the tested range of $D$, the additive stochastic noise changes $\kappa(\Delta\omega_\tau)$ and $\sigma_\omega$ substantially, whereas the intermediate-time decay and late-time level of the spatially normalized autocorrelation $C_\omega(\tau)$ change comparatively little.
	Therefore, the noise has a much stronger effect on rapid local temporal fluctuations than on the persistence of the long-time behavior and decay of $C_\omega(\tau)$.
	
	There is no contradiction between these observations because the two statistics measure different properties. The increment statistics quantify rapid temporal changes of the frequency at individual nodes. In contrast, $C_\omega(\tau)$ compares the frequency at the reference time $t_0$ with the frequency at a later time $t_0+\tau$. Additive noise can substantially broaden the local frequency fluctuations and modify their increment statistics while the large-scale spatial ordering of the node frequencies remains correlated with its earlier configuration.
	
	However, the tested white-noise forcing does not reproduce the empirical combination of observables. The measurements exhibit a modestly non-Gaussian increment distribution together with sparse extreme events and a broadband, multiscale PSD. In the present HU387 analysis, independent noise drives the increment statistics toward Gaussianity without generating the missing empirical low-frequency structure. White noise may still be one component of the physical forcing, but these simulations show that it is not sufficient by itself. A time-dependent and possibly correlated modulation of the nodal injections is therefore a natural next ingredient.
	
\begin{figure*}[tp]
		\centering
		\includegraphics[width=\textwidth]{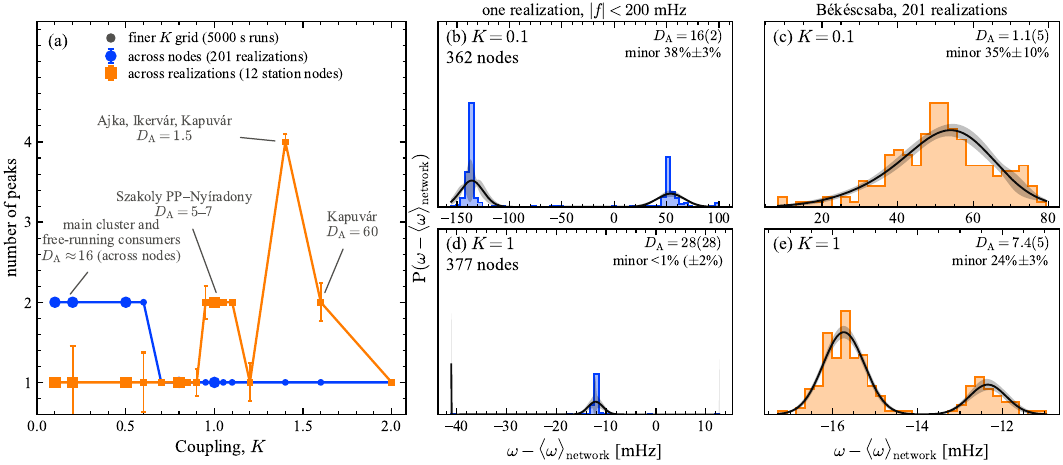}
		\caption{Two forms of multimodality in the noise-free HU387 model, $D=0$. We study the steady-state mean frequency of each node in $201$ realizations of the quenched dispatch disorder, measured relative to the network mean. (a) Number of peaks versus $K$ (Sec.~\ref{subsec:multimodality_methods}), as the median over the realizations across nodes (circles) and over the twelve station-associated nodes across realizations (squares), with $1\sigma$ bootstrap errors. Large markers are the realizations described above; small markers are the same $201$ realizations rerun for $5000\,\mathrm{s}$ on a finer grid of couplings. The annotations name the island whose locking selects the branches and give the median $D_{\rm A}$ of the station nodes. (b,d) Representative distributions across nodes at $K=0.1$ and $K=1$. (c,e) Distribution of the B\'ek\'escsaba node across realizations at the same couplings. Black curves show the fitted mixtures with $1\sigma$ bootstrap bands.}
		\label{fig:hu387_multimodality}
	\end{figure*}

\begin{figure}[H]
		\centering
		\includegraphics[width=\columnwidth]{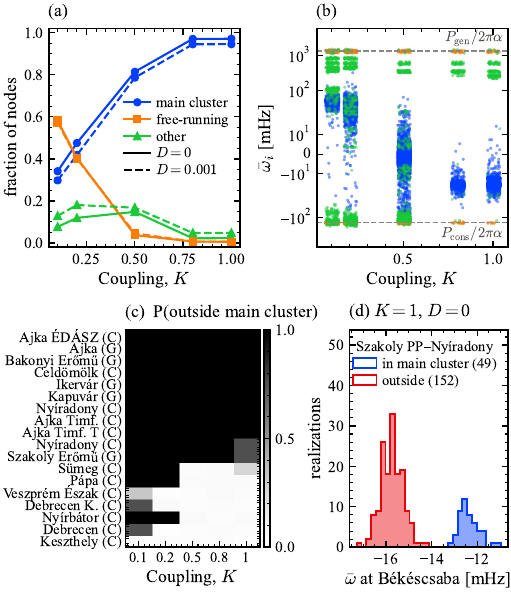}
		\caption{Frequency locking in HU387. Each node of each of the $201$ realizations is classified by its late-time frequency $\bar\omega_i$ as belonging to the main cluster, free-running (close to its uncoupled frequency), or other (separately locked islands and partially entrained nodes); see the text for the precise criteria. (a) Fraction of nodes in each class versus $K$ at $D=0$ (solid) and $D=10^{-3}$ (dashed). (b) Late-time node frequencies for $20$ realizations per coupling at $D=0$, coloured by class; dashed lines mark the uncoupled frequencies of generators and consumers. (c) Probability that a node lies outside the main cluster at $D=0$ for the nodes that remain outside most frequently at strong coupling. (d) Late-time frequency at B\'ek\'escsaba for $K=1$, $D=0$, separated according to whether the Szakoly--Ny\'iradony pair is locked to the main cluster.}
		\label{fig:hu387_locking}
	\end{figure}

\begin{figure*}[htp!]
		\centering
		\includegraphics[width=\textwidth]{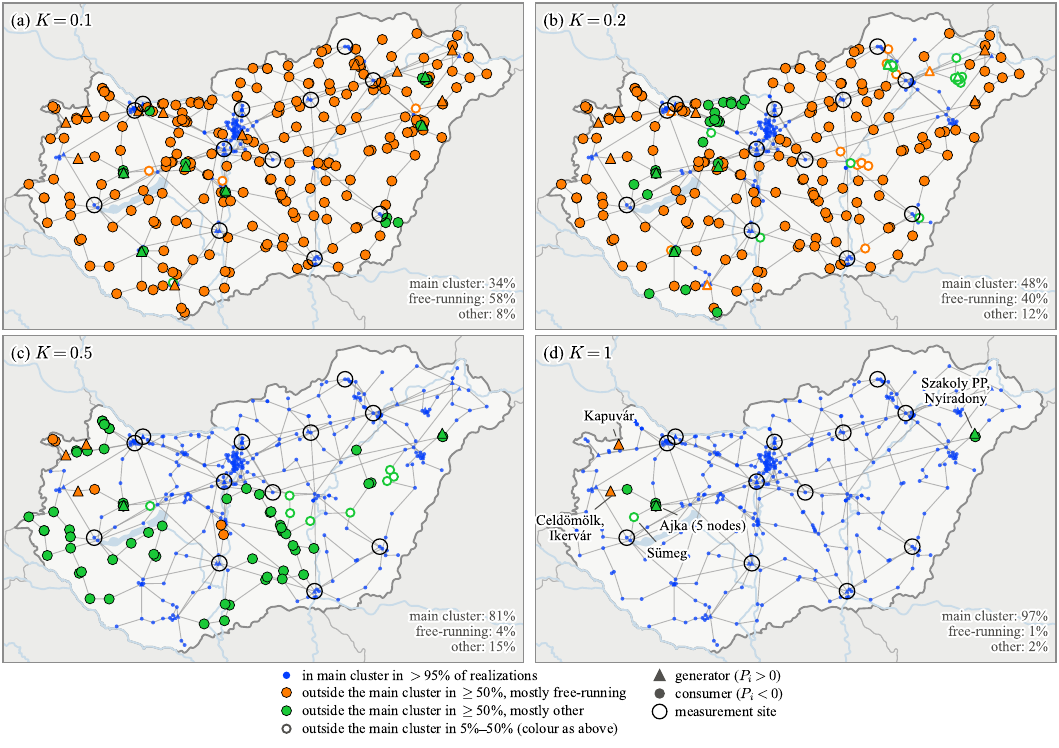}
		\caption{Spatial organization of frequency locking in HU387 at (a) $K=0.1$, (b) $K=0.2$, (c) $K=0.5$, and (d) $K=1$, with $D=0$ averaged over $201$ realizations. In every realization, each node is assigned from its late-time mean frequency $\bar\omega_i$, to one of three classes, as in Fig.~\ref{fig:hu387_locking}: the main cluster ($\bar\omega_i$ within $5\,\mathrm{mHz}$ of the most common late-time frequency), free-running ($\bar\omega_i$ within $10\%$ of its uncoupled frequency $P_i/\alpha$), or other (neither; separately locked islands and partially entrained nodes). Small blue markers: nodes in the main cluster in more than $95\%$ of the realizations. Filled markers: nodes outside the main cluster in at least $50\%$ of the realizations, coloured by the more frequent of their two outside classes (orange: free-running, green: other). Open markers: nodes outside the main cluster in $5$--$50\%$ of the realizations, with the same colour code. The numbers in the lower right corner of each panel give the share of all node--realization pairs in each class. Triangles denote generators and circles consumers; black rings mark the twelve nodes associated with measurement sites. Grey lines show the transmission network.}
		\label{fig:hu387_locking_map}
	\end{figure*}

	The second major empirical feature is the occurrence of multi-peaked frequency  distributions~\cite{hartmann_12models,WenSCIREP2025}. The HU387 model produces multimodality through two distinct mechanisms. At weak coupling, the instantaneous frequency distribution across the $387$ nodes can separate into different spatial populations. We quantify this with Ashman's statistic $D_{\rm A}$, Eq.~(\ref{eq:ashman_d}), and with the number of peaks of a fitted Gaussian mixture, as described in Sec.~\ref{subsec:multimodality_methods}. Across all nodes of a realization, bimodality is found in all $201$ realizations at $K=0.1$ and $0.2$, in $75(3)\%$ at $K=0.5$, and in none at $K=0.8$ and $1$. Counting the peaks of the fitted mixture density directly, Fig.~\ref{fig:hu387_multimodality}(a), gives two peaks across nodes in every realization up to $K=0.5$, in $80\%$ of them at $K=0.6$, and a single peak from $K=0.7$ on. Thus, the across-node multimodality disappears as the network becomes increasingly frequency locked. This happens at $K\approx0.6$--$0.7$, well above the coupling range of the slowly decaying correlations, and tracks the entrainment of the peripheral nodes, Fig.~\ref{fig:hu387_locking}(a). Additive noise does not remove it either (Appendix Sec.~\ref{supp:noise_multimodality}).
	
	A different form of multimodality occurs when a single node is examined across realizations with different quenched injection patterns. At $K=1$ and $D=0$, the distribution of the per-realization late-time mean frequency is bimodal at all twelve station-associated nodes, with the two branches separated by approximately $3.2$--$3.5\,\mathrm{mHz}$. An example for B\'ek\'escsaba is 
	\begin{figure}[H]
		\centering
		\includegraphics[width=\columnwidth]{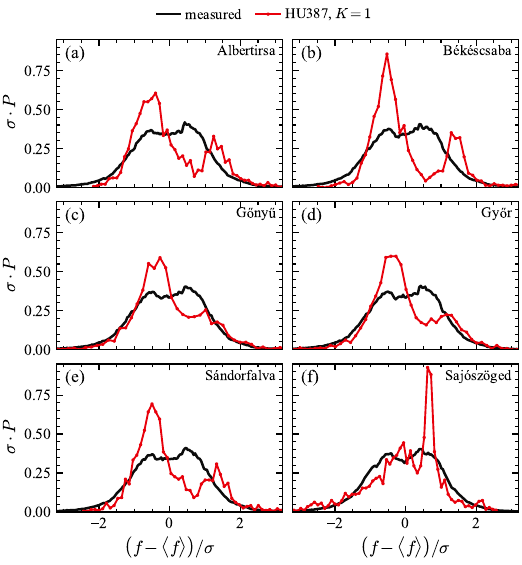}
		\caption{Frequency distributions at six measurement sites (black) and at the corresponding HU387 nodes (red), each scaled by its own standard deviation $\sigma$. Measured: $5\,\mathrm{s}$ window averages pooled over the month. Simulated: frequencies at $K=1$ and $D=0$ pooled over $201$ realizations of the quenched dispatch disorder. The measured width is approximately $\sigma=22\,\mathrm{mHz}$, compared with simulated widths of $1.7$--$3.7\,\mathrm{mHz}$. The measured Ashman statistics are $D_{\rm A}=1.40$--$1.42$, while the simulated values range from $1.59$ to $3.84$. The simulated distributions can therefore develop more clearly separated branches, but with substantially smaller absolute widths.}
		\label{fig:multipeak_compare}
	\end{figure}
	
	\noindent shown in Fig.~\ref{fig:hu387_multimodality}(e), where $D_{\rm A}=7.4(5)$. This across-realization multimodality depends non-monotonically on $K$, Fig.~\ref{fig:hu387_multimodality}(a). At weak coupling, two peaks occur at only four of the twelve station nodes at $K=0.2$ and at none at $K=0.1$ and $0.5$; the larger number of nodes with $D_{\rm A}$ above the threshold at these couplings reflects shoulders and flat-topped single peaks rather than separate peaks. A finer coupling grid shows that above $K=0.8$ the multi-peaked distributions are confined to narrow windows: two peaks for $0.95\leq K\leq1.1$, a single peak at $K=0.8$--$0.9$ and at $K=1.2$, four peaks at $K=1.4$, two at $K=1.6$, and a single peak again at $K=2$. Each window corresponds to a peripheral island whose locking to the main cluster depends on the quenched dispatch, as discussed below and in Appendix Sec.~\ref{supp:locking_scan}. It therefore does not arise from different nodes within a realization, but from different long-time dynamical configurations selected by the quenched initial condition. Figure~\ref{fig:multipeak_compare} compares these strong-coupling ensemble distributions with the measured short-window frequency distributions at corresponding sites.
	
	After rescaling each distribution by its own width, the empirical and simulated profiles can show a qualitatively similar two-peaked structure. However, this comparison concerns the shape only: before rescaling, the measured distributions are considerably broader. This discrepancy persists over the parameter range tested. In the noise-free synchronized regime, the simulated station-node distributions remain substantially narrower than the measured $\sim22\,\mathrm{mHz}$ width. At weaker coupling, the width can approach the empirical scale, but it is then dominated by the separation between frequency-locked nodes and nodes rotating with the same common frequency. Additive noise increases the width and brings it close to the measured value, but at the same time suppresses the across-realization two-peak structure. Among the tested values of $K$ and $D$, we therefore find no parameter set that reproduces both the empirical width and a comparable bimodal structure.
	
	Both forms of multimodality can be understood through a frequency-locking mechanism, a similar phenomenon to phase locking~\cite{Bronski2021}, of HU387, using the classification of the nodes into the main cluster, free-running nodes, and other nodes defined in Sec.~\ref{subsec:multimodality_methods}. This behaviour is reminiscent of Chimera-like patterns. At $D=0$ and $K=0.1$, only $34\%$ of the nodes belong to the main cluster, whereas $58\%$ remain close to their uncoupled frequencies. The synchronized fraction increases with coupling and reaches approximately $97\%$ for $K\geq0.8$, while fewer than $1\%$ of the nodes remain free-running. Noise with $D=10^{-3}$ changes these fractions only weakly.
	
	The spatial organization of these states is shown in Fig.~\ref{fig:hu387_locking_map}. At weak coupling, the main synchronized cluster is concentrated in the densely connected core of the network, while many peripheral consumer nodes remain close to their uncoupled frequencies. Whether a node is entrained depends less on its degree and more on the total weight of its lines (strength of a node), $s_i=\sum_j W_{ij}$, i.e.\ on how strongly it is coupled to its neighbours. At $K=0.1$, the quarter of the nodes with the smallest $s_i$ lie outside the main cluster in $98\%$ of the realizations, and the quarter with the largest $s_i$ is outside of the main cluster in $38\%$ of the realizations. The degree separates the nodes much less clearly: nodes of degree $1$, $2$ and $3$ are outside of the main cluster (behave like outliers or frozen nodes, oscillating with a different frequency than the bulk of the system) in $60\%$, $79\%$ and $70\%$ of the realizations. Nodes of degree $4$ or more are not in the main cluster in $36\%$ of the cases. Dangling nodes, i.e.\ the $26\%$ of the nodes on tree-like branches attached to the meshed part of the network, are not unlocked more often than the others ($64\%$ versus $67\%$). A degree-one node attached to a well-connected substation (degree $\geq4$) is outside the main cluster in $40\%$ of the realizations, but one attached to a low-degree node in $87\%$. At weak coupling, the unlocked nodes are therefore mainly low-degree consumers connected by weak lines; about half of them have degree two. The across-node bimodality at $K=0.1$ and $0.2$ therefore reflects the coexistence of these two populations. As the coupling increases, the peripheral nodes become entrained, and this bimodality disappears.
	
	At strong coupling, almost all nodes belong to the main synchronized cluster, but a few small peripheral islands remain only marginally entrained. For $K\geq0.8$, $11$ nodes lie outside the main cluster in at least half of the realizations. They form four groups, each containing a generator, plus one consumer between two of them (node~45, Celld\"om\"olk): two single generators that rotate close to their uncoupled frequency, one of degree one (node~150, Kapuv\'ar) and one of degree two (node~131, Ikerv\'ar); a generator--consumer pair of two degree-two nodes; and six nodes around a generator substation that is itself well connected (degree eight) but tied to the rest of the grid only by weak lines. Generators make up $9\%$ of the nodes but are outside the main cluster about seven times more often than consumers ($13$--$14\%$ versus $2\%$ of the realizations). Only $4$ of the $11$ nodes are dangling. The common feature is therefore not a low degree as such, but a large net injection relative to the weight of the lines that connect the group to the rest of the network. Each group is attached by one to four lines with $W_{ij}=0.4$--$0.8$, below the median line weight $1.66$ of the network, while the lines inside the groups are considerably stronger. Such a group can lock to the main cluster only if the lines connecting it to the rest can carry its net injection, $K\sum_{\rm cut}W_{ij}\geq\left|\sum_{i}P_i\right|$, where the sum runs over the lines leaving the group. For the degree-one generator, the degree-two generator, the pair and the six-node group, this requires $K\geq1.06$, $0.59$, $0.71$ and $0.77$, respectively. The bound is necessary but not sufficient: with the weak damping used here, a drifting state can remain stable above it, so that the degree-two generator is still free-running at $K=1$. A large injection on a weak attachment thus keeps the group drifting even when the rest of the network is synchronized. Of these groups, only the generator--consumer pair (the Szakoly--Ny\'iradony pair) locks in some realizations at $K\leq1$ and remains in a persistent drifting state in others, depending on the quenched initial condition (power dispatch). This switching between two long-time configurations is the origin of the strong-coupling bimodality across realizations.
	
	The branch is selected during the initial transient: within the first $\sim100\,\mathrm{s}$ the pair either joins the main cluster or settles into its drifting state, and the frequency at B\'ek\'escsaba then stays on one of the two branches until the end of the run (Appendix Sec.~\ref{supp:locking_scan}).
	
	When the pair is unlocked, its nonzero mean frequency must be compensated by a small shift of the remaining nodes. For the observed drifting frequency of approximately $620\,\mathrm{mHz}$, the expected displacement of the remaining $385$ nodes is
	\begin{equation}
		\frac{2\times620\,\mathrm{mHz}}{387}\simeq3.2\,\mathrm{mHz},
	\end{equation}
	in close agreement with the measured branch separation of approximately $3.4\,\mathrm{mHz}$. Consequently, all station-associated nodes exhibit the same two branches even though they are geographically distant from the Szakoly--Ny\'iradony pair. At $K=0.8$ the pair remains unlocked in all realizations, and no across-realization bimodality appears. Thus, the strong-coupling bimodality in HU387 is a global signature of realization-dependent locking of a peripheral, chimera-like island~\cite{10.1063/5.0154581}, rather than a local property of the measured node.
	
	Additive noise suppresses this mechanism. At $D=10^{-3}$, the stochastic broadening of the nodal mean frequency is already larger than the $\sim3.4\,\mathrm{mHz}$ separation of the two branches, causing the across-realization bimodality to disappear (Appendix Fig.~\ref{fig:supp_noise_multimodality}).
	
	\subsection{Kuramoto simulation results on two-dimensional lattices}
	\label{subsec:lattice}
	
	The results on the HU387 network, which has spectral dimension $d_s\simeq2$\cite{hartmann_12models}, establish that KM dynamics on low-dimensional networks can generate slowly decaying autocorrelations. To test this in a controllable way, allowing finite-size analysis, we finally consider KM on periodic square lattices~\cite{OdorDengENTROPY2023,AcebronREVMODPHYS2005}. The following lattice results are therefore used as a controlled reference system rather than as a direct quantitative model of the measured grid, as they lack effects from topological heterogeneity. On the other hand, the effects of stochastic noise can be better identified as in the more complex HU387 simulations. They also allow a controlled way of studying finite-size effects, which introduce a breakdown of PL-s: exponential cutoff or saturation; see the $L=20$ sized 
	\begin{figure}[H]
		\centering
		\includegraphics[width=1\columnwidth]{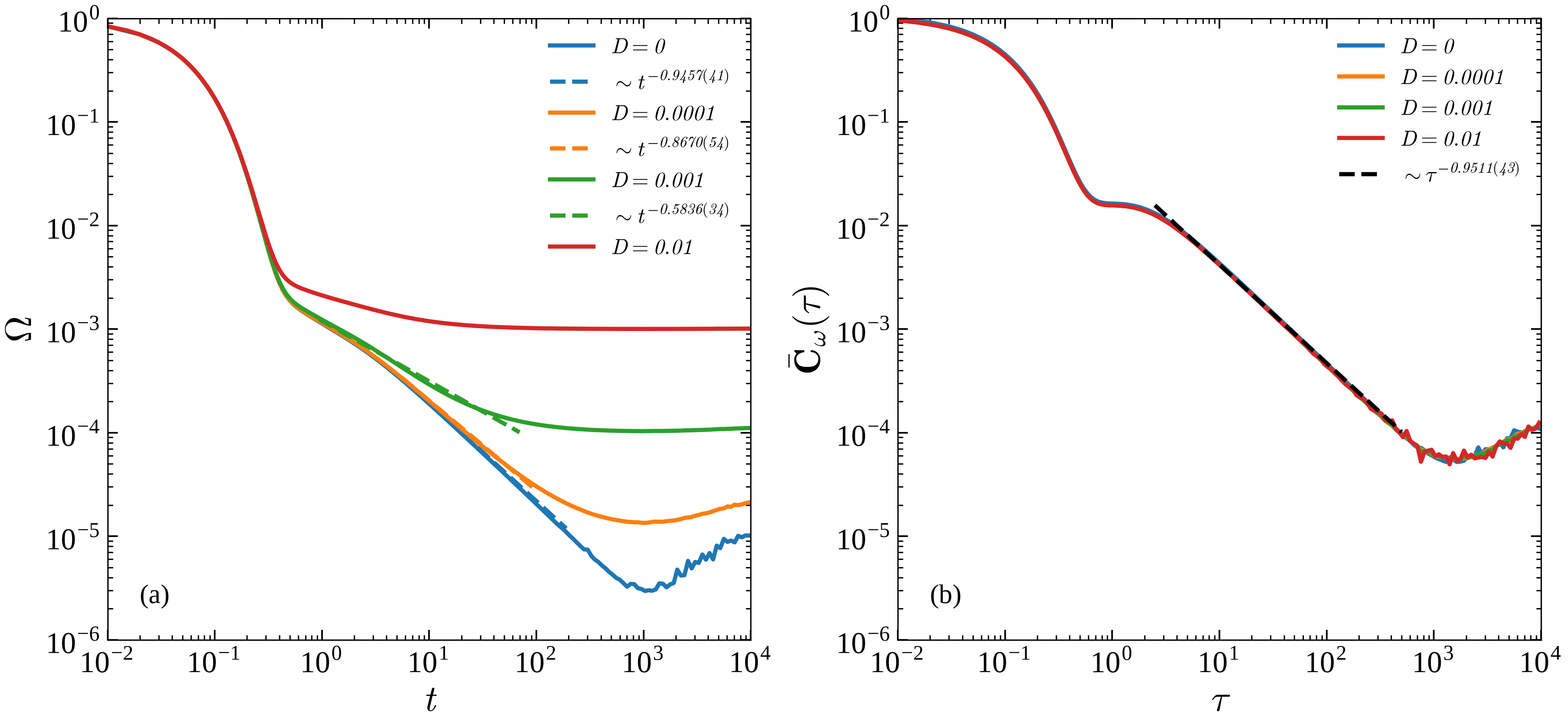}
		\caption{Effect of Gaussian white noise on 2D lattices with $L=300$ in the weak coupling regime: $m=1$, $K=2$ and $\alpha=10$, averaged over $200$ random realizations with independently drawn self-frequencies.
			(a) frequency variance and (b) frequency autocorrelation $\bar C_\omega$, defined by~Eq.~\ref{eq:cbar}.
			Red dashed lines indicate finite-range power-law fits.}
		\label{fig:L2d300_noise_K2_fq_var_ac}
	\end{figure}
	\noindent analysis in the Appendix\ref{supp:lattice_noise}. Thus, we can understand different crossovers and see differences from the mean-field, aggregated single-oscillator behavior.
	
	At first parameter values are set to be $m=1,~\alpha=10,~K=2$, and the initial phases of the oscillators are kept $1$, that is, phase-ordered initial conditions. These parameter values have been chosen arbitrarily in the phase-desynchronized phase. With this setup we have integrated the system Eq.(\ref{model_eq}) up to $t=10^4$, by checking visually appearance of the steady state. The step size of the integration has been taken as $\delta t=0.01$.
	
	\begin{figure}[H]
		\centering
		\includegraphics[width=1\linewidth]{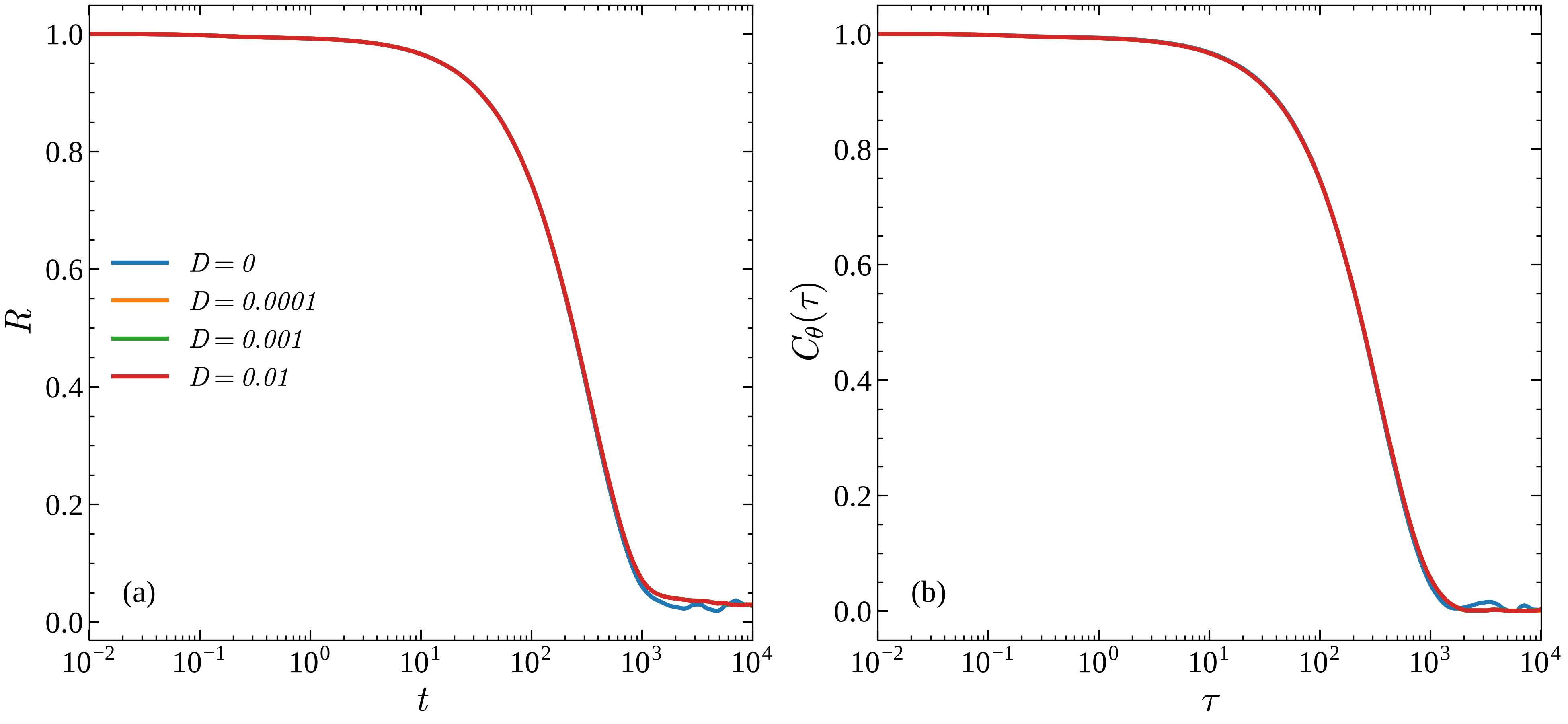}
		\caption{Effect of Gaussian white noise on 2D lattices with $L=300$ (a) the order parameter and (b) the phase auto-correlations defined by (\ref{Adef}), corresponding to Fig.~\ref{fig:L2d300_noise_K2_fq_var_ac}. Parameters: $m=1$, $K=2$ and $\alpha=10$. The quantities are averaged over $200$ realizations with independently drawn self-frequencies. At the specified parameter values, the system remains in an incoherent state after inducing noise. Also, the noise does not affect the phase autocorrelation. }
		\label{fig:L2d_300_K2_R}
	\end{figure}
	
	We have presented the simulation results for frequency variance ($\Omega$) and frequency autocorrelation ($\bar C_\omega(\tau)$) 
	at different noise strength with parameters shown in Fig.\ref{fig:L2d300_noise_K2_fq_var_ac} for phase-ordered initial conditions. 
	Note that at this parameter set, the system stays in the phase-incoherent state. Results for large coupling, $K=20$, order parameter and phase 
	auto-correlations are plotted in Fig.(\ref{fig:L2d_300_K2_R}). 
	It has been observed from Fig.\ref{fig:L2d300_noise_K2_fq_var_ac} that for no noise both the frequency variance and frequency autocorrelation, 
	defined by~Eq.(\ref{eq:cbar}) follow power-law decay with exponents $\simeq 0.95$ in agreement with former large sized 2 dimensional KM  results~\cite{OdorDengENTROPY2023}. The up-veering, large-time part of the $D=0$ curve tells us what happens in smaller systems. \footnote{For very small sizes $L=20$ see Appendix~\ref{supp:lattice_noise}.}
	Note that here, the autocorrelation was calculated by normalizing by the variance of the starting time. As we increase the noise strength, the window size where PL could be fitted shrinks considerably, suggesting a crossover scaling effect: although the fitted exponents seem to decrease, they should stay the same for $L\to\infty$ and the $\Omega(t)$ curves turn to constant. In contrast, $\bar C_\omega(\tau)$ curves keep the size of the scaling region, but start fluctuating with an increase in noise. \footnote{Note, that for $C_\omega(\tau)$
		defined by~Eq.\eqref{eq:cv} there is a $D$-dependent crossover is visible in the auto-correlations, see Fig~\ref{fig:acf_nonstationary}.} 
	
	\begin{figure}[H]
		\centering
		\includegraphics[width=1\linewidth]{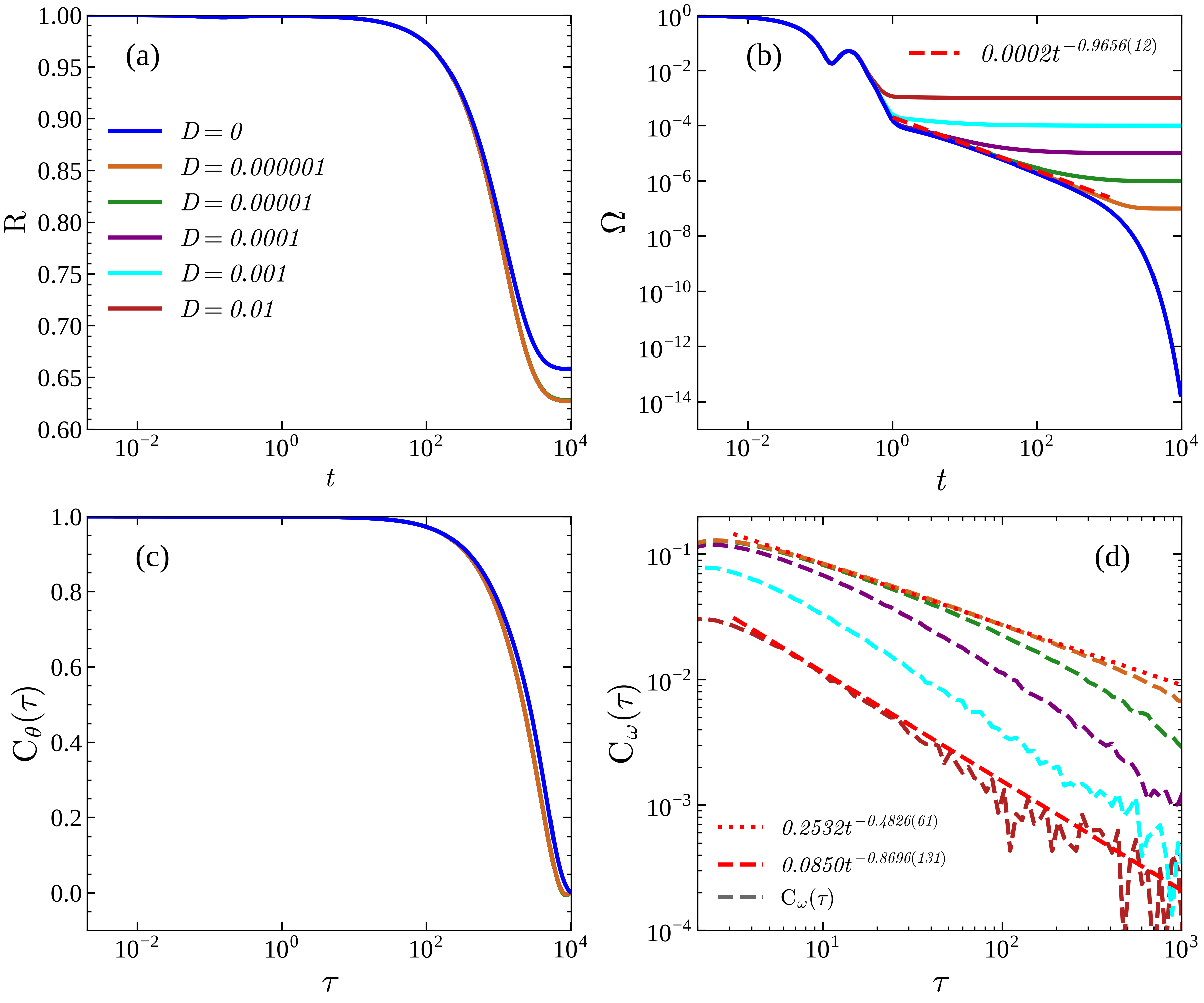}
		\caption{Effect of Gaussian white noise on second-order KM on 2D lattice with $L=300$. (a) the order parameter, (b) frequency variance, (c) phase autocorrelations 
			and (d) two-frequency autocorrelation at $K=20$ for $m=1$ and $\alpha=10$. Solid color lines correspond to different $D$-values shown by the legend of (a). The quantities are averaged over $200$ realizations. Red dotted, and dashed curves indicate representative finite-range power-law fits and show the crossover from the noiseless to noisy scaling. }
		\label{fig:L2d300_noise_K20}
	\end{figure}
	
	To check the effect of noise at the synchronized state, we have increased the value of the coupling strength ($K$) and present the results in Fig.\ref{fig:L2d300_noise_K20}. Starting from $D=0$, we increase its value up to $D=0.01$. For all $D$ the system is in a partially phase synchronization state. Here we have found the frequency autocorrelation by two methods, i.e $\bar C_\omega(\tau)$ and $C_\omega(\tau)$.
	
	We noticed that the lattice system shows similar behavior for $\Omega$ and $\bar C_\omega(\tau)$ as in the case of $K=2$. It is evident from the figure that very small Gaussian noise affects the fitted effective decay exponent of frequency variance more than in the case of $K=2$, but finite sizes cause breakdown of scaling already from $\tau>100$. Again, the autocorrelations $\bar C_\omega$ and $C_\omega$ exhibit different behavior. Small noises do not affect the power-law exponent of $\bar C_\omega$ for $\tau\le 100$, however, fluctuations increase with $D$. On the other hand, the tail exponent $\rho$ of $C_\omega(\tau)\propto\tau^{-\rho}$ increases from $\approx 0.5$ to $\approx 1$ with increasing noise strength. This figure clarifies the effect of the normalization quantity on computing the autocorrelation function. 
	
	\subsubsection{Dependence on the initial phases}
	
	From Fig.\ref{fig:L2d300_noise_K20} we have seen that at $K=20$, the system remains in partial synchronized state with respect to phases, when the system starts from an phase ordered initial condition ($\theta_i=1$). At the same time, we observed that if the phases are initially randomly distributed uniformly, i.e the phases are in $\theta_i\in(0,2\pi)$, then the system stays in an incoherent state. 
	\begin{figure}[H]
		\centering
		\includegraphics[width=1\linewidth]{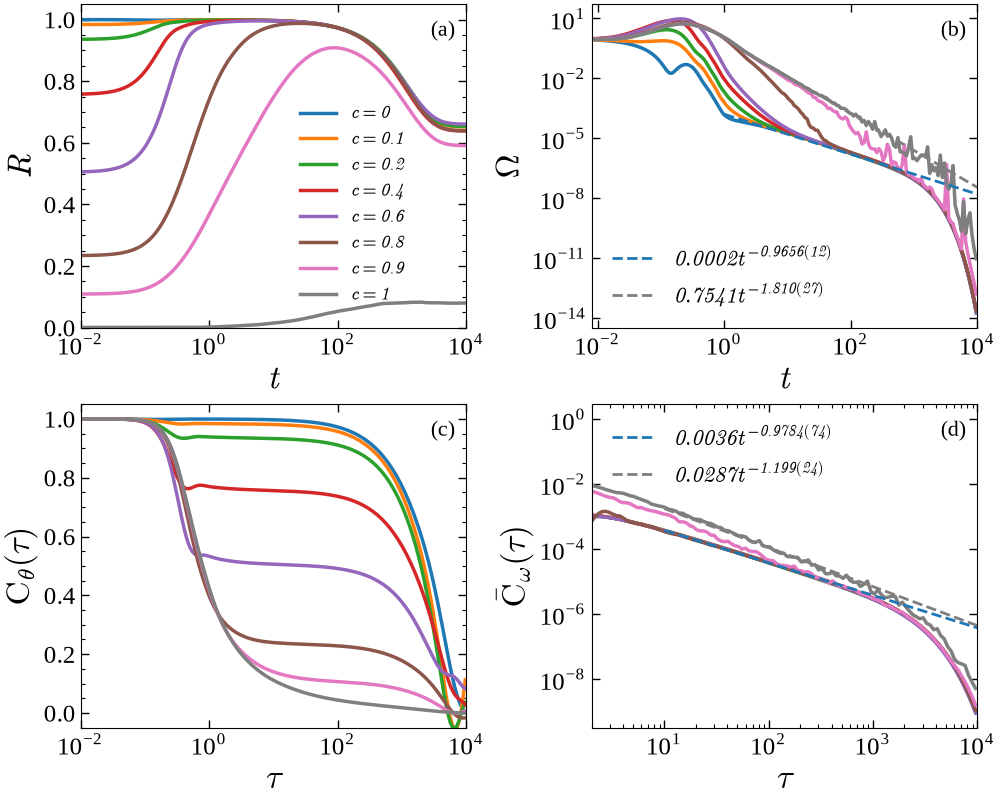}
		\caption{Effect of initial conditions on (a) Order parameter, (b) frequency variance, (c) phase autocorrelation, and (d) frequency autocorrelation of second-order KM on a 2D lattice with $L=300$, $K=20$, $\alpha=10$. The quantities are measured by averaging $200$ realizations for different frequency sets. The red and blue dashed lines show power-law fitting. The colors used are the same for all the sub-figures.}
		\label{fig:L2d300_diff_init_K20}
	\end{figure}
	Therefore, at $K=20$, the system shows bistability between incoherent and partially synchronized states. To check the effect of the initial conditions, we have introduced a parameter $c$, which specifies the initial spread of the phases in $(0,c 2\pi)$. In Fig.\ref{fig:L2d300_diff_init_K20} we have presented the behavior of the considered system for different $c$ values. From the order parameter curves in Fig.\ref{fig:L2d300_diff_init_K20}(a) we have noticed that for $c=0.1$, i.e., when the initial phases are spread in a small regime, the oscillators show partial synchronization. As the $c$ value increases, i.e the range of the initial phases increases, the oscillators start from a low synchronization level, but with increasing time, they reach the fixed point at the partial synchronization state. However, for large values of $c~(>0.9)$, the oscillators stay in an incoherent state. From Fig.\ref{fig:L2d300_diff_init_K20}(b) we have seen that $\Omega$ decays can be fitted in the $1<t<1000$ interval with different exponents, roughly:
	$\rho \propto 1.85$ and $\rho\propto 1$, for the incoherent and partial synchronization state respectively. The latter value agrees with the linear approximation law: $\Omega\propto t^{-d/2}$~\cite{OdorDengENTROPY2023}. Besides that, we have noticed that for larger $c$ values, the curve $\Omega$ is not smooth; but fluctuates and the $\Omega\propto t^{-1}$ decay crosses over to $\Omega\propto t^{-1.85}$ before breaking down at a finite size cutoff $t \simeq 1000$. Then in Fig.\ref{fig:L2d300_diff_init_K20}(c), we have shown how $C_\theta(\tau)$ changes with time for different $c$. Further, Fig.\ref{fig:L2d300_diff_init_K20}(d) shows 
	that as $\Omega$, $\bar C_\omega(\tau)$ also goes through a crossover to a smaller exponent for large $c$. 
	But in this case the exponent value reduces to $\rho\simeq1.2$, which is smaller than the exponent for $\Omega$.
	
	\subsubsection{Two-dimensional lattice power spectrum}

\begin{figure}[H]
		\centering
		\includegraphics[width=\columnwidth]{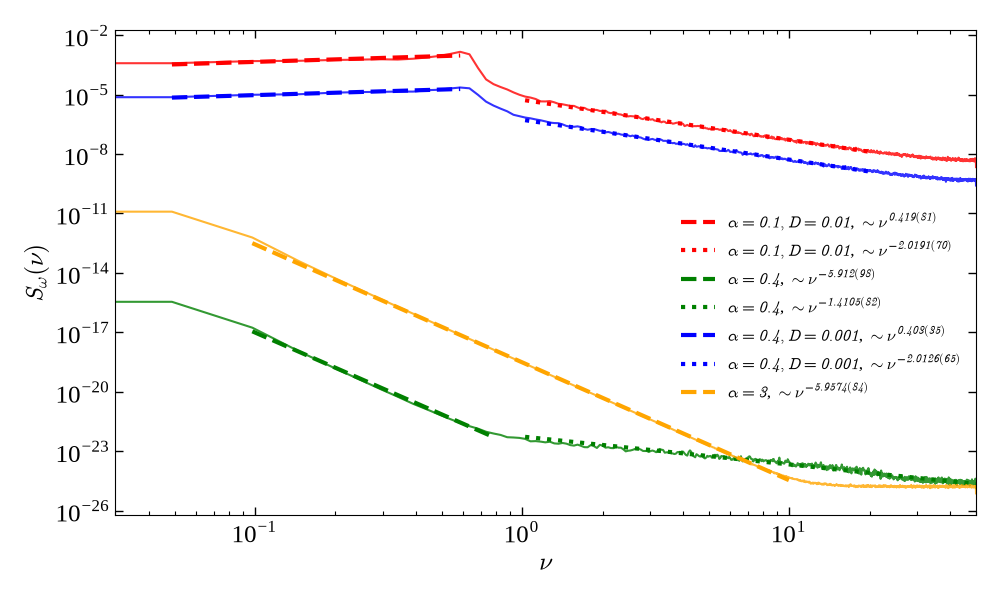}
		\caption{Node-frequency power spectra of the two-dimensional KM solutions on $L=300$ sized periodic lattices, for $K=4$ and $\alpha=\{0.1,0.4,3\}$ in the steady-state. The lower curves correspond to the noiseless solution, the upper ones to $D=0.001,0.01$. Labels show the parameters and the PL fitting forms corresponding to the dashed lines.}
		\label{fig:lattice_psd}
	\end{figure}

	Figure~\ref{fig:lattice_psd} shows the power spectral density of the KM-generated frequencies for $N=300 \times 300$ lattices, with parameters $K=4.0, c=0.0$, and $\alpha = \{0.1,0.4,3.0\}$, in the steady-state ($t > 1000$).  For small damping factors, we can observe a PL decay tail, which can be fitted with $S_\omega(\nu)\propto \nu^{-\gamma}$, with exponent $\gamma=1.410(8)$ with respect to the empirical observations at high frequencies: $\gamma > 2$. But when $\alpha$ is large, the damping will cause $\gamma\simeq 6$. In the case of additional white noise with amplitudes $D=0.001,0.01$ the PSDs exhibit similar behavior to the HU387 case: initial almost constant region crossing over to  $S_\omega(\nu)\propto \nu^{-2}$ at $\nu\simeq 1$ without adding Brownian noise to the KM.
	
	For the noisy 2D lattice case, similarly to the HU387 simulations, the relation between the exponents of the low-frequency PSD and the long-time ACF, $1=\rho+\gamma$, originating from the Wiener-Khinchin theorem,  seems to hold within numerical error, since $\gamma\simeq 0$ and $\rho\simeq 1$ (see Fig.~\ref{fig:hu387_cv}(b)). Deviations might come from frequencies not sampled from the fully steady state.
	
	\subsection{Comparison across empirical data and spatial models}
	\label{subsec:comparison}
	
	\begin{table*}[!ht]
		\centering
		\small
		\setlength{\tabcolsep}{6pt}
		\renewcommand{\arraystretch}{1.18}
		
		\begin{tabular}{@{}
				>{\RaggedRight\arraybackslash}p{0.12\textwidth}
				>{\RaggedRight\arraybackslash}p{0.25\textwidth}
				>{\RaggedRight\arraybackslash}p{0.29\textwidth}
				>{\RaggedRight\arraybackslash}p{0.24\textwidth}
				@{}}
			\toprule
			\textbf{Observable} &
			\textbf{Measurements} &
			\textbf{HU387 model} &
			\textbf{Square lattice} \\
			\midrule
			
			\textbf{Temporal PSD}
			& Two finite scaling bands,
			$\gamma=1.628(5)$
			($10^{-3}$--$2\times10^{-2}\,$Hz)
			and $2.23(16)$ ($0.1$--$1\,$Hz),
			site means; market lines at $1/15\,$min
			and harmonics; $99.85\%$ of the variance
			common to all substations
			& $D=0$: network mean vanishes identically,
			Eq.~(\ref{eq:hom}).
			
			\medskip
			$D>0$: network mean Lorentzian, flat below
			$\alpha/2\pi$ and $\gamma=1.92$
			above ($0.02$--$2\,$Hz) at every $K$
			and $D$; three to five decades below
			the measured common motion at $10^{-3}\,$Hz.
			
			\medskip
			Both: node spectra dominated by swing-mode
			peaks of width $\simeq\alpha/2\pi$
			& $D=0$: $\gamma\simeq 1.4$
			for low $\alpha$ and high frequencies.
			
			\medskip 
			$10^{-3}$:
			$\gamma\simeq 2$ for low $\alpha$ and high frequencies .
			
			\medskip
			Larger $D$: flat stochastic background \\
			\addlinespace[4pt]
			\midrule
			\addlinespace[4pt]
			
			\textbf{Auto-correlation}
			& Increments: $1/e$ time $d=0.61\,$s,
			then oscillation about zero.
			
			\medskip
			Inverse variance:
			$C_\beta\propto \tau^{-\rho_\beta}$, $\rho_\beta=0.26(2)$,
			over $20$--$900\,$s
			& $D=10^{-3}$, $0.15\leq K\leq0.3$:
			$\lvert C_\omega\rvert\propto\tau^{-\rho}$
			with $\rho\simeq0.25$--$0.52$
			up to $\sim500\,$s, then a late-time plateau.
			
			\medskip
			$D=0$: decay and plateau change little
			& $L=2000$, $K>K_c$: tail
			$\lvert C_\omega\rvert\propto\tau^{-\rho}$
			with $\rho=0.50(2)$ at $D=0$
			and $\rho=1.0(1)$ for any $D>0$ \\
			\addlinespace[4pt]
			\midrule
			\addlinespace[4pt]
			
			\textbf{One-second increment PDF}
			& Modestly non-Gaussian resolved bulk
			plus sparse extreme events
			& $D=0$ ($K=0.8$): sub-Gaussian,
			$\kappa\simeq1.88$.
			
			\medskip
			$D>0$: $\kappa\to3$ (Gaussian)
			with increasing $D$; $\sigma_\omega$
			reaches the measured width $\sigma_f$ slightly
			above $D=10^{-3}$
			& Not analysed on directly comparable
			stationary records \\
			\addlinespace[4pt]
			\midrule
			\addlinespace[4pt]
			
			\textbf{Variance mixture}
			& Operational $T/d\simeq36(5)$;
			Gaussian mixture describes much
			of the resolved bulk
			& Not applicable at any tested $D$:
			the increments are never leptokurtic,
			so there is no excess kurtosis
			for a variance mixture to explain for $K=0.8$
			& Not tested \\
			\addlinespace[4pt]
			\midrule
			\addlinespace[4pt]
			
			\textbf{Multiple peaks}
			& Broad, flat-topped $5\,$s-window PDFs,
			the same at every site
			($\sigma\simeq22\,$mHz,
			$D_{\rm A}\simeq1.4$,
			below the two-peak threshold of $2$);
			$D_{\rm A}\simeq0.9$ after
			$5\,$min averaging
			& $D=0$: two peaks across nodes
			up to $K=0.5$; across realizations,
			two peaks in narrow windows
			($0.95\leq K\leq1.1$, $K=1.6$)
			set by island locking; widths only
			$1.7$--$3.7\,$mHz.
			
			\medskip
			$D=10^{-3}$: across-realization
			bimodality removed
			& Frequency PDFs not analysed. 
			In the lack of network heterogeneity, the Gaussian
			$P_i$ self-frequencies are not expected to cause multi-peaks,
			in the sample averages.
			Bistability between
			incoherent and partially synchronized
			states, selected by the initial
			phase spread \\
			
			\bottomrule
		\end{tabular}
		
		\caption{Summary of the empirical and model observables.
			$D$ is the strength of the additive white noise;
			$D=0$ is the deterministic model.
			Numerical spectral slopes and correlation exponents
			refer only to the stated finite fitting ranges
			and model settings.}
		\label{tab:exponents}
	\end{table*}
	
	Table~\ref{tab:exponents} summarizes the hierarchy of comparisons. Spatial coupling is sufficient to generate slow relaxation, heterogeneous frequency fields, and multi-peaked states without introducing annealed stochastic forcing~\cite{hartmann_12models,HartmannSEGAN2024,OdorDengENTROPY2023}. These features are therefore not uniquely diagnostic of external noise. However, neither the deterministic lattice nor the present HU387 simulations reproduce the empirical temporal spectrum and increment statistics simultaneously.
	
	The measurements instead indicate a combination of spatial dynamics and slowly varying temporal forcing. The superstatistical analysis shows that the local fluctuation variance evolves on a scale well separated from the fast increments, whereas the deterministic network contains only quenched injection heterogeneity. The natural next extension is therefore not to replace the spatial model by a one-node stochastic process, but to retain the network and introduce time-dependent nodal injections whose temporal statistics are constrained independently by data.

	\section{Conclusion}
	\label{sec:conclusion}
	
	We have compared one month of subsecond-rate sampled synchronous frequency measurement data from sixteen sites of the Hungarian transmission grid with those from second-order Kuramoto dynamics on the HU387 transmission-network model and on two-dimensional periodic lattices. Previous empirical studies have already demonstrated that power-grid frequency fluctuations display non-Gaussian statistics, spatial heterogeneity, and non-trivial temporal correlations~\cite{schaefer2018nongaussian,Gorjao_2021,KraljicIEEETRANS2022}. Our results show that the high-resolution Hungarian data likewise contain a combination of spatial heterogeneity and non-stationary temporal dynamics that cannot be reduced to a single simple stochastic process.
	
	At short averaging times, the measured frequency distributions retain a pronounced site dependence and can develop broad or multi-peaked profiles, whereas averaging over several minutes produces predominantly unimodal distributions centered close to the nominal frequency. Such spatially dependent frequency statistics have also been observed in other synchronous-grid measurements~\cite{Gorjao_2021,WenSCIREP2025}. The temporal statistics show a similarly multiscale structure. Signed frequency increments decorrelate rapidly, while the window-wise inverse variance evolves on a substantially slower scale and exhibits persistent 
	PL decaying correlations. The separation $T/d\gg1$, together with the leptokurtic increment statistics, supports a superstatistical description in which fast, approximately local-Gaussian fluctuations are modulated by a slowly evolving variance~\cite{BeckCONTMECH2004,schaefer2018nongaussian,Gorjao_2021}. The corresponding Gaussian mixture accounts for much of the well-resolved non-Gaussian increment distribution, although the sparse extreme events remain outside this description.
	
	The power spectra provide an independent signature of this separation of scales. The measured PSD cannot be represented by a single power law: the low-frequency range follows an exponent $\gamma\simeq1.63$, whereas the higher-frequency spectrum approaches an approximately inverse-square form. The latter is compatible, over a finite frequency interval, with Brownian-like or dynamically filtered stochastic forcing, but the low-frequency sector and the pronounced operational periodicity 
	demonstrate that such a description is incomplete.
	The empirical low-frequency PSD exponent $\gamma\simeq 1.63$ entails non-stationary behavior even for long times, where the Wiener-Khinchin theorem breaks down and a generalized scaling law: $3=\gamma+\rho$ should be applicable~\cite{Nonstat-WK}. This provides an estimate for the autocorrelation tail decay exponent $ \rho\simeq 1.37$, which we could not confirm due to the large oscillations caused by the periodic 15-min power trading. However, for UK and SA, about half of this exponent is reported~\cite{KraljicIEEETRANS2022}, which is possible due to the way of their frequency autocorrelation is defined. Similar departures from simple stationary stochastic descriptions have motivated more elaborate statistical models of grid-frequency dynamics~\cite{KraljicIEEETRANS2022,KruseIEEE2020}. In addition, deterministic frequency deviations associated with scheduled generation and market operation introduce characteristic time scales that are not captured by an autonomous noise process~\cite{ENTSOe2019dfd}. The empirical frequency dynamics therefore contain slow collective modulation in addition to fast stochastic fluctuations.
	
	The KM on $\simeq 2$ dimensional graphs clarify which parts of this behavior can arise from the network dynamics themselves. Kuramoto-type oscillator models provide a minimal description of synchronization and loss of synchrony in coupled power networks~\cite{FilatrellaEtAlEPJB2008,DorflerPNAS2013,AcebronREVMODPHYS2005} but provide an understanding of the relevant physical processes in a controllable manner. Solutions of KM on the HU387 power-grid network, quenched heterogeneity and nonlinear coupling generate chaotic noise, maximal at $K_c$,  persistent Chimera-like spatial frequency patterns, discrete spectral modes, and multi-peaked states even without annealed noise. In agreement with our previous analysis of increasingly heterogeneous Hungarian grid models~\cite{hartmann_12models,HartmannSEGAN2024}, the simulated multi-peaked distributions demonstrate that spatial structure and different injection configurations alone can generate qualitative features resembling to the empirical frequency PDFs. In HU387, we identified two such mechanisms: the coexistence of synchronized and unlocked low-degree nodes at weak coupling, and the realization-dependent locking of weakly attached peripheral islands at strong coupling. None of the tested parameter sets, however, reproduces the measured width and bimodality simultaneously; the profiles are similar only after rescaling. Quantitatively, however, the deterministic model underestimates the measured fluctuation width and does not reproduce the broadband empirical spectrum. The very close frequency values of the empirical measurement substations locate this discrepancy: the measured spectrum is almost entirely characterized by the rotation of the oscillators with the common frequency, and if there is a frequency change, then all nodes change together in the same direction. In the case of the HU387, it does not depend on the network and is only weakly forced, whereas the simulated node spectrum is dominated by swing modes of nodes oscillating against each other. Adding independent Gaussian noise broadens the simulated frequency distribution, drives short-time increment statistics toward Gaussianity, and smears the multi-peak distributions. However, the tested model with white-noise forcing does not reproduce the measured combination of spectral, correlation, and distributional properties exactly, but results in similar functional forms. Note that the 15 min periodic power I/O has not been taken into account in our modeling.
	
	The square-lattice results show that slow relaxation and non-trivial spectral structure are not specific to the detailed Hungarian topology and allow for a finite-size scaling study to identify the basic statistical physics effects. Note that the spectral dimension of the HU387 is $d_s\simeq 2$~\cite{hartmann_12models}. The second-order Kuramoto model on finite-dimensional lattices is known to exhibit synchronization behavior that differs from the mean-field limit and can display strong finite-dimensional, finite-size and initial-condition effects~\cite{OdorDengENTROPY2023}. Here we show that algebraic relaxation of the auto-correlations appears in the locally coupled, finite-dimensional system for $K>K_c$ and with PL exponent depending on the noise and on the initial phase synchronization state. The lattice therefore identifies generic consequences of spatial coupling with respect to the mean-field or one-oscillator models used, whereas the HU387 calculations demonstrate how these mechanisms are modified by realistic topology and heterogeneous injections. 
	
	The finite-size scaling study has proven that to observe PL decaying correlations, we need sufficiently large systems. On the $L=20$ lattice, as well as on the 387-node HU-HV network, PL-s are suppressed, but in reality the HU387 is just a small part of the huge European power grid and a network with lower voltages; thus, our large-system simulation result should be able to explain several complex features even without special noise assumptions. Note that we have not seen PL frequency ACF-s in our Hungarian data, like those in the UK and SA data~\cite{KraljicIEEETRANS2022,lonardi2026understandingcomplexityfrequencyphase}; probably the periodic oscillations suppress them. But we do observe them in the smeared $\beta(t)$ functions, with exponent $\rho_\beta=0.26(2)$.
	
	Taken together, the results suggest that a realistic stochastic power-grid model requires both spatial network dynamics and explicitly time-dependent driving. A natural next step is therefore to retain the heterogeneous transmission network while replacing purely quenched or independent white-noise forcing by temporally evolving nodal injections constrained by measured load, generation, dispatch and control statistics. This is consistent with the broader move toward stochastic grid-frequency models that incorporate correlated and state-dependent forcing rather than assuming a single stationary Gaussian source~\cite{KraljicIEEETRANS2022,Oberhofer_2023,GorjaoEUROPHYS2025}. Such a model should be tested simultaneously against several observables---the node-frequency PSD, increment PDFs, variance correlations and synchronized spatial measurements---rather than tuned to a single exponent or distribution. In particular, simultaneous measurements at the available substations would allow the spatial correlation observables of the model to be constructed directly from data. This would provide a more stringent test of whether the slow variance modulation and the observed multiscale frequency statistics originate from switching between spatial network states, externally driven operating conditions, or their interaction.

	\section*{Acknowledgements} The authors thank for the useful discussions with Pere Colet and Damià Gomila. We thank access to the Hungarian national supercomputer network via hpc.dkf.hu.
	
	\section*{CRediT authorship contribution statement}
	
	\textbf{K.B.: }Conceptualization, Methodology, Software, Validation, Formal Analysis, Investigation, Data Curation, Writing and Visualization; \textbf{M.C.:} Writing - Original draft, Visualization, Investigation and Formal Analyses, Data curation; \textbf{S.D.: } Validation, Formal Analysis, Investigation, Writing – Original Draft; \textbf{A.B.:} Software, Writing – Original Draft ; 
	\textbf{B.H.:} Data, Validation ; \textbf{G.O.:} Conceptualization, Methodology, Investigation and Formal Analyses, Validation, Writing, Supervision, Project Administration 
	\\All authors have read and reviewed the manuscript before submission.
	
	\section*{Competing interests} 
	The authors declare that they have no known competing interests that could influence the work reported in this article.
	
	\section*{Data Availability} 
	Data and code are available upon request from the corresponding author.
	
	\section*{Research Funding}
	This work was supported by the National Research, Development and Innovation Oﬀice NKFIH under Grant No. K146736.
	
	\FloatBarrier
	\bibliographystyle{unsrt}
	\bibliography{sample}

@article{Gorjao_2021,
doi = {10.1088/1367-2630/ac08b3},
url = {https://doi.org/10.1088/1367-2630/ac08b3},
year = {2021},
month = {jul},
publisher = {IOP Publishing},
volume = {23},
number = {7},
pages = {073016},
author = {Gorjão, Leonardo Rydin and Schäfer, Benjamin and Witthaut, Dirk and Beck, Christian},
title = {Spatio-temporal complexity of power-grid frequency fluctuations},
journal = {New Journal of Physics}
}

@article{xt4m-bkp8,
  title = {Effects of inertia on the asynchronous state of a disordered Kuramoto model},
  author = {Kati, Yagmur and Toenjes, Ralf and Lindner, Benjamin},
  journal = {Phys. Rev. E},
  volume = {112},
  issue = {4},
  pages = {044301},
  numpages = {11},
  year = {2025},
  month = {Oct},
  publisher = {American Physical Society},
  doi = {10.1103/xt4m-bkp8},
  url = {https://link.aps.org/doi/10.1103/xt4m-bkp8}
}

@Article{e22060666,
AUTHOR = {Ódor, Géza and Hartmann, Bálint},
TITLE = {Power-Law Distributions of Dynamic Cascade Failures in Power-Grid Models},
JOURNAL = {Entropy},
VOLUME = {22},
YEAR = {2020},
NUMBER = {6},
ARTICLE-NUMBER = {666},
URL = {https://www.mdpi.com/1099-4300/22/6/666},
PubMedID = {33286438},
ISSN = {1099-4300},
DOI = {10.3390/e22060666}
}

@article{RONNBERG20171,
title = {On waveform distortion in the frequency range of 2kHz–150kHz—Review and research challenges},
journal = {Electric Power Systems Research},
volume = {150},
pages = {1-10},
year = {2017},
issn = {0378-7796},
doi = {https://doi.org/10.1016/j.epsr.2017.04.032},
url = {https://www.sciencedirect.com/science/article/pii/S0378779617301864},
author = {Sarah K. Rönnberg and Math H.J. Bollen and Hortensia Amaris and Gary W. Chang and Irene Y.H. Gu and Łukasz H. Kocewiak and Jan Meyer and Magnus Olofsson and Paulo F. Ribeiro and Jan Desmet}
}

@article{AshmanARXIV1994,
  title={Detecting bimodality in astronomical datasets},
  author={Ashman, Keith M and Bird, Christina M and Zepf, Steven E},
  journal={arXiv preprint astro-ph/9408030},
  year={1994}
}

@article{HARTMANN2025117095,
title = {Quantitative comparison of power grid reinforcements},
journal = {Chaos, Solitons \& Fractals},
volume = {200},
pages = {117095},
year = {2025},
issn = {0960-0779},
doi = {https://doi.org/10.1016/j.chaos.2025.117095},
url = {https://www.sciencedirect.com/science/article/pii/S0960077925011087},
author = {Bálint Hartmann and Géza Ódor and Kristóf Benedek and István Papp and Michelle T. Cirunay}
}

@article{KraljicIEEETRANS2022,
  author={Kraljic, David},
  journal={IEEE Transactions on Power Systems}, 
  title={Towards Realistic Statistical Models of the Grid Frequency}, 
  year={2023},
  volume={38},
  number={1},
  pages={256-266},
  doi={10.1109/TPWRS.2022.3163336}}

@Article{RydinGorjão2020,
author={Rydin Gorj{\~a}o, Leonardo
and Jumar, Richard
and Maass, Heiko
and Hagenmeyer, Veit
and Yalcin, G. Cigdem
and Kruse, Johannes
and Timme, Marc
and Beck, Christian
and Witthaut, Dirk
and Sch{\"a}fer, Benjamin},
title={Open database analysis of scaling and spatio-temporal properties of power grid frequencies},
journal={Nature Communications},
year={2020},
month={Dec},
day={11},
volume={11},
number={1},
pages={6362},
issn={2041-1723},
doi={10.1038/s41467-020-19732-7},
url={https://doi.org/10.1038/s41467-020-19732-7}
}

@misc{lonardi2026understandingcomplexityfrequencyphase,
      title={Understanding the complexity of frequency and phase angle fluctuations in power grids}, 
      author={Alessandro Lonardi and Jacques M. Maritz and Leonardo Rydin Gorjão and Christian Beck},
      year={2026},
      eprint={2604.03133},
      archivePrefix={arXiv},
      primaryClass={physics.soc-ph},
      url={https://arxiv.org/abs/2604.03133}, 
}

@inproceedings{Oberhofer_2023,
   title={Non-linear, bivariate stochastic modelling of power-grid frequency applied to islands},
   url={http://dx.doi.org/10.1109/PowerTech55446.2023.10202986},
   DOI={10.1109/powertech55446.2023.10202986},
   booktitle={2023 IEEE Belgrade PowerTech},
   publisher={IEEE},
   author={Oberhofer, Ulrich and Gorjao, Leonardo Rydin and Yalcin, G. Cigdem and Kamps, Oliver and Hagenmeyer, Veit and Schafer, Benjamin},
   year={2023},
   month=june, pages={1–1} }

@article{schaefer2018nongaussian,
  author  = {Schäfer, Benjamin and Beck, Christian and Aihara, Kazuyuki and Witthaut, Dirk and Timme, Marc},
  title   = {Non-Gaussian power grid frequency fluctuations characterized by Lévy-stable laws and superstatistics},
  journal = {Nature Energy},
  volume  = {3},
  number  = {2},
  pages   = {119--126},
  year    = {2018},
  doi     = {10.1038/s41560-017-0058-z},
  url     = {https://doi.org/10.1038/s41560-017-0058-z}
}

@article{GorjaoEUROPHYS2025,
  title={Fluctuations and Control: Understanding the Frequency of Power Grids},
  author={Rydin Gorj{\~a}o, Leonardo and Maritz, Jacques M and Sch{\"a}fer, Benjamin and Witthaut, Dirk},
  journal={Europhysics News},
  volume={56},
  number={5},
  pages={17--20},
  year={2025},
  publisher={EDP Sciences}
}

@article{WenSCIREP2025,
  title={Nonstandard power grid frequency statistics across continents},
  author={Wen, Xinyi and Anvari, Mehrnaz and Rydin Gorj{\~a}o, Leonardo and Yalcin, G Cigdem and Hagenmeyer, Veit and Sch{\"a}fer, Benjamin},
  journal={Scientific Reports},
  volume={15},
  number={1},
  pages={38470},
  year={2025},
  publisher={Nature Publishing Group UK London}
}

@article{HartmannSEGAN2024,
  title={Dynamical heterogeneity and universality of power-grids},
  author={Hartmann, B{\'a}lint and {\'O}dor, G{\'e}za and Papp, Istv{\'a}n and Benedek, Krist{\'o}f and Deng, Shengfeng and Kelling, Jeffrey},
  journal={Sustainable Energy, Grids and Networks},
  volume={39},
  pages={101491},
  year={2024},
  publisher={Elsevier}
}

@article{KruseIEEE2020,
  title={Predictability of power grid frequency},
  author={Kruse, Johannes and Sch{\"a}fer, Benjamin and Witthaut, Dirk},
  journal={IEEE access},
  volume={8},
  pages={149435--149446},
  year={2020},
  publisher={IEEE}
}

@article{UhlenbeckPR1930,
  title={On the theory of the Brownian motion},
  author={Uhlenbeck, George E and Ornstein, Leonard S},
  journal={Physical review},
  volume={36},
  number={5},
  pages={823},
  year={1930},
  publisher={APS}
}

@article{SchaferPRE2017,
  title={Escape routes, weak links, and desynchronization in fluctuation-driven networks},
  author={Sch{\"a}fer, Benjamin and Matthiae, Moritz and Zhang, Xiaozhu and Rohden, Martin and Timme, Marc and Witthaut, Dirk},
  journal={Physical Review E},
  volume={95},
  number={6},
  pages={060203},
  year={2017},
  publisher={APS}
}

@article{AcebronREVMODPHYS2005,
  title={The Kuramoto model: A simple paradigm for synchronization phenomena},
  author={Acebr{\'o}n, Juan A and Bonilla, Luis L and P{\'e}rez Vicente, Conrad J and Ritort, F{\'e}lix and Spigler, Renato},
  journal={Reviews of modern physics},
  volume={77},
  number={1},
  pages={137--185},
  year={2005},
  publisher={APS}
}

@article{BeckCONTMECH2004,
  title={Superstatistics: theory and applications},
  author={Beck, Christian},
  journal={Continuum mechanics and thermodynamics},
  volume={16},
  number={3},
  pages={293--304},
  year={2004},
  publisher={Springer}
}

@article{OdorDengENTROPY2023,
  title={Synchronization transition of the second-order Kuramoto model on lattices},
  author={{\'O}dor, G{\'e}za and Deng, Shengfeng},
  journal={Entropy},
  volume={25},
  number={1},
  pages={164},
  year={2023},
  publisher={MDPI}
}

@article{FilatrellaEtAlEPJB2008,
  title={Analysis of a power grid using a Kuramoto-like model},
  author={Filatrella, Giovanni and Nielsen, Arne Hejde and Pedersen, Niels Falsig},
  journal={The European Physical Journal B},
  volume={61},
  number={4},
  pages={485--491},
  year={2008},
  publisher={Springer}
}

@article{DorflerPNAS2013,
  title={Synchronization in complex oscillator networks and smart grids},
  author={D{\"o}rfler, Florian and Chertkov, Michael and Bullo, Francesco},
  journal={Proceedings of the National Academy of Sciences},
  volume={110},
  number={6},
  pages={2005--2010},
  year={2013},
  publisher={National Academy of Sciences}
}

@techreport{ENTSOe2019dfd,
  title={Deterministic Frequency Deviations--Root Causes and Proposals for Potential Solutions},
  author={{ENTSO-E}},
  institution={European Network of Transmission System Operators for Electricity},
  year={2019}
}

@article{Nonstat-WK,
  title = {Wiener-Khinchin Theorem for Nonstationary Scale-Invariant Processes},
  author = {Dechant, Andreas and Lutz, Eric},
  journal = {Phys. Rev. Lett.},
  volume = {115},
  issue = {8},
  pages = {080603},
  numpages = {6},
  year = {2015},
  month = {Aug},
  publisher = {American Physical Society},
  doi = {10.1103/PhysRevLett.115.080603},
  url = {https://link.aps.org/doi/10.1103/PhysRevLett.115.080603}
}

@ARTICLE{Pere-Damia-Canaria,
  author={Martínez-Barbeito, María and Gomila, Damià and Colet, Pere},
  journal={IEEE Transactions on Sustainable Energy}, 
  title={Dynamical Model for Power Grid Frequency Fluctuations: Application to Islands With High Penetration of Wind Generation}, 
  year={2023},
  volume={14},
  number={3},
  pages={1436-1445},
  doi={10.1109/TSTE.2022.3231975}}

@article{PhysRevResearch.6.013194,
  title = {Improving power-grid systems via topological changes or how self-organized criticality can help power grids},
  author = {{\'O}dor, G{\'e}za and Papp, Istv{\'a}n and Benedek, Krist{\'o}f and Hartmann, B{\'a}lint},
  journal = {Phys. Rev. Res.},
  volume = {6},
  issue = {1},
  pages = {013194},
  numpages = {10},
  year = {2024},
  month = {Feb},
  publisher = {American Physical Society},
  doi = {10.1103/PhysRevResearch.6.013194},
  url = {https://link.aps.org/doi/10.1103/PhysRevResearch.6.013194}
}

@article{hartmann_12models,
    author = {Hartmann, B. and Ódor, G. and Benedek, K. and Papp, I.},
    title = {Studying power-grid synchronization with incremental refinement of model heterogeneity},
    journal = {Chaos: An Interdisciplinary Journal of Nonlinear Science},
    volume = {35},
    number = {1},
    pages = {013138},
    year = {2025},
    month = {01},
    issn = {1054-1500},
    doi = {10.1063/5.0237050},
    url = {https://doi.org/10.1063/5.0237050},
    eprint = {https://pubs.aip.org/aip/cha/article-pdf/doi/10.1063/5.0237050/20351864/013138_1_5.0237050.pdf},
}

@article{10.1063/5.0154581,
    author = {Deng, Shengfeng and Ódor, Géza},
    title = {Chimera-like states in neural networks and power systems},
    journal = {Chaos: An Interdisciplinary Journal of Nonlinear Science},
    volume = {34},
    number = {3},
    pages = {033135},
    year = {2024},
    month = {03},
    issn = {1054-1500},
    doi = {10.1063/5.0154581},
    url = {https://doi.org/10.1063/5.0154581},
    eprint = {https://pubs.aip.org/aip/cha/article-pdf/doi/10.1063/5.0154581/19846492/033135_1_5.0154581.pdf},
}

@article{Zeraati2022Timescales,
  author  = {Zeraati, Roxana and Engel, Tatiana A. and Levina, Anna},
  title   = {A flexible Bayesian framework for unbiased estimation of timescales},
  journal = {Nature Computational Science},
  volume  = {2},
  pages   = {193--204},
  year    = {2022},
  doi     = {10.1038/s43588-022-00214-3}
}

@article{Gardiner1985Stochastic,
  title={Handbook of stochastic methods for physics, chemistry and the natural sciences},
  author={Gardiner, Crispin W},
  journal={Springer series in synergetics},
  year={1985},
  publisher={Springer Berlin Heidelberg}
}

@article{GorjaoIEEE2020,
  title={Data-driven model of the power-grid frequency dynamics},
  author={Gorj{\~a}o, Leonardo Rydin and Anvari, Mehrnaz and Kantz, Holger and Beck, Christian and Witthaut, Dirk and Timme, Marc and Sch{\"a}fer, Benjamin},
  journal={IEEE access},
  volume={8},
  pages={43082--43097},
  year={2020},
  publisher={IEEE}
}

@article{Bronski2021,
author={Bronski, Jared C.
and Wang, Lan},
title={Partially Phase-Locked Solutions to the Kuramoto Model},
journal={Journal of Statistical Physics},
year={2021},
month={Jun},
day={09},
volume={183},
number={3},
pages={46},
issn={1572-9613},
doi={10.1007/s10955-021-02783-5},
url={https://doi.org/10.1007/s10955-021-02783-5}
}

@article{SchaferCOMPLEXITY2023,
  title={Microscopic Fluctuations in Power-Grid Frequency Recordings at the Subsecond Scale},
  author={Sch{\"a}fer, Benjamin and Rydin Gorj{\~a}o, Leonardo and Yalcin, G Cigdem and F{\"o}rstner, Ellen and Jumar, Richard and Maass, Heiko and K{\"u}hnapfel, Uwe and Hagenmeyer, Veit},
  journal={Complexity},
  volume={2023},
  number={1},
  pages={2657039},
  year={2023},
  publisher={Wiley Online Library}
}

@article{VanderPlas_2018_lomb_scargle,
doi = {10.3847/1538-4365/aab766},
url = {https://doi.org/10.3847/1538-4365/aab766},
year = {2018},
month = {may},
publisher = {The American Astronomical Society},
volume = {236},
number = {1},
pages = {16},
author = {VanderPlas, Jacob T.},
title = {Understanding the Lomb–Scargle Periodogram},
journal = {The Astrophysical Journal Supplement Series}
}

@article{welch,
  author={Welch, P.},
  journal={IEEE Transactions on Audio and Electroacoustics}, 
  title={The use of fast Fourier transform for the estimation of power spectra: A method based on time averaging over short, modified periodograms}, 
  year={1967},
  volume={15},
  number={2},
  pages={70-73},
  doi={10.1109/TAU.1967.1161901}}

@book{essenwanger1986elements,
  author    = {Essenwanger, Oskar M.},
  title     = {Elements of Statistical Analysis},
  publisher = {Elsevier},
  year      = {1986},
  address   = {Amsterdam},
  isbn      = {9780444424266}
}

@article{Kasdin1995,
  author  = {Kasdin, N. Jeremy},
  title   = {{Runge-Kutta} algorithm for the numerical integration of stochastic differential equations},
  journal = {Journal of Guidance, Control, and Dynamics},
  year    = {1995},
  volume  = {18},
  number  = {1},
  pages   = {114-120},
  doi     = {10.2514/3.56665}
}
	
	\clearpage
	\section*{Appendix}
	\raggedbottom

This section gives the data-handling rules, estimator definitions, simulation settings and fitting details used in the comparisons reported in the main text.

\section{Empirical data handling}
\label{supp:data}

The empirical data consist of synchronous 10-Hz frequency measurements from sixteen Hungarian 400-kV measurement sites between 31 January and 28 February 2025. Complete daily records were analysed separately before aggregation. Consequently, the number of daily records entering a site-wise statistic depends on data availability at that site. The record P\'ecs--Ernestinovo~II is not used: on every one of its 29 days it holds its value for $78$--$100\%$ of the $0.1\,\mathrm{s}$ steps, with a median daily standard deviation of $2.6\,\mathrm{mHz}$ against $20.8\,\mathrm{mHz}$ for P\'ecs--Ernestinovo~I on the same bus, and it is uncorrelated with that record at every lag.

For each record, the physical frequency deviation is defined relative to the nominal grid frequency as

\begin{equation}
    x_i(t)=f_i(t)-50\,\mathrm{Hz}.
\end{equation}
For autocorrelation and spectral estimation, the temporal mean of the individual analysed segment is removed separately. Thus subtraction of the nominal frequency and statistical mean-centring are treated as distinct operations.

The lag-$\tau$ increment is
\begin{equation}
    \Delta f_{i,\tau}(t)
    =
    f_i(t+\tau)-f_i(t).
\end{equation}

\section{Power spectral density estimation}
\label{supp:psd}

For uniformly sampled records, the one-sided PSD is estimated with Welch's method~\cite{welch} after subtracting the temporal mean of the analysed segment.

For the empirical 10-Hz data, the Welch spectra are computed using Hann-windowed segments of $65536=2^{16}$ samples ($6553.6\,\mathrm{s}$) with $50\%$ overlap, and a constant trend is removed from each segment. The resulting frequency resolution, $\Delta\nu=1/6553.6\,\mathrm{s}\approx1.53\times10^{-4}\,\mathrm{Hz}$, is chosen so that the low-frequency fitting band ($\nu\geq10^{-3}\,\mathrm{Hz}$) is resolved, while a full day still provides about 25 segments to average. A Hann window~\cite{essenwanger1986elements} is used with an overlap of $35999$ samples, and a constant trend is removed from each segment.

Because the empirical daily records differ in length, the PSD of each daily file is computed separately. The resulting spectra are interpolated in log--log space onto a common frequency grid before aggregation. Frequencies represented by fewer than half of the daily records are discarded. Site-wise spectra are reported as pointwise medians across daily records to reduce sensitivity to isolated high-amplitude events.

For the densely sampled HU387 records, the same Welch procedure is used with the simulation output cadence of $1\,\mathrm{s}$.

Welch's method is not applied to the logarithmically sampled lattice or network relaxation records because those records do not have a fixed sampling interval. Their spectra are instead estimated with a Lomb--Scargle periodogram~\cite{VanderPlas_2018_lomb_scargle} and are treated as a separate analysis from the uniformly sampled Welch spectra.

Where a finite-band spectral fit is reported, the fitted form is either
\begin{equation}
\label{eq:supp_psd_power}
    S(\nu)=A\nu^{-\gamma},
\end{equation}
or, only in a frequency range where the spectrum approaches an approximately constant floor,
\begin{equation}
\label{eq:psd_power_law_floor}
    S(\nu)=A\nu^{-\gamma}+S_0.
\end{equation}
Fits are performed by least squares on $\log_{10}S$, so that relative rather than absolute deviations are weighted approximately uniformly over spectra spanning several decades. The quoted residual is the root-mean-square residual in $\log_{10}S$. The constant $S_0$ is omitted from the low-frequency fits, where a free offset would absorb crossover curvature and bias $\gamma$. A fitted $S_0$ is a phenomenological floor and does not identify its origin: additive measurement noise, the frequency-estimation procedure and unresolved fast grid fluctuations cannot be separated using the frequency record alone.

\section{Kuramoto simulation setup}
\label{supp:simulations}

\subsection{Technical details of the HU387 simulations}
\label{supp:hu387_runs}

All HU387 KM simulations use scenario~9 of the HU387 model~\cite{hartmann_12models}: $N=387$ nodes, $640$ symmetric lines with weights $W_{ij}$, and net injections $P_i$, positive for producers and negative for consumers. Every node, producer or consumer, follows Eq.~(\ref{model_eq}) with $m_i=1$ and $\alpha_i=\alpha=0.1\,\mathrm{s}^{-1}$, and $K$ is fixed during a run.
In each realization, every injection is perturbed once, $P_i\to P_i(1+0.05\,\zeta_i)$, with independent random variable $\zeta_i\sim\mathcal N(0,1)$ drawn from the seed of the realization, and all $P_i$ are then shifted by a common constant so that $\sum_iP_i=0$. The runs start from $\theta_i(0)=0$ and $\dot\theta_i(0)=P_i$.

For details of calculating the integration step, choosing the integrators and validating them against each other, see Supplementary Sec.~\ref{supp:hu387_runs_more}.

\subsubsection{HU387 KM simulations by varying noise}

The noise sweep at $K=0.8$ is evaluated on the nodes recorded every $1\,\mathrm{s}$. The first $5000$ samples of every record are discarded, since they still contain a transient with markedly larger increments; after this cut, the standard deviation agrees between the four quarters of a record at the percent level.

\subsubsection{HU387 runs for the spectral comparison}

The spectra of Fig.~\ref{fig:psd_comparison} use the network-mean frequency and the frequencies of the nodes of the measured substations, after discarding the first $5000\,\mathrm{s}$. They are Welch estimates with the settings of the measurement ($65536$-sample Hann segments). An earlier coupling sweep sampled every $1\,\mathrm{s}$ overestimated the power at $0.1$--$0.3\,\mathrm{Hz}$ by up to a factor of $36$ at $K=1$ through aliasing and is not used.

\subsubsection{Spatial snapshot runs}

Some HU387 runs record all 387 nodes on a logarithmic time schedule. These records are not used for increment statistics because their temporal sampling is nonuniform. They are used only for spatial snapshot analyses, for which the last recorded state of each realization is considered.

For the instantaneous distribution across nodes, the density is estimated with a Gaussian kernel of bandwidth $0.08$ times the sample standard deviation. A maximum is counted when its height exceeds $5\%$ of the global maximum. The reported number of modes is stable against halving or doubling the kernel bandwidth. The HU387 edge list contains 640 edges, corresponding to a mean degree of approximately $3.28$.

\subsection{Square-lattice simulations}

The square-lattice KM calculations are done on two-dimensional square lattices, with periodic boundary conditions using zero-centered Gaussian-distributed quenched nodal frequencies, uniform inertia and damping. All quantities are dimensionless and calculated in a co-moving phase reference frame that can be gauged out from KM solutions. Since we are dealing with large lattices, which require a lot of memory during simulation, to keep the numerical solutions efficient, the state of the system has been saved at logarithmic time steps given by
\begin{equation}
    t_k=1.08^k.
\end{equation}
for $k=1,2,3....$, and calculate required quantities. This is standard method in case of handling dynamical models exhibiting long-time relaxations.

\section{Robustness of superstatistical estimates}
\label{supp:robust}

The empirical increment kurtosis is sensitive to a very small number of extreme observations. Across the measurement sites, the raw kurtosis spans approximately $4.7$--$44.6$. Winsorizing \blue{???} the outermost $0.01\%$ of $|\Delta f_\tau|$ reduces the values to approximately $3.07$--$3.25$, demonstrating that the raw fourth moment is dominated by rare events rather than by the bulk of the distribution.

Direct inspection of increments larger than $15\,\mathrm{mHz}$ shows that these events are rare compared with the full record length and typically occur in short clusters. The largest excursions are sufficiently extreme that individual samples can materially alter the sample kurtosis. Kurtosis is therefore used primarily as a scale-selection and non-Gaussianity diagnostic rather than as a precise characterization of the tail shape.

The estimated superstatistical coarse-graining time is substantially more stable than the raw kurtosis. For the pooled records reported in the analysis, representative sites with very different raw kurtoses give similar values of $T$ near $20.4\,\mathrm{s}$. The mean and coefficient of variation of $\beta$ change more moderately under winsorization than the raw fourth moment, indicating that the inferred bulk variation of the local variance is not controlled by the most extreme increments.

The distributional comparisons are therefore restricted to histogram bins containing at least 100 observations. This restriction prevents unresolved extreme bins from dominating fitted tail parameters or comparisons between Gaussian, $q$-Gaussian and superstatistical models.

\section{Bimodality criterion and peak counting}
\label{supp:multimodality}

The bimodality criterion of the main text, Sec.~\ref{subsec:multimodality_methods}, compares Ashman's $D_{\rm A}$ of the two-component fit with its $95$th percentile for standard-normal samples of the same size. The second condition, a minor-component weight of at least $5\%$, is needed because $D_{\rm A}$ ignores the weights: on a locked cluster a fraction of a millihertz wide, a single outlying node gives a very large $D_{\rm A}$ without forming a genuine second mode. Across nodes, only nodes with $|\bar\omega_i|<200\,\mathrm{mHz}$ are used; the remaining $10$--$26$ nodes drift at $0.3$--$1.5\,\mathrm{Hz}$ and are separated from the rest by an empty gap.

The number of peaks is obtained from a Gaussian mixture with one to five components, of which the one with the lowest Bayesian information criterion (BIC) is kept. Since two overlapping components can describe a single skewed peak, the peaks are the local maxima of the fitted mixture density, and a peak counts only if the probability mass of its basin, between the neighbouring minima of the density, is at least $2\%$ of the sample. Neighbouring maxima are merged, shallowest valley first, while the valley is higher than half of the lower maximum. With this setting, $60$ samples each of $201$ and of $377$ values give a second peak for at most $3\%$ of gamma- and lognormal-distributed samples and for none of the normal ones, whereas two Gaussians separated as the two branches at $K=1$ give two peaks in all of them. Errors are bootstrap errors of the median over the realizations (across nodes) and over the realizations with the twelve station nodes refitted in each draw (across realizations).

\section{Detailed empirical PSD at B\'ek\'escsaba}
\label{supp:bekescsaba_psd}

\begin{figure}[tbp]
    \centering
    \includegraphics[width=\columnwidth]{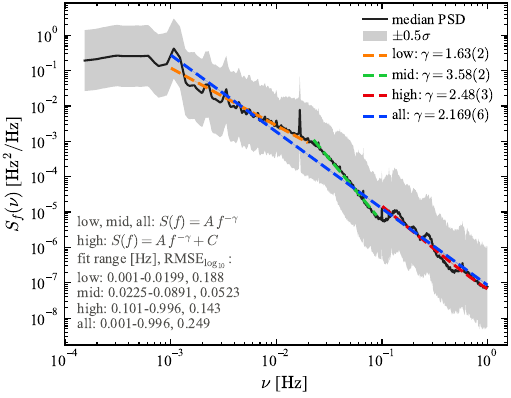}
    \caption{Median Welch power spectral density of the grid-frequency deviations measured at the B\'ek\'escsaba 400-kV site. Dashed lines indicate finite-band power-law fits. The low- and high-frequency intervals are treated separately because a single power law does not describe the complete spectrum. The constant $S_0$ is included only in the high-frequency fit.}
    \label{fig:supp_bekescsaba_psd}
\end{figure}

Figure~\ref{fig:supp_bekescsaba_psd} shows the median Welch PSD of the measured frequency deviations at the B\'ek\'escsaba 400-kV site, computed from the complete daily records as described in Sec.~\ref{supp:psd}. The low-frequency band, $0.001\lesssim\nu\lesssim0.02\,\mathrm{Hz}$, is fitted with Eq.~\eqref{eq:supp_psd_power}, and the high-frequency band, $0.1\lesssim\nu\lesssim1\,\mathrm{Hz}$, where the spectrum approaches a constant floor, with Eq.~\eqref{eq:psd_power_law_floor}; the exponents are discussed in the main text. The crossover band between them is not assigned an independent scaling exponent: its steep local slope depends strongly on the selected interval and characterizes the crossover between the two regimes rather than a scale-free process.

\section{Topological investigation of frequency-locked outlier nodes}
\label{supp:topinvest}

\begin{table*}[tb]
    \centering
    \caption{Islands that remain outside the main cluster in HU387 at strong coupling, $D=0$. $\sum P_i$: net nominal injection of the island. $W_{\rm cut}$: weights of the lines leaving the island. $K_{\rm min}=|\sum P_i|/\sum W_{\rm cut}$: lower bound of the coupling at which the island can lock. Last two columns: fraction of the $201$ realizations in which the island is outside the main cluster at $K=0.8$ and $K=1$. The consumer at Celld\"om\"olk (node~45, outside in all realizations at both couplings) is connected only to the Ajka island and to Ikerv\'ar, by lines with $W=0.41$ and $0.72$.}
    \label{tab:supp_islands}
    \begin{tabular}{llccccc}
        \hline
        Island & Node IDs & $\sum P_i$ & $W_{\rm cut}$ & $K_{\rm min}$ & $K=0.8$ & $K=1$ \\
        \hline
        Ajka & 2, 3, 16, 252, 353, 354 & $1.36$ & $0.41,0.51,0.44,0.40$ & $0.77$ & $1.00$ & $1.00$ \\
        Szakoly--Ny\'iradony & 239, 306 & $0.77$ & $0.67,0.40$ & $0.71$ & $1.00$ & $0.76$ \\
        Ikerv\'ar & 131 & $0.85$ & $0.72,0.74$ & $0.59$ & $1.00$ & $1.00$ \\
        Kapuv\'ar & 150 & $0.85$ & $0.81$ & $1.06$ & $1.00$ & $1.00$ \\
        \hline
    \end{tabular}
\end{table*}

At strong coupling, the same $11$ nodes stay outside the main frequency cluster in practically every realization: at $K=0.8$ and $D=0$, each of them is outside in all $201$ realizations and no other node is outside in more than half of them. At $K=1$, nine of them are still outside in every realization, while the Szakoly--Ny\'iradony pair is outside in $76\%$ of them, since it locks in the remaining $24\%$ (Sec.~\ref{supp:locking_scan}). These are the nodes that form the separate frequency bands in Fig.~\ref{fig:HU_sc9_freq_ts}(d) and the filled markers in Fig.~\ref{fig:hu387_locking_map}(d). Following the lines between them, they make up four islands, listed in Table~\ref{tab:supp_islands}: a group of six nodes around the Ajka generator substation, the Szakoly power plant with the neighbouring Ny\'iradony consumer, and the single generators at Ikerv\'ar and Kapuv\'ar. The consumer at Celld\"om\"olk (node~45) is the eleventh node. Its only two lines lead to Ajka and to Ikerv\'ar, so it has no direct line to the main cluster. Node~252 is listed under Ny\'iradony in the substation table, but its only line leads to node~354 of the Ajka group, so it belongs to that island. Node numbers here are the $1$-based identifiers of the network files.

\begin{figure*}[!htbp]
    \centering
    \includegraphics[width=\textwidth]{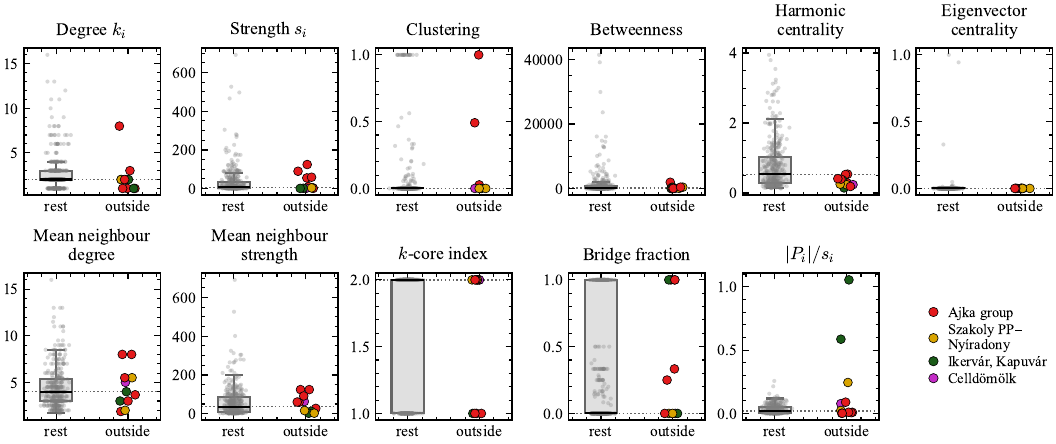}
    \caption{Single-node measures of the weighted HU387 graph for the $11$ nodes that remain outside the main cluster at strong coupling (right in each panel, coloured by island) and for the other $376$ nodes (left; box: quartiles and median, whiskers: $1.5$ interquartile ranges, dotted line: median of the rest). Parallel circuits are merged into one link with the summed weight; betweenness and harmonic centrality use the path length $1/W_{ij}$. The bridge fraction is the fraction of a node's links whose removal disconnects the network, and $|P_i|/s_i$ is the coupling at which a node can carry its own injection when its neighbours are locked.}
    \label{fig:supp_branch_metrics}
\end{figure*}

\begin{figure*}[!hb]
    \centering
    \includegraphics[width=.9\textwidth]{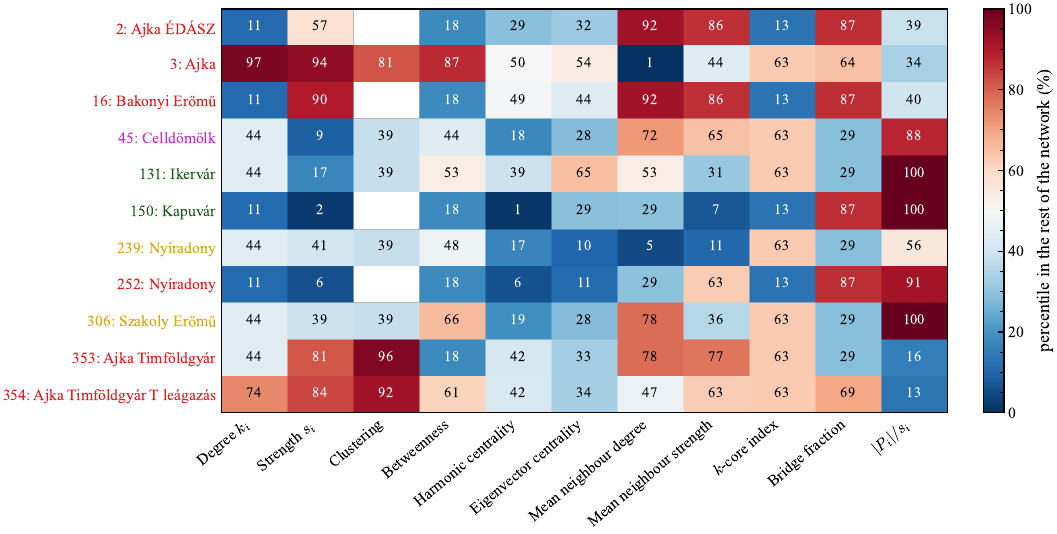}
    \caption{Percentile of each of the $11$ outlier nodes within the distribution of the other $376$ nodes, for the measures of Fig.~\ref{fig:supp_branch_metrics} (ties counted half). Node labels are coloured by island as in Fig.~\ref{fig:supp_branch_metrics}. Blank cells: clustering is undefined for degree-one nodes.}
    \label{fig:supp_branch_percentiles}
\end{figure*}

To test whether these nodes are structurally exceptional, they are compared with the other $376$ nodes on eleven single-node measures of the weighted HU387 graph. The $640$ lines of the network file include parallel circuits; these are merged into $515$ links, each carrying the sum of the weights $W_{ij}$ of its circuits, as in the coupling term of the model. The measures are the degree $k_i$ and the strength $s_i=\sum_jW_{ij}$; the weighted local clustering coefficient; the betweenness and the harmonic centrality, both with path length $1/W_{ij}$; the eigenvector centrality of $W$; the mean degree and mean strength of the neighbours; the $k$-core index; the bridge fraction, i.e.\ the fraction of the node's links whose removal disconnects the network; and the ratio $|P_i|/s_i$ of the injected or consumed power to the strength. The last ratio is the smallest coupling at which a single node can carry its own injection through its lines when all of its neighbours are locked, $K s_i\geq|P_i|$. It is the single-node version of the cut criterion of the main text. Figure~\ref{fig:supp_branch_metrics} shows the distributions of all eleven measures, and Fig.~\ref{fig:supp_branch_percentiles} gives the percentile of each outlier node within the rest of the network.

As a group, the outlier nodes are not distinguished by any of the topological measures. Their median degree is $2$, the same as for the rest of the network, and a two-sided Mann--Whitney test gives $p\geq0.06$ for ten of the eleven measures. The exception is the harmonic centrality, whose median is $0.29$ for the outliers against $0.53$ for the rest ($p=0.015$). The outliers therefore lie on the electrical periphery of the grid, but so do many nodes that lock. Two of the measures carry little information in this network. The eigenvector centrality is concentrated on the G\"ony\H{u} hub and is close to zero elsewhere (median $2\times10^{-8}$). The $k$-core index is at most $2$, so it only separates tree-like branches from the meshed part of the grid.

The islands are, moreover, structurally different from each other. The Ajka group is strongly coupled internally. Its hub, Ajka (node~3), is in the $97$th percentile in degree, the $94$th in strength and the $87$th in betweenness. Its degree-one neighbours, Bakonyi Er\H{o}m\H{u} (node~16) and Ajka \'ED\'ASZ (node~2), are in the $90$th and $57$th percentiles in strength, and the two Ajka Timf\"oldgy\'ar nodes (nodes~353 and 354) are in the $81$st--$96$th percentiles in strength and clustering. The lines inside the group have $W_{ij}$ up to $45$. In contrast, the four lines that connect the group to the rest of the grid have $W_{ij}=0.40$--$0.51$, while the median link weight of the network is $1.66$, and the group carries a net injection $\sum P_i=1.36$ from its two generators. No single node of the group is extreme in $|P_i|/s_i$ ($13$th--$91$st percentile), and the group stays outside the main cluster only because of this combination of a large net injection and a weak connection. The single generators and the pair, on the other hand, are extreme in $|P_i|/s_i$. Kapuv\'ar ($1.06$) and Ikerv\'ar ($0.59$) have the two largest values in the network, and the Szakoly power plant ($0.25$) has the fourth largest. Kapuv\'ar is also in the lowest percentiles in strength and harmonic centrality ($2$nd and $1$st). The third-largest value, $0.26$, belongs to a node that is in the main cluster in every realization at $K\geq0.8$, so $|P_i|/s_i$ alone does not decide whether a node locks.

The appropriate criterion is therefore not a property of a single node but the condition $K\sum_{\rm cut}W_{ij}\geq|\sum_iP_i|$ for the whole island, Table~\ref{tab:supp_islands}. It requires $K\geq0.77$ for the Ajka group, $0.71$ for the Szakoly--Ny\'iradony pair, $0.59$ for Ikerv\'ar and $1.06$ for Kapuv\'ar. Every island is attached by one to four lines with $W_{ij}=0.40$--$0.81$, and at $K=1$ all of them except the Kapuv\'ar generator are above this bound, yet they remain outside the main cluster in most or all realizations. As discussed in the main text, the bound is necessary but not sufficient with the weak damping used here. In summary, neither a low degree nor a dangling position characterises the outlier nodes, since only four of the eleven are dangling and the Ajka hub is among the best-connected nodes of the grid. What they have in common is a large net injection on a group of nodes that is attached to the rest of the network by weak lines. This is a joint property of the topology, the line weights, and the dispatch, and it cannot be identified from the unweighted topology alone. This analysis supports our previous KM simulation results on different power-grid networks, where we showed that stability, vulnerability, and cascade sizes do not only depend on the topology, but also on the synchronization dynamics of the system~\cite{PhysRevResearch.6.013194,hartmann_12models,HARTMANN2025117095}

\subsection{Frequency-locking ranges in the coupling of the peripheral islands}
\label{supp:locking_scan}

To find the coupling dependence of the across-realization multimodality at which the islands of Table~\ref{tab:supp_islands} join the main cluster, we repeated the noise-free HU387 simulations on a finer coupling grid, $0.6\leq K\leq2$, for $5000\,\mathrm{s}$ each. We used the same $201$ random seeds, so every realization has the same noise in $P_i$ as before. Whether an island locks, i.e.\ joins the main cluster, is decided within the first $\sim100\,\mathrm{s}$. At $K=0.8$ and $K=1$, the short runs give the same results as the full $96\,\mathrm{h}$ runs. We also restarted the runs at $0.7\leq K\leq1$ from the final states of the $96\,\mathrm{h}$ runs at $K=1$ and at $K=0.8$.

\begin{figure}[H]
    \centering
    \includegraphics[width=.95\columnwidth]{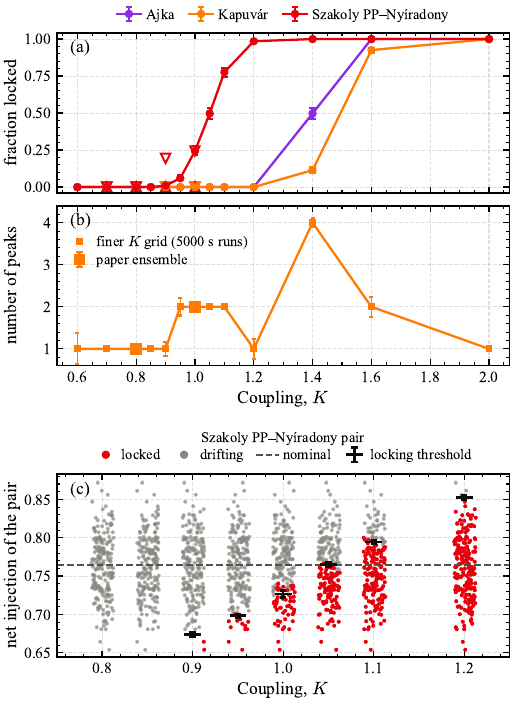}
    \caption{Locking of the peripheral HU387 islands on a fine coupling grid, $D=0$, $201$ realizations ($5000\,\mathrm{s}$ runs with the seeds of the main ensemble). (a) Fraction of realizations in which the Ajka island, the Kapuv\'ar generator and the Szakoly--Ny\'iradony pair are locked to the main cluster, from zero initial phases (circles and lines; $1\sigma$ Wilson errors). Open triangles in the colour of each island mark the runs at $0.7\leq K\leq1$ restarted from the late-time states of the $96\,\mathrm{h}$ runs at $K=1$ (pointing down) and at $K=0.8$ (pointing up). Most of them lie at zero or on the circles; the only clear departure is the Szakoly--Ny\'iradony pair restarted from $K=1$, which stays locked in $19\%$ of the realizations at $K=0.9$. (b) Median number of peaks of the station-node distributions across realizations, as in Fig.~\ref{fig:hu387_multimodality}(a); large markers are the $96\,\mathrm{h}$ runs at $K=0.8$ and $1$, small markers the $5000\,\mathrm{s}$ runs. (c) Quenched net injection of the Szakoly--Ny\'iradony pair in each realization, coloured by whether the pair locks, with the separating threshold and its $1\sigma$ bootstrap error; the dashed line marks the nominal injection.}
    \label{fig:supp_locking_scan}
\end{figure}

The Szakoly--Ny\'iradony pair locks in $1\%$ of the realizations at $K=0.9$, $6\%$ at $K=0.95$, $24\%$ at $K=1$, $50\%$ at $K=1.05$, $78\%$ at $K=1.1$ and $99\%$ at $K=1.2$, Fig.~\ref{fig:supp_locking_scan}(a). In each realization, the outcome depends on the net power of the pair, which changes slightly between realizations because of the $5\%$ random perturbation of the initial injections. The pair locks when this power is below a threshold that grows roughly in proportion to the coupling, from $0.674(3)$ at $K=0.9$ to $0.852(3)$ at $K=1.2$, i.e.\ about $0.73\,K$, Fig.~\ref{fig:supp_locking_scan}(c). This single threshold predicts the outcome in at least $94\%$ of the realizations. The nominal net power of the pair is $0.765$, and the perturbation spreads it by $0.041$. Only for $0.9\lesssim K\lesssim1.2$ does the threshold fall inside this spread, so only there do some realizations lock and others not. At $K=0.8$, locking would need a net power four standard deviations below the nominal value. 
\begin{figure}[H]
    \centering
    \includegraphics[width=.95\columnwidth]{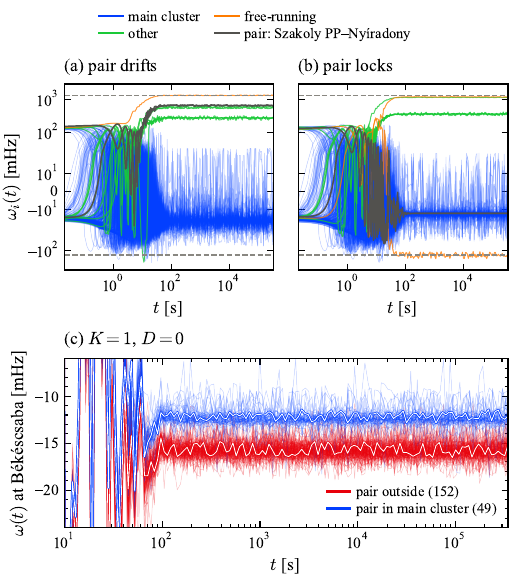}
    \caption{Selection of the strong-coupling branch in HU387 at $K=1$ and $D=0$. The Szakoly--Ny\'iradony pair is the generator--consumer chain of two degree-two nodes at a dangling end of the network (Sec.~\ref{subsec:hu387}); whether it locks to the main cluster decides on which of the two branches of Fig.~\ref{fig:hu387_locking}(d) the station nodes end. Frequency traces from zero initial phases to the end of the $96\,\mathrm{h}$ run (logarithmically spaced samples). (a) Frequencies $\omega_i(t)$ of all nodes in a realization in which the Szakoly--Ny\'iradony pair keeps drifting and (b) in one in which it locks, coloured by the late-time class of each node as in Fig.~\ref{fig:hu387_locking}, on a symmetric logarithmic axis; the pair is drawn in dark grey and dashed lines mark the uncoupled frequencies of generators and consumers. (c) Frequency at B\'ek\'escsaba in all $201$ realizations, coloured by whether the pair ends in the main cluster; the realizations of (a) and (b) are highlighted. The two branches separate within the first $\sim100\,\mathrm{s}$, and each realization stays on its branch for the rest of the run.}
    \label{fig:hu387_locking_traces}
\end{figure}
\noindent The Ajka island and the Kapuv\'ar generator start to lock only at $K\approx1.4$. The station nodes have several peaks across realizations exactly in these coupling ranges, Fig.~\ref{fig:supp_locking_scan}(b) and Fig.~\ref{fig:hu387_multimodality}(a). When restarted from the final states at $K=1$, the pair stays locked in $19\%$ of the realizations at $K=0.9$, but in none at $K\leq0.8$. When restarted from the unlocked final states at $K=0.8$, it never locks, even at $K=1$. For $0.9\lesssim K\lesssim1$, both outcomes are therefore possible at the same coupling, and the start from zero phases decides which one is reached.

Figure~\ref{fig:hu387_locking_traces} shows the statistics and corresponding time series when the Szakoly--Ny\'iradony pair locks and drifts in the $96\,\mathrm{h}$ runs at $K=1$. All nodes start with zero phase at their own uncoupled frequencies. Within the first $\sim100\,\mathrm{s}$, most nodes join the main cluster, and the pair either joins as well or keeps turning at its own frequency, Fig.~\ref{fig:hu387_locking_traces}(a,b). After that, the frequency at B\'ek\'escsaba stays at one of two levels for the rest of the run, Fig.~\ref{fig:hu387_locking_traces}(c).

\section{Noise dependence of the HU387 multimodality}
\label{supp:noise_multimodality}

\begin{figure}[H]
    \centering
    \includegraphics[width=\columnwidth]{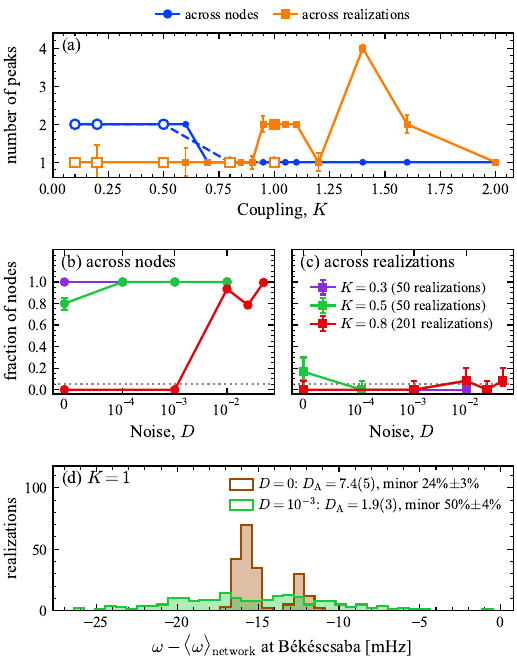}
    \caption{Effect of additive white noise on the two kinds of HU387 multimodality, with the bimodality criterion of Fig.~\ref{fig:hu387_multimodality}. (a) Median number of peaks against $K$, across nodes (circles) and across realizations (squares), as in Fig.~\ref{fig:hu387_multimodality}(a). Solid lines and filled markers are $D=0$, dashed lines and open markers $D=10^{-3}$, with $201$ realizations and the same seeds at both $D$; where the two values coincide, the open marker covers the filled one. Small markers are the $D=0$ runs on the finer $K$ grid of Sec.~\ref{supp:locking_scan}. (b) Across nodes and (c) across realizations, fraction of bimodal samples against $D$ at $K=0.3$ and $0.5$ ($50$ realizations) and $K=0.8$ ($201$ realizations); the legend in (c) applies to both panels. Error bars are $1\sigma$ Wilson intervals; dotted lines mark the $5\%$ rate expected for unimodal samples. (d) Distribution of the B\'ek\'escsaba node over the realizations at $K=1$ for $D=0$ and $D=10^{-3}$, with Ashman's $D_{\rm A}$ and the weight of the minor mixture component ($1\sigma$ bootstrap errors).}
    \label{fig:supp_noise_multimodality}
\end{figure}

Figure~\ref{fig:supp_noise_multimodality} repeats the bimodality test of the main text, Fig.~\ref{fig:hu387_multimodality}, for every noise level available for the HU387 runs. With noise, the late-time frequencies are snapshots of a fluctuating signal; as at $D=0$, their average over the trailing samples, taken from the second half of each run, is used. Additive noise removes the bimodality across realizations wherever it is present at $D=0$ and does not create it elsewhere, whereas the bimodality across nodes persists and, at $K=0.8$, appears for $D\geq10^{-2}$.

\section{Effect of noise on the synchronization transition results on small, $N=20\times 20$ lattices}

\label{supp:lattice_noise}
\begin{figure}[H]
    \centering
    \includegraphics[width=1\linewidth]{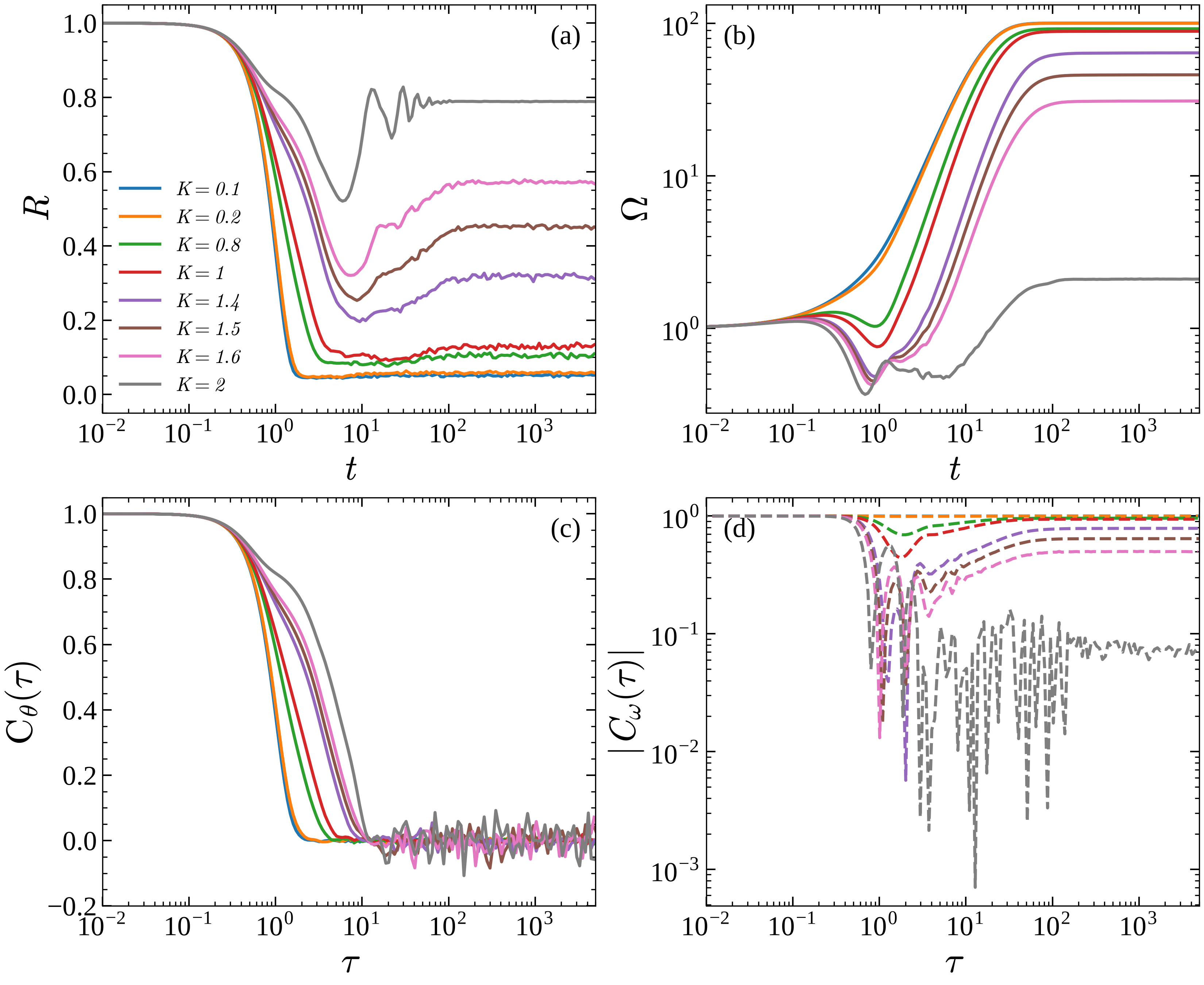}
    \caption{$K$ dependence of KM solutions on periodic $20\times20$ lattice, $m=1$, $\alpha=0.1$, $D=0$, averaged over $200$ realizations of simulation for random self-frequencies. (a) Order parameter $R$ (b) frequency variance $\Omega$ (c) phase autocorrelation (d) frequency autocorrelation  $C_\omega$ (dash-dotted) are plotted, where different colors correspond to different $K$-s shown in the legend of (a).
    }
    \label{fig:l2d20_D0}.
\end{figure}

\begin{figure}[H]
    \centering
    \includegraphics[width=.875\linewidth]{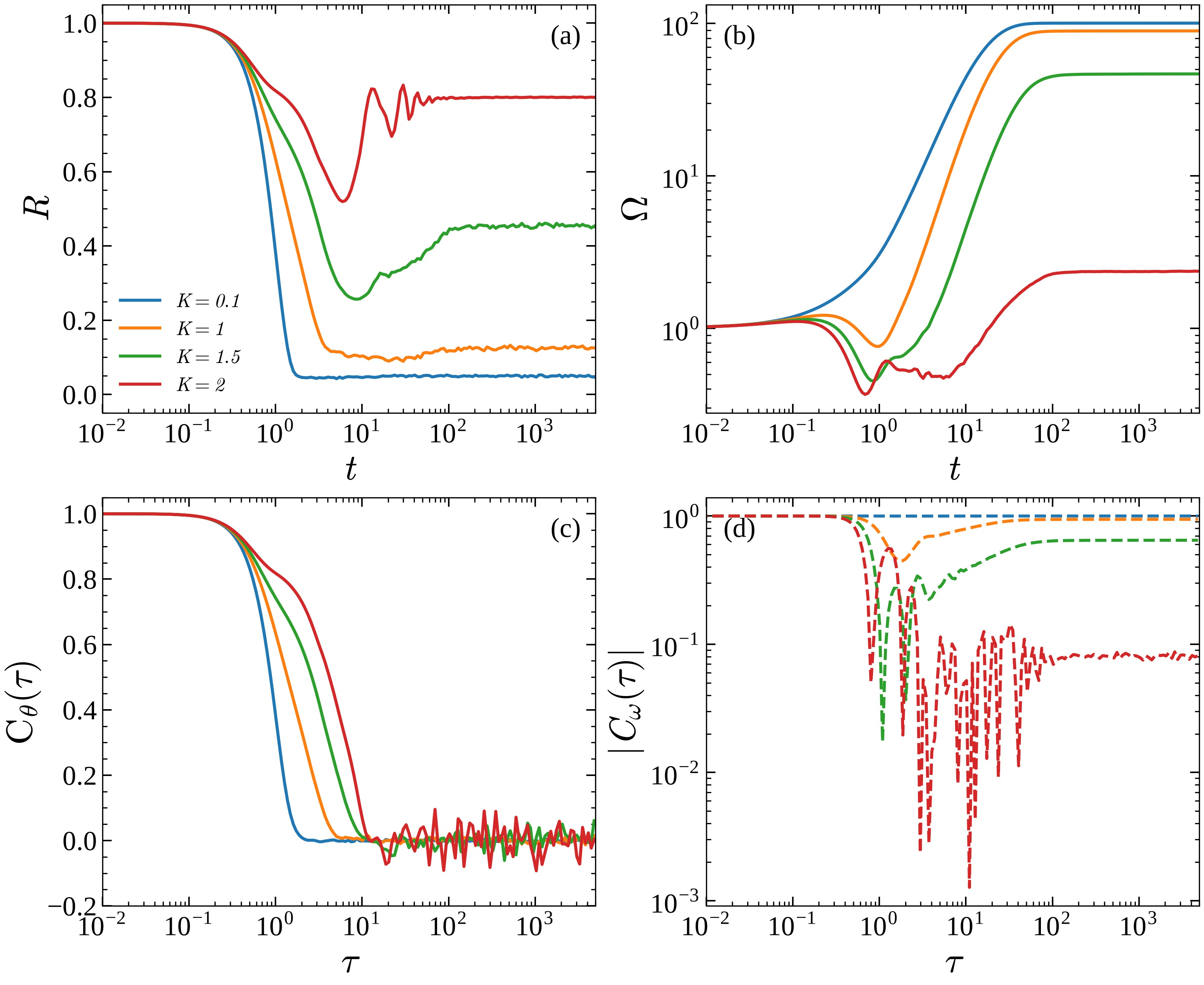}
    \caption{$K$ dependence of KM solutions on periodic $20\times20$ lattice, $m=1$, $\alpha=0.1$, $D=0.001$, averaged over $200$ realizations of simulation for random self-frequencies. (a) Order parameter $R$ (b) frequency variance $\Omega$ (c) phase autocorrelation (d) frequency autocorrelation  $C_\omega$ (dash-dotted) are plotted, where different colors correspond to different $K$-s shown in the legend of (a).}
    \label{fig:l2d20_D0p001}
\end{figure}

\begin{figure}[!ht]
    \centering
    \includegraphics[width=.875\linewidth]{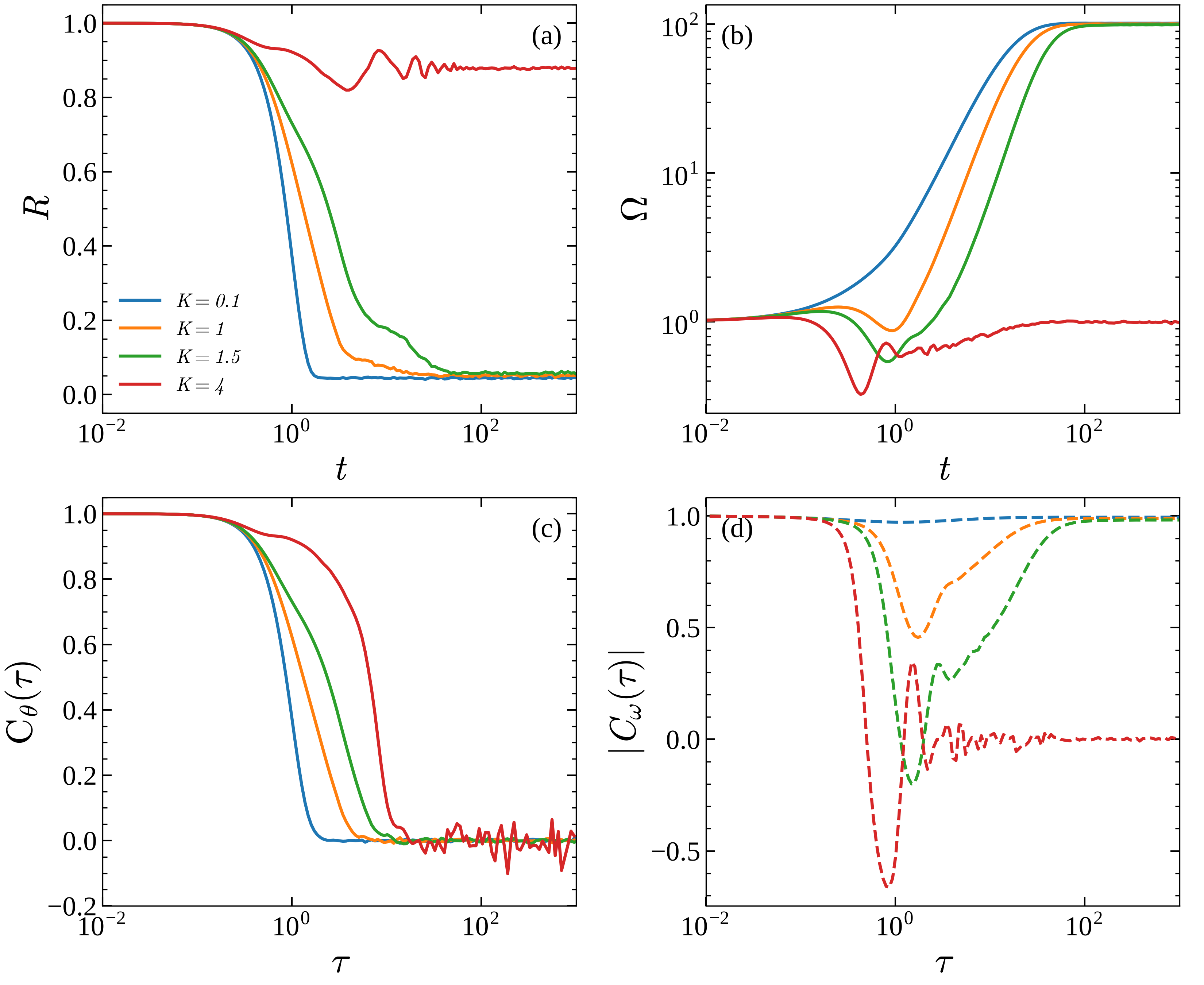}
    \caption{$K$ dependence of KM solutions on periodic $20\times20$ lattice, $m=1$, $\alpha=0.1$, $D=0.1$, averaged over $200$ realizations of simulation for random self-frequencies. (a) Order parameter $R$ (b) frequency variance $\Omega$ (c) phase autocorrelation (d) frequency autocorrelation  $C_\omega$ (dash-dotted) are plotted, where different colors correspond to different $K$-s shown in the legend of (a). }
    \label{fig:l2d20_D0p1}
\end{figure}

To test how additive noise shifts the synchronization transition at the damping used for HU387, $\alpha=0.1$, Figs.~\ref{fig:l2d20_D0}--\ref{fig:l2d20_D0p1}
).
\noindent show the relaxation of a periodic $20\times20$ lattice ($N=400$) with $m=1$ for $D=0$, $10^{-3}$ and $10^{-1}$. The natural frequencies $P_i$ are Gaussian with zero mean and unit variance; all oscillators start with equal phase value and the frequencies start with $\dot\theta_i(0)=P_i$. The quantities are measured by taking the average over $200$ random realizations with independently drawn self-frequencies. These parameters are selected to match the case for HU387. 

Without noise (Fig. \ref{fig:l2d20_D0}), the order parameter values grow gradually with the coupling in the steady state; in (a) the R values increase with the increase in the coupling strength, from $R\approx0.06$ at $K\leq0.2$ to $0.13$ at $K=1$, $0.45$ at $K=1.5$ and to $0.8$ at $K=2$. This identifies the critical coupling for phase synchronization near $K=1.5$. On the other hand, the frequency variance decreases correspondingly from $\Omega\approx100$, the variance $1/\alpha^2$ of the uncoupled frequencies $P_i/\alpha$, i.e.\ free-running nodes, to $\Omega\approx2$ at $K=2$.
The $C_\theta(t)$ curves reflect how phases are correlated with the increase in coupling strength. At the synchronized state, it starts fluctuating around $0$. Besides, the frequency autocorrelation decreases with an increase in the $K$ value. Although due to finite size effect it does not follow the power-law decay, rather it saturates to some finite value.

From Fig. \ref{fig:l2d20_D0p001}, it is apparent that at $D=10^{-3}$, the steady-state values of $R$ and $\Omega$ coincide with those at $D=0$; noise of this size changes the relaxation, not the synchronized fraction. But a further increase in the noise strength affects the synchronization behavior of the system. At $D=0.1$, the lattice stays incoherent at $K=1$ and $1.5$ ($R\approx0.05$, $\Omega\approx100$), and phase order appears only at high coupling values, $K=4$ ($R\approx0.88$). Noise of this strength thus shifts the onset of phase synchronization from $K\approx1.5$ to above $K=1.5$, consistent with the desynchronization of HU387 at $D=0.1$, Fig.~\ref{fig:HU_sc9_freq_ts}. 
But the small sizes and $\alpha$-s make it impossible to see $\Omega(t)$-s ACF-s with PL tails, even though we
know their existence at larges sizes.

	\clearpage
	\setcounter{section}{0}
	\setcounter{figure}{0}
	\setcounter{table}{0}
	\setcounter{equation}{0}
	
	\renewcommand{\thesection}{S\arabic{section}}
	\renewcommand{\thesubsection}{\thesection.\arabic{subsection}}
	\renewcommand{\thefigure}{S\arabic{figure}}
	\renewcommand{\thetable}{S\arabic{table}}
	\renewcommand{\theequation}{S\arabic{equation}}
	
	\setcounter{affn}{0}
\maketitle

\section{Autocorrelation estimator for empirical, stationary time series}
\label{supp:acf}

For a mean-centred discrete time series $x_t$ assuming stationary state, 
the normalized autocorrelation at lag $\ell$ is estimated as

\begin{equation}
\label{eq:supp_acf}
    C(\ell)
    =
    \frac{1}{(M-\ell)\sigma^2}
    \sum_{t=1}^{M-\ell}x_t x_{t+\ell},
\end{equation}
where $M$ is the number of samples and $\sigma^2$ is the variance of the analysed segment. For a uniformly sampled record with sampling interval $\Delta t$, the physical lag is $\tau=\ell\Delta t$.

The same estimator is applied to the increment series $\Delta f_\tau(t)$. For uniformly sampled data, the correlations are evaluated with an FFT-based implementation. The characteristic fast time scale $d$ used in the superstatistical analysis is the smallest positive lag at which the increment autocorrelation falls below $1/e$.

For nonuniformly sampled records, the FFT-based estimator is not used. Such records are analysed with an autocorrelation estimator appropriate to their actual time stamps.

\section{Two-time correlation and nonstationarity}
\label{supp:twotime}

To retain the dependence on the reference time, we compute a two-time correlation using pairs of finite windows. For a reference time $t_0$, a window of duration $T_{\rm c}$ is
\begin{equation}
    x_{\mathrm{ref}}(t_0)
    =
    \{x(t)\}_{t=t_0}^{t_0+T_{\rm c}},
\end{equation}
with local mean $\mu_{\mathrm{ref}}$ and standard deviation $\sigma_{\mathrm{ref}}$. For a lag $\tau$, the shifted window is
\begin{equation}
    x_{\mathrm{shift}}(t_0,\tau)
    =
    \{x(t)\}_{t=t_0+\tau}^{t_0+\tau+T_{\rm c}},
\end{equation}
with corresponding mean $\mu_{\mathrm{shift}}$ and standard deviation $\sigma_{\mathrm{shift}}$. The normalized two-time correlation is
\begin{equation}
\label{eq:twotime}
    C(t_0,\tau)
    =
    \frac{
    \left\langle
    (x_{\mathrm{ref}}-\mu_{\mathrm{ref}})
    (x_{\mathrm{shift}}-\mu_{\mathrm{shift}})
    \right\rangle
    }{
    \sigma_{\mathrm{ref}}\sigma_{\mathrm{shift}}
    }
\end{equation}

Windows for which $t_0+\tau+T_{\rm c}$ exceeds the available record are excluded, as are windows with vanishing variance. The resulting $C(t_0,\tau)$ retains both the lag dependence and slow temporal modulation of the local correlation structure.

\section{Power Spectral Density $S_{x}(\nu)$}

In the case of the lattice and HU387 networks, the macroscopic time series $x(t_n)$ at uniform time steps $t_n$ is obtained by averaging over the $N$ spatial nodes (Eq.~\ref{eqn:average}). Here, $x(t_n)$ may refer to the empirical frequency $f(t)$ or the KM-generated frequency $\omega(t)$.
\begin{equation}
\label{eqn:average}
    x(t_n) = \frac{1}{N} \sum_{i=1}^{N} x_i(t_n).
\end{equation}

To analyze the frequency scaling of the system, initial transients are removed by filtering out data points for $t \le 1000.0$.

The Power Spectral Density (PSD), $S_x(\nu)$, as a function of frequency $\nu$ is defined via the Fourier transform of the signal $x(t)$ over a finite time window $T$:
\begin{equation}
    S_x(\nu) = \lim_{T \to \infty} \frac{1}{T} \left| \int_{-T/2}^{T/2} x(t) e^{-i 2\pi \nu t} \, dt \right|^2,
\end{equation}
which in practice is estimated using Welch's method on the discrete time series.

\section{Superstatistical coarse graining}
\label{supp:superstats}

For a chosen increment lag $\tau$, the record is divided into non-overlapping windows of duration $T$. In each window the local inverse variance is estimated as
\begin{equation}
    \beta
    =
    \left[\operatorname{Var}_{T}(\Delta f_\tau)\right]^{-1}.
\end{equation}
The superstatistical approximation assumes that the increment distribution is approximately stationary within a window while the variance changes on longer time scales.

The coarse-graining duration $T$ is selected using the local-kurtosis criterion: $T$ is chosen from the scale dependence of the mean kurtosis of the increment distribution evaluated within windows, using the scale at which the local increment statistics are closest to Gaussian.

For a centred Gaussian conditional distribution with inverse variance $\beta$,

\begin{equation}
    p(\Delta f_\tau\mid\beta)
    =
    \sqrt{\frac{\beta}{2\pi}}
    \exp\!\left[-\frac{\beta(\Delta f_\tau)^2}{2}\right],
\end{equation}
the predicted marginal distribution is
\begin{equation}
    p(\Delta f_\tau)
    =
    \int_0^\infty
    p(\Delta f_\tau\mid\beta)\,
    g(\beta)\,\mathrm{d}\beta.
\end{equation}
We evaluate this expression using both the empirical distribution of $\beta$ and parametric models of $g(\beta)$.

The fast scale $d$ is defined independently as the first $1/e$ crossing of the autocorrelation of $\Delta f_\tau$. The ratio $T/d$ is therefore used as a diagnostic of the separation between fast increment dynamics and slow variance modulation. The autocorrelation of $\beta(t)$ is measured separately and is not used to define $T$.

\section{Candidate superstatistical kernels}
\label{supp:kernels}

The following positive distributions are used as candidate models for $g(\beta)$: Gamma, inverse-Gamma, lognormal, $F$, and Gaussian truncated to $\beta>0$. Their standard probability-density functions are used with the parameterization implemented in the analysis code.

The Gamma candidate corresponds to $\chi^2$ superstatistics; when inserted into the Gaussian superstatistical mixture it yields a $q$-Gaussian marginal. The other kernels are included as alternative positive distributions with different tail behaviour. We do not infer a microscopic mechanism solely from the identity of the best-fitting kernel.


\section{Distribution fitting and model comparison}
\label{supp:modelsel}

Candidate distributions are fitted to the empirical binned probability densities by nonlinear least squares using the SciPy \texttt{curve\_fit} routine. For a candidate density $p(x;\boldsymbol{\phi})$ with parameter vector $\boldsymbol{\phi}$ and empirical bin values $y_j$, the fitted parameters minimize
\begin{equation}
    \mathrm{RSS}
    =
    \sum_{j=1}^{n_{\rm b}}
    \left[
        y_j-p(x_j;\boldsymbol{\phi})
    \right]^2.
\end{equation}

The cumulative-distribution agreement is quantified by the Kolmogorov--Smirnov distance
\begin{equation}
    D_{\mathrm{KS}}
    =
    \sup_x
    \left|
    F_n(x)-F(x;\hat{\boldsymbol{\phi}})
    \right|.
\end{equation}
Because the model parameters are estimated from the same data, the standard distribution-free Kolmogorov--Smirnov significance levels are not used; $D_{\mathrm{KS}}$ is treated only as a relative measure of agreement.

Under the approximation of independent Gaussian residuals in the fitted binned density, model comparison is additionally summarized by
\begin{equation}
    \mathrm{AIC}
    =
    n_{\rm b}\ln\left(\frac{\mathrm{RSS}}{n_{\rm b}}\right)+2n_{\rm p},
\end{equation}
and
\begin{equation}
    \mathrm{BIC}
    =
    n_{\rm b}\ln\left(\frac{\mathrm{RSS}}{n_{\rm b}}\right)+n_{\rm p}\ln n_{\rm b},
\end{equation}
where $n_{\rm p}$ is the number of fitted parameters and $n_{\rm b}$ is the number of fitted histogram points. Lower values indicate better agreement within this least-squares comparison. When KS, AIC and BIC do not identify the same candidate, the disagreement is reported rather than selecting a model from a single criterion.

Distributional fits are restricted to histogram bins containing at least 100 samples. This prevents very sparsely sampled extreme bins from dominating the comparison.

\section{Further technical details of the HU387 KM simulations}
\label{supp:hu387_runs_more}

\textit{Integration.} All runs, noisy and noiseless, are integrated with the fixed-step symplectic Euler--Maruyama scheme
\begin{equation}
\label{eq:supp_em}
    \dot\theta_i \leftarrow \dot\theta_i+\delta t\,F_i(\boldsymbol{\theta},\dot{\boldsymbol{\theta}})+\sqrt{2D\,\delta t}\;\eta_i,
    \qquad
    \theta_i \leftarrow \theta_i+\delta t\,\dot\theta_i ,
\end{equation}
where $F_i=P_i-\alpha\dot\theta_i+K\sum_jW_{ij}\sin(\theta_j-\theta_i)$, the phase is advanced with the updated frequency, and the $\eta_i\sim\mathcal N(0,1)$ are independent for every node and step. This realizes $\langle\xi_i(t)\xi_j(t')\rangle=2D\delta_{ij}\delta(t-t')$, with $D$ in $\mathrm{rad^2\,s^{-3}}$. In the noiseless runs, $D=0$, and the step reduces to the symplectic Euler method. An adaptive Runge--Kutta solver is not suited to the noisy equation, as it would redraw the noise at each internal sub-step and scale it with $\delta t$ instead of $\sqrt{\delta t}$. The fixed-step scheme is, moreover, more than $30$ times faster than the adaptive Runge--Kutta--Fehlberg (RKF45) solver of the same code.

\textit{Step size.} The step has to resolve the fastest swing mode of the network, $\Omega_{\max}=\sqrt{K\lambda_{\max}}$, where $\lambda_{\max}=1123$ is the largest eigenvalue of the weighted network Laplacian; at $K=1$, $\Omega_{\max}/2\pi=5.3\,\mathrm{Hz}$. The scheme is stable for $\Omega_{\max}\delta t<2$, i.e.\ for $\delta t<0.060\,\mathrm{s}$ at $K=1$ and $\delta t<0.042\,\mathrm{s}$ at $K=2$, the largest coupling used, and it reproduces the mode frequencies with a relative error of about $(\Omega_{\max}\delta t)^2/24$. With $\delta t=0.02\,\mathrm{s}$ this error is below $2\%$ for $K\leq1$. Reducing the step to $0.005\,\mathrm{s}$ changes a noiseless trajectory at $K=1$ by $1.4\%$ of its standard deviation over $2\times10^4\,\mathrm{s}$ and leaves its spectral lines unchanged.

\textit{Comparison with RKF45.} The scheme was compared with RKF45 (relative tolerance $10^{-12}$) on the noiseless equations, with the same network, dispatch, seeds and initial state. Over the first $\sim20\,\mathrm{s}$, single trajectories agree to $0.3$--$6\%$ of the frequency standard deviation at $K=0.1$--$1$, and the difference shrinks in proportion to $\delta t$. At $K\leq0.5$ the trajectories then separate through the chaotic divergence of the weakly coupled dynamics, as those of any two integrators do, so that only statistics can be compared beyond this point. For $20$ realizations at each of $K=0.1$, $0.3$, $0.5$ and $1$, integrated for $5000\,\mathrm{s}$, the late-time frequency variance $\Omega$ and the plateau of $\lvert C_\omega\rvert$ agree within $1\sigma$, and the widths of the $0.1\,\mathrm{s}$ and $1\,\mathrm{s}$ frequency increments within $1.3\sigma$. The node-averaged spectral power in the bands $0.01$--$0.1$, $0.1$--$1$ and $1$--$5\,\mathrm{Hz}$ agrees within $5\%$, apart from the lowest band at $K=0.1$, where it differs by $11(11)\%$, and at $K=1$. The mean number of nodes outside the main cluster is also the same. The remaining differences come from bistable nodes that end on the other branch in single realizations, in at most $0.2\%$ of the node realizations and in either direction; at $K=1$ such a node also accounts for the difference in the very weak low-frequency power of the locked network. The fixed-step scheme thus reproduces the statistics of RKF45, the solver of our earlier deterministic simulations.

\textit{Output.} Records are either logarithmically spaced in time or uniformly sampled. Logarithmically spaced records store all nodes and are used only for quantities taken across nodes at fixed times: $C_\omega$, with $t_0$ the first stored snapshot, and the late-time frequencies $\bar\omega_i$. Uniform records store selected nodes, or all nodes, every $1\,\mathrm{s}$ or $0.1\,\mathrm{s}$, and are used for all temporal statistics. Spectra use the $0.1\,\mathrm{s}$ records, whose Nyquist frequency of $5\,\mathrm{Hz}$ lies above all swing modes at $K\leq0.5$ and all but one at $K=1$; records sampled every $1\,\mathrm{s}$ would fold $20$--$32\%$ of the modes into the band below $0.5\,\mathrm{Hz}$. Runs last between $5000\,\mathrm{s}$ and $345600\,\mathrm{s}$ ($96\,\mathrm{h}$), with $10$ to $201$ realizations per parameter setting.
	
\end{document}